\pdfoutput=1
\documentclass[%
reprint,
superscriptaddress,
bibnotes,
longbibliography,
amsmath,amssymb,
pra, 
floatfix,
]{revtex4-1}
\usepackage{graphicx}
\usepackage{dcolumn}
\usepackage{bm}
\usepackage{hyperref}

\usepackage{natbib}
\usepackage{mathrsfs,mathtools,color,wasysym,dsfont,here}
\usepackage{amsthm}
\usepackage{physics}
\usepackage[all]{xy}
\usepackage{tikz}
\usepackage{amsfonts}
\usepackage{array,booktabs}
\usepackage{comment}
\usetikzlibrary{calc}
\usetikzlibrary{arrows}
\usetikzlibrary{decorations.markings,decorations.pathmorphing}
\usepackage{cases}
\usepackage{siunitx}
\usepackage{tablefootnote}

\AtBeginDocument{
\heavyrulewidth=.08em
\lightrulewidth=.05em
\cmidrulewidth=.03em
\belowrulesep=.65ex
\belowbottomsep=0pt
\aboverulesep=.4ex
\abovetopsep=0pt
\cmidrulesep=\doublerulesep
\cmidrulekern=.5em
\defaultaddspace=.5em
}

\newcommand{\Hamzero}{{\mathcal{H}^{(0)}}}
\newcommand{\Hamone}{{\mathcal{H}^{(1)}}}
\newcommand{\Hamtwo}{{\mathcal{H}^{(2)}}}

\begin{document}

\title{Electric field induced thermal Hall effect of triplons in the quantum dimer magnets \textit{X}CuCl$_{3}$ (\textit{X} = Tl,\ K)}

\author{Nanse Esaki}
\email{esaki-nanse0428@g.ecc.u-tokyo.ac.jp}
\affiliation{Department of Physics, Graduate School of Science, The University of Tokyo, 7-3-1 Hongo, Tokyo 113-0033, Japan}
\author{Yutaka Akagi}
\affiliation{Department of Physics, Graduate School of Science, The University of Tokyo, 7-3-1 Hongo, Tokyo 113-0033, Japan}
\author{Hosho Katsura}
\affiliation{Department of Physics, Graduate School of Science, The University of Tokyo, 7-3-1 Hongo, Tokyo 113-0033, Japan}
\affiliation{Institute for Physics of Intelligence, The University of Tokyo, 7-3-1 Hongo, Tokyo 113-0033, Japan}
\affiliation{Trans-scale Quantum Science Institute, The University of Tokyo, 7-3-1, Hongo, Tokyo 113-0033, Japan}
\begin{abstract}
We theoretically propose the electric field induced thermal Hall effect of triplons in the quantum dimer magnets \textit{X}CuCl$_{3}$ (\textit{X} $=$ Tl,\ K), which exhibit ferroelectricity in the Bose-Einstein condensation phase of triplons. The interplay between ferroelectricity and magnetism in these materials leads to the magnetoelectric effect, 
i.e., an electric-field induced Dzyaloshinskii-Moriya (DM) interaction between spins on the same dimer. We argue that this intra-dimer DM interaction breaks the symmetry of the system in the absence of an electric field and gives rise to the thermal Hall effect, which can be detected in experimentally accessible electric and magnetic fields. We also show that the thermal Hall effect can be controlled by changing the strength or direction of the electric field. 
\end{abstract}

\maketitle

{\it Introduction.}--- Quantum spin systems exhibit a variety of interesting properties that are not present in their classical counterparts. Quantum dimer magnets are such examples, where the neighboring $S = 1/2$ spins form dimers with a spin-singlet 
ground state and triplet 
bosonic excitations called triplons. The triplons undergo Bose-Einstein condensation (BEC) when the magnetic field exceeds a critical value \cite{Nikuni2000, Tanaka2001, Matsumoto2002, Oosawa2002, Matsumoto2004, yamada2007, Giamarchi2008, cavadini2001, cavadini2002, ruegg2003}. In the BEC phase, the ground state of an individual dimer is a coherent superposition of the singlet and triplet states, which breaks 
the inversion symmetry and can lead to the spontaneous polarization on dimers.
In particular, \textit{X}CuCl$_{3}$ (\textit{X} $=$ Tl,\ K) is known to exhibit ferroelectricity in the BEC phase, whereas these materials have inversion centers at the center of dimers in the weak magnetic field regime \cite{Kimura2016, Kimura2017, Kimura2018, Kimura2020}. When one applies an electric field in the BEC phase, the spin-dependent polarization can couple with the electric field, inducing the intradimer Dzyaloshinskii-Moriya (DM) interaction \cite{ramesh2007multiferroics, cheong2007multiferroics, bibes2008towards, liu2011electric, tokura2014multiferroics, yang2018controlling, zhang2018electrical, srivastava2018large, rana2019, li2021, mankovsky2021electric, huang2021tuning, richter2022}. It is thus natural to ask whether the transport of triplons in these materials can be significantly affected or controlled by the electric field.

Various transverse transport phenomena associated with the Berry curvature have been proposed for bosonic excitations such as magnons \cite{Katsura2010, Onose2010, Matsumoto2011, Ideue2012, Shindou2013, Zhang2013, Matsumoto2014, Mook2014edge, Mook2014, hirschberger2015, chisnell2015, Xu2016, cheng2016, zyuzin2016, kim2016realization, shiomi2017, laurell2017, nakata2017magnonic, nakata2017, owerre2017, murakami2017thermal, laurell2018magnon, cookmeyer2018, doki2018, mcclarty2018, joshi2018, zyuzin2018, Mook2019, kondoz2, Kawano2019, kondo3D, Kim2019, hwang2020, Akagi2020, kondo2020, nakata2021, kondodirac, Fujiwara2022, kim2022, mcclarty2022, Neumann2022, go2023, zhuo2023topological, zhang2023}, photons \cite{Raghu2008, Petrescu2012, Rechtsman2013, Hafezi2013, Ben-Abdallah2016}, phonons \cite{Strohm2005, Sheng2006, Kagan2008, Zhang2010, Zhang2011, Qin2012}, and triplons \cite{Romhanyi2015, Malki2017, McClarty2017, joshi2019, Sun2021, Bhowmick2021, Suetsugu2022, thomasen2021}. Of particular interest is the thermal Hall effect of magnons induced by the DM interactions that has been observed experimentally \cite{Onose2010, Ideue2012}.
By contrast, the thermal Hall effect of triplons has yet to be detected experimentally \cite{Suetsugu2022}, despite the theoretical prediction for SrCu$_{2}$(BO$_{3}$)$_{2}$ \cite{Romhanyi2015, Malki2017, Sun2021, Bhowmick2021}.

In this Letter, we propose the electric field induced thermal Hall effect of triplons in 
\textit{X}CuCl$_{3}$. The magnetic properties of these materials are well described by the isotropic Heisenberg Hamiltonian with DM interactions \cite{Matsumoto2002, Oosawa2002, saha2002comparative, Matsumoto2004}. In the absence of an electric field, the system possesses an effective PT symmetry and does not exhibit the thermal Hall effect. We find that an electric field induces intradimer DM interactions breaking this symmetry, thereby leading to the thermal Hall effect.
We also show that the magnitude (direction) of the thermal Hall current can be controlled by manipulating the strength (direction) of the electric field. Our numerical results for TlCuCl$_{3}$ suggest that the thermal Hall effect in \textit{X}CuCl$_{3}$ can be observed in experimentally attainable electric and magnetic fields.

\medskip

\begin{figure}[H]
 \centering
  \includegraphics[width=1.0\linewidth]{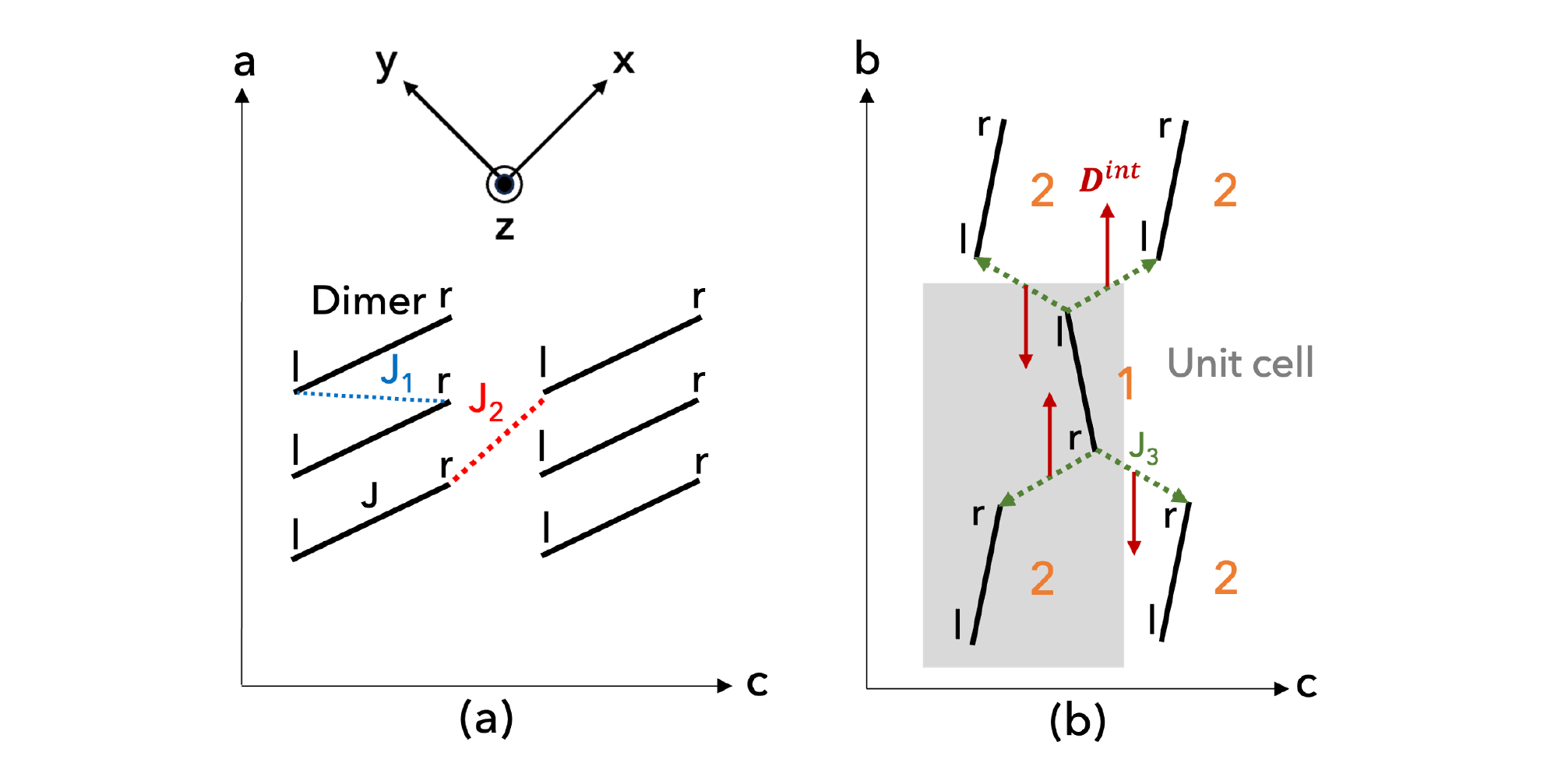}
  \caption{A schematic picture of dimers and relevant interactions in $X$CuCl$_{3}$: (a) $a$-$c$ plane; (b) $b$-$c$ plane. The symbols $l$ and $r$ denote the left and right spins of each dimer. Thick black lines indicate the intradimer exchange coupling $J$, whereas the dotted blue, red, and green lines denote interdimer exchange couplings $J_1$, $J_2$, and $J_3$, respectively. In (a), $x$, $y$, and $z$ axes are indicated. In (b), $1$ and $2$ are the sublattice indices. The solid brown (dotted green) arrows represent the direction of (sign convention for) $\bm{D}^{\mathrm{int}}$.}
  \label{fig:structure}
\end{figure}

{\it The model.}--- \textit{X}CuCl$_{3}$ is a three-dimensional interacting dimer system where the $S=1/2$ spins of Cu$^{2+}$ ions form dimers due to the strong intradimer interactions \cite{Matsumoto2002, Oosawa2002, saha2002comparative, Matsumoto2004} [see Fig. \ref{fig:structure} and Supplemental Material \footnotemark[1]]. The unit cell contains two equivalent dimers, which belong to two different sublattices labeled as $1$ and $2$ in the following. The spin-$1/2$ operators $\bm{S}^{m}_{l}(\bm{R})$ and $\bm{S}^{m}_{r}(\bm{R})$ denote the left and right spins of the dimer in the unit cell at the position $\bm{R}$ on the sublattice $m(=1,2)$, respectively. The lattice unit vectors $\hat{a}$, $\hat{b}$, and $\hat{c}$ correspond to the $a$, $b$, and $c$ axes, respectively. The Hamiltonian of the system in a magnetic field $\bm{H}\parallel b$ and an electric field $\bm{E}$ is given by
\begin{align}\label{hamiltonian}
    \mathcal{H} & = \mathcal{H}_{\rm Hei} + \mathcal{H}_{\rm DM} + \mathcal{H}_{\rm ext}, \\
    \label{HHei}
   \mathcal{H}_{\rm Hei} &\!=\!\frac{1}{2}\!\sum_{\bm{R}, \bm{R}'} \sum_{\alpha,\beta} \sum_{m, n}
    J^{mn}_{\alpha \beta} (\bm{R}'\!-\!\bm{R}) \bm{S}^m_\alpha (\bm{R}) \cdot \bm{S}^n_\beta (\bm{R}'), \\
    \label{HDM}
    \mathcal{H}_{\rm DM} & \!=\! \frac{1}{2}\!\sum_{\bm{R}, \bm{R}'} \sum_{\alpha} \bm{D}^{\mathrm{int}}_\alpha (\bm{R}'\!-\!\bm{R}) \cdot [\bm{S}^1_\alpha (\bm{R}) \times \bm{S}^2_\alpha (\bm{R}')], \\
    \label{Hext}
    \mathcal{H}_{\rm ext} & \!=\! -\sum_{\bm{R}}\sum_{\alpha = l,r} [g\mu_{B}\bm{H}\cdot \bm{S}^{m}_{\alpha}(\bm{R})] + \bm{E}\cdot \bm{P}^{m}(\bm{R}),
\end{align}
where the sums in Eqs. (\ref{HHei}), (\ref{HDM}), and (\ref{Hext}) are taken over $\alpha, \beta = l, r$ and the sublattice indices $m, n = 1,2$ ($m\leq n$).

In the Hamiltonian $\mathcal{H}_{\rm Hei}$, $J^{mm}_{lr} (\bm{0}) = J$ describes the antiferromagnetic intradimer exchange coupling, whereas $J^{mm}_{lr} (\hat{a}) =J_{1}$ and $J^{mm}_{lr}(2\hat{a}+\hat{c}) = J_{2}$ are exchange couplings between the spins belonging to the same sublattices [see Fig. \ref{fig:structure} (a)]. The model also includes Heisenberg interactions between the spins on the diferrent sublattices: $J^{12}_{rr}(\bm{0})=J^{12}_{rr}(2\hat{a}+\hat{c})=J^{12}_{ll}(\hat{b})=J^{12}_{ll}(2\hat{a}+\hat{b}+\hat{c})=J_{3}$ [see Fig. \ref{fig:structure} (b)]. The Hamiltonian $\mathcal{H}_{\rm Hei}$ has been studied as a minimal model of \textit{X}CuCl$_{3}$ \cite{Matsumoto2002, Oosawa2002, Matsumoto2004}. 
The Hamiltonian $\mathcal{H}_{\rm DM}$ in Eq. (\ref{HDM}) describes the symmetry-allowed interdimer DM interactions, where $\bm{D}^{\mathrm{int}}_{r}(\bm{0})=\bm{D}^{\mathrm{int}}_{l}(2\hat{a}+\hat{b}+\hat{c})=D^{\mathrm{int}}\hat{b}$ and $\bm{D}^{\mathrm{int}}_{r}(2\hat{a}+\hat{c})=\bm{D}^{\mathrm{int}}_{l}(\hat{b})=-D^{\mathrm{int}}\hat{b}$ are interdimer DM vectors parallel to the $b$ axis [see Fig. \ref{fig:structure} (b)] \footnote{See Supplemental Material for details}. Here, we do not consider the other components of the interdimer DM vectors allowed by crystal symmetry since their contribution to the thermal Hall effect is negligible. The remaining interactions in Eqs. (\ref{HHei}) and (\ref{HDM}) are zero. The experimental values of the above mentioned parameters are listed in Table \ref{interactionvalues}. Equation (\ref{Hext}) describes the Zeeman and polarization terms of \textit{X}CuCl$_{3}$, where $g=2.06$ for $\bm{H}\parallel b$ \cite{yamada2007}, $\mu_{B}$ is the Bohr magneton, 
and $\bm{P}^{m}(\bm{R})$ is the local polarization on each dimer. To simplify the analysis, we use the coordinate system as $x\parallel \hat{a}+\hat{c}/2$, $y\parallel \hat{a}-\hat{c}/2$, $z\parallel \hat{b}$ where $\hat{a}\perp \hat{b}\perp \hat{c}$, $2a\sim c$, and $2\sqrt{2}a\sim b$ hold approximately for these materials \cite{Tanaka2001} [see Fig. \ref{fig:structure} (a)].

The polarization term in Eq. (\ref{Hext}) can be interpreted as the electric field-induced intradimer DM interaction \footnotemark[1]
\begin{equation}
\label{effectiveDM}
-\bm{E}\cdot\bm{P}^{m}(\bm{R}) = \bm{D}^{\mathrm{ext},m}\cdot[\bm{S}^{m}_{l}(\bm{R})\times\bm{S}^{m}_{r}(\bm{R})]. 
\end{equation}
Here the intradimer DM vector $\bm{D}^{\mathrm{ext},m}$ can be written in terms of the polarization tensor of each sublattice $\tilde{C}^{m}$ that has nine independent components $C^{m}_{\mu\nu}$ ($\mu,\nu =x,y,z$) \cite{Kaplan2011, Kimura2017, Kimura2018, Kimura2020}
\begin{align} 
\label{effectiveDMdetails}
    D^{\mathrm{ext},m}_{\nu} = -E_{\mu^{\prime}} C^{m}_{\mu^{\prime}\nu},
\end{align}
where repeated indices are summed over. In the above expression, $D^{\mathrm{ext},m}_{\nu}$ and $E_{\nu}$ ($\nu=x,y,z$) are the $\nu$-component of $\bm{D}^{\mathrm{ext},m}$ and $\bm{E}$. The two tensors $C^{1}_{\mu\nu}$ and $C^{2}_{\mu\nu}$ are related by \footnotemark[1]
\begin{align} \label{C1andC2}
    C^{2}_{\mu\nu} =-\gamma_{\mu\mu^{\prime}}C^{1}_{\mu^{\prime}\nu^{\prime}}\gamma_{\nu\nu^{\prime}},
\end{align}
where $\gamma = \mathrm{diag}(1,1,-1)$. 
The experimental values of $C^{1}_{\mu\nu}$ obtained in the previous studies \cite{Kimura2020} are listed in Table \ref{polarizationtensor}. We ignore the $z$ component of the electric field-induced intradimer DM interaction term in the later analysis \footnote{In the later analysis, we only focus on the lowest triplon mode. Since the $z$-component of the electric field-induced intradimer DM interaction term barely affects the lowest mode, we can ignore the $z$ component in the present analysis.}. In the Supplemental Material \footnotemark[1], we provide a qualitative picture of how the electric field induces the thermal Hall effect in relation to the no-go condition for magnons \cite{Katsura2010, Ideue2012, Kawano2019}.

\begin{table}[H]
    \centering
    \begin{tabular}{ccccccc}\hline\hline
    Parameter & $J$ & $J_{1}$ & $J_{2}$ & $J_{3}$ & $D^{\mathrm{int}}$ \\
    \hline
    Energy & $5.5$ & $0.43$ & $3.16$ & $0.91$ & $-$ \\ \hline\hline
    \end{tabular}
    \caption{Experimental values of the interactions (in \si{meV}) for TlCuCl$_{3}$ \cite{Matsumoto2004}. The value of $D^{\mathrm{int}}$ 
    remains undetermined \cite{Matsumoto2002, Oosawa2002, saha2002comparative, Matsumoto2004}.}
    \label{interactionvalues}
\end{table}

\begin{table}[H]
    \centering
    \begin{tabular}{ccccccccc}\hline\hline
    $C_{\mu\nu}^{1}$ & $C_{xx}^{1}$ & $C_{xy}^{1}$ & $C_{yx}^{1}$ & $C_{yy}^{1}$ & $C_{zx}^{1}$ &$C_{zy}^{1}$ &$C_{zz}^{1}$ \\
    \hline
    Values & $-27.5$ & $-5$ & $-32.5$ & $124.5$ & $-$ & $-$ &$2.5$ \\ \hline\hline
    \end{tabular}
    \caption{Experimental values of the polarization tensor $C^{1}_{\mu\nu}$ (in \si{\mu C/m^{2}}) for TlCuCl$_{3}$ \tablefootnote{The values in Table \ref{polarizationtensor} are taken from \cite{Kimura2020}. Although the values in Table \ref{polarizationtensor} look different from the values in \cite{Kimura2020}, they are consistent if we use the same coordinate system as in \cite{Kimura2020}.}. The values of $C_{zx}^{1}$ and $C_{zy}^{1}$ are undetermined 
    \cite{Kimura2016, Kimura2017, Kimura2020}. See Supplemental Material for details. The values of $C_{xz}^{1}$ and $C_{yz}^{1}$ are not used in our study.}
    \label{polarizationtensor}
\end{table}

{\it Methods.}--- To study the excitation spectrum of the system, we introduce bond operators $s^{m\dag}_{\bm{R}}$ and $t^{m\dag}_{\bm{R}\alpha}$ ($\alpha = +,0,-$) that create the singlet state $\ket{s}^{m}_{\bm{R}}$ and the three triplet states $\ket{t_{\alpha}}^{m}_{\bm{R}}$ out of the vacuum $\ket{0}^{m}_{\bm{R}}$ on each dimer \cite{Sachdev1990, Matsumoto2002, Oosawa2002, Matsumoto2004}: 
\begin{equation}
    \begin{split}
        &\ket{s}^{m}_{\bm{R}} = s^{m\dag}_{\bm{R}}\ket{0}^{m}_{\bm{R}}=\frac{1}{\sqrt{2}}(\ket{\uparrow\downarrow}^{m}_{\bm{R}}-\ket{\downarrow\uparrow}^{m}_{\bm{R}}),\\
        &\ket{t_{+}}^{m}_{\bm{R}} = t^{m\dag}_{\bm{R}+}\ket{0}^{m}_{\bm{R}}=-\ket{\uparrow\uparrow}^{m}_{\bm{R}},\\ &\ket{t_{0}}^{m}_{\bm{R}} = t^{m\dag}_{\bm{R}0}\ket{0}^{m}_{\bm{R}}=\frac{1}{\sqrt{2}}(\ket{\uparrow\downarrow}^{m}_{\bm{R}}+\ket{\downarrow\uparrow}^{m}_{\bm{R}}),\\ &\ket{t_{-}}^{m}_{\bm{R}}= t^{m\dag}_{\bm{R}-}\ket{0}^{m}_{\bm{R}}=\ket{\downarrow\downarrow}^{m}_{\bm{R}},
    \end{split}\label{bondoperator}
\end{equation}
where $\bm{R}$ and $m$ denote the position of the unit cell and the sublattice index. These obey Bose statistics and are subject to the constraint $s^{m\dag}_{\bm{R}}s^{m}_{\bm{R}}+\sum_{\alpha=+,0,-} t^{m\dag}_{\bm{R}\alpha}t^{m}_{\bm{R}\alpha}=1$ on each dimer. In the BEC phase, the ground state is well represented by a coherent superposition of the singlet and triplet states on each dimer \cite{Matsumoto2004, Kimura2016} 
\begin{equation}
\label{variationalGS}
\ket{\mathrm{GS}}^{m}_{\bm{R}}=\cos{\theta_{m}}\ket{s}^{m}_{\bm{R}}+
    \sin{\theta_{m}}\exp(i\phi_{m})\ket{t_{+}}^{m}_{\bm{R}},
\end{equation}
where $\theta_{m}$ and $\phi_{m}$ are variational parameters for each sublattice $m$. In Eq. (\ref{variationalGS}) and the following analysis, we focus on the high magnetic field regimes $H\geq\SI{40}{T}$ for \textit{X} = Tl and $H\geq\SI{25}{T}$ for \textit{X} = K in \textit{X}CuCl$_{3}$, where the contribution of the other two triplet modes to the ground state (\ref{variationalGS}) can be neglected \cite{Matsumoto2004}.

To analyze the excited states, we perform the following unitary transformation
\begin{gather}
    a^{m\dag}_{\bm{R}} = \cos{\theta_{m}}s^{m\dag}_{\bm{R}}+\sin{\theta_{m}}\exp(i\phi_{m})t^{m\dag}_{\bm{R}+},\nonumber\\
    b^{m\dag}_{\bm{R}+} = -\sin{\theta_{m}}s^{m\dag}_{\bm{R}}+\cos{\theta_{m}}\exp(i\phi_{m}) t^{m\dag}_{\bm{R}+}, \nonumber\\
    b^{m\dag}_{\bm{R}0}=t^{m\dag}_{\bm{R}0},\nonumber\\
    b^{m\dag}_{\bm{R}-}=t^{m\dag}_{\bm{R}-},
\label{unitary transform}
\end{gather}
which preserves the particle number constraint, i.e., $a^{m\dag}_{\bm{R}}a^{m}_{\bm{R}}+\sum_{\alpha=+,0,-} b^{m\dag}_{\bm{R}\alpha}b^{m}_{\bm{R}\alpha}=1$. We follow the standard procedure and replace $a^{m\dag}_{\bm{R}}a^{m}_{\bm{R}}$ with $1-(1/N) \sum_{\bm{R},\alpha}b^{m\dag}_{\bm{R},\alpha}b^{m}_{\bm{R},\alpha}$, where $N$ is the number of dimers on the sublattice $m$. This assumption is justified at low temperatures. 
By introducing the Fourier transform: $b^{m\dag}_{\bm{R}\alpha}=1/N\sum_{\bm{k}\alpha}b^{m\dag}_{\bm{k}\alpha}e^{i\bm{k}\cdot\bm{R}^{m}}$
($\bm{R}^{1}=\bm{R}$, and $\bm{R}^{2}=\bm{R}-(\hat{a}+\hat{b}/2+\hat{c}/2)$)
and retaining only up to quadratic order in $b^{\dag m}_{\bm{k}\alpha}$ and $b^{m}_{\bm{k}\alpha}$, the Hamiltonian (\ref{hamiltonian}) takes the form $\mathcal{H}=\Hamzero + \Hamone + \Hamtwo$. Here the constant term $\Hamzero$ represents the energy of the variational ground state and $\Hamone$ ($\Hamtwo$) is the linear (quadratic) term in bosonic operators. The linear term $\Hamone$ vanishes when we choose the parameters $\theta_{m}$, $\phi_{m}$ to minimize $\Hamzero$ \footnotemark[1]. The quadratic term $\Hamtwo$ represents the bosonic Bogoliubov-de Gennes (BdG) Hamiltonian. 
\medskip

{\it Low-energy effective model.}---
Here, we construct the low-energy effective model for ease of analysis. When the magnetic field is strong, the energy of the lowest excitation mode and those of the other two modes are sufficiently separated \footnotemark[1]. For this reason, we consider only the operators $b^{\dag}_{+}$ and $b_{+}$ to discuss the thermal Hall effect in the high magnetic field and low-temperature regimes. As a result, 
we obtain the BdG Hamiltonian of the form
\begin{equation}
\label{BdG}
\Hamtwo \simeq \frac{1}{2}\sum_{\bm{k}}\bm{b}^{\dag}_{\bm{k}} H_{\mathrm{BdG}}(\bm{k})\bm{b}_{\bm{k}},
\end{equation}
with a vector $\bm{b}_{\bm{k}}=(b^{1}_{\bm{k}+},b^{2}_{\bm{k}+},b^{1\dag}_{-\bm{k}+},b^{2\dag}_{-\bm{k}+})^{T}$ and the $4\times4$ matrix
\begin{equation}\label{b2}
H_{\mathrm{BdG}}(\bm{k})=
    \begin{pmatrix}
    \Xi(\bm{k}) & \Pi(\bm{k}) \\
    \Pi^{*}(-\bm{k})&\Xi^{*}(-\bm{k})\\
    \end{pmatrix}.
\end{equation}
The explicit expression of the matrix (\ref{b2}) is given in the Supplemental Material \footnotemark[1].

To preserve the bosonic commutation relations, the BdG Hamiltonian (\ref{b2}) has to be diagonalized using a paraunitary matrix $T(\bm{k})$. The matrix satisfies $T^{\dag}(\bm{k})\Sigma_{z}T(\bm{k})=\Sigma_{z}$, where $\Sigma_{z}=\mathrm{diag}(1,1,-1,-1)$. The BdG Hamiltonian is diagonalized as
\begin{equation}
\label{diagonalization}
\Sigma_{z} H_{\mathrm{BdG}}(\bm{k})T(\bm{k})=T(\bm{k})\Sigma_{z}E(\bm{k}),
\end{equation}
where the diagonal matrix $E(\bm{k})$ takes the form $E(\bm{k})=\mathrm{diag}(E_{1}(\bm{k}), E_{2}(\bm{k}), E_{1}(-\bm{k}), E_{2}(-\bm{k}))$. The positive energies $E_{1}(\bm{k})$ and $E_{2}(\bm{k})$ correspond to the upper and lower particle bands, respectively. In order to calculate the thermal Hall conductivity, it is sufficient to consider these two bands \cite{Shindou2013, Matsumoto2014}.

\medskip

{\it Thermal Hall effect.}---
We here provide the expression of the three-dimensional thermal Hall conductivity in the $z$-$x$ plane \cite{Matsumoto2014}
\begin{equation}
\label{thermalHalleffect}
\kappa_{zx}=-\frac{k_{\rm B}^{2}T}{\hbar}\sum_{n=1}^{2}\int_{\rm BZ} \frac{d^{3}k}{(2\pi)^{3}}\left[c_{2}(\rho(E_{n}(\bm{k})))-\frac{\pi^{2}}{3}\right]\Omega^{y}_{n}(\bm{k}),
\end{equation} 
where $\rho(E_{n}(\bm{k}))=1/(e^{\beta E_{n}(\bm{k})}-1)$ is the Bose distribution function with $\beta$ being the inverse temperature. The explicit form of $c_{2}(\rho)$ is given by $c_{2}(\rho)=(1+\rho)\left(\log\frac{1+\rho}{\rho}\right)^{2}-(\log \rho)^{2}-2\mathrm{Li}_{2}(-\rho)$, where $\mathrm{Li}_{2}(x)$ is the dilogarithm function. The Berry curvature of the $n$th band $\Omega^{y}_{n}(\bm{k})$ is defined as $\Omega^{y}_{n}(\bm{k})=-2\mathrm{Im}\left[\Sigma_{z}\frac{\partial T^{\dag}(\bm{k})}{\partial k_{z}}\Sigma_{z}\frac{\partial T(\bm{k})}{\partial k_{x}}\right]_{nn}$.

\medskip

\begin{figure}[H]
 \centering
  \includegraphics[width=1.0\linewidth]{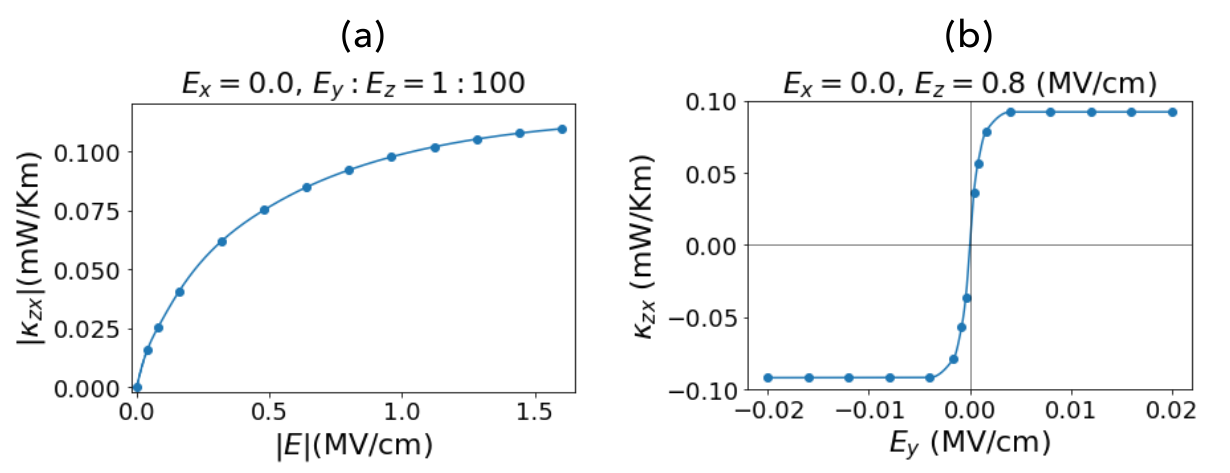}
  \caption{(a) The absolute value of the thermal Hall conductivity $|\kappa_{zx}|$ as a function of $|\bm{E}|$ with $E_{x}=\SI{0.0}{MV/cm}$ and $E_{y}: E_{z}=1:100$. (b) $\kappa_{zx}$ as a function of $E_{y}$ with $E_{x}=\SI{0.0}{MV/cm}$ and $E_{z}=\SI{0.8}{MV/cm}$. The parameters used for (a) and (b) are listed in Table \ref{interactionvalues} and \ref{polarizationtensor}. We set the moderate values for the undetermined parameters as $D^{\mathrm{int}}=\SI{0.091}{meV}$, $C_{zx}^{1} = \SI{-16.3}{\mu C/m^{2}}$, and $C_{zy}^{1}=\SI{62.3}{\mu C/m^{2}}$.
  The magnetic field and temperature are $H=\SI{42}{T}$ and $T=\SI{10}{K}$, respectively.}
  \label{fig:thermal Hall conductivity}
  \end{figure}

{\it Results.}--- Figure \ref{fig:thermal Hall conductivity} shows the numerical results of 
$\kappa_{zx}$ in Eq. (\ref{thermalHalleffect}). In the numerics \footnote{We used the method of Ref. \cite{fukui2005}.}, we set the moderate values of $|\bm{D}^{\mathrm{int}}|$, $C_{zx}^{1}$, and $C_{zy}^{1}$ whose values are unknown [see the caption of Fig. \ref{fig:thermal Hall conductivity}]

From Fig. \ref{fig:thermal Hall conductivity} (a), we find that increasing $|\bm{E}|$ 
leads to an enhancement of $|\kappa_{zx}|$. This behavior is consistent with the approximate expression of $|\kappa_{zx}|$, which is discussed in the next section. The obtained values in Fig. \ref{fig:thermal Hall conductivity} (a) are comparable to the experimental values of the thermal Hall conductivity of magnons and phonons \cite{Strohm2005, Onose2010, Ideue2012, hirschberger2015, doki2018} and thus are expected to be experimentally accessible. In addition, the applied electric field whose strength is of the order of $\SI{0.1}{MV/cm}$ is realizable in experiments \cite{Lottermoser2004}.

Fig. \ref{fig:thermal Hall conductivity} (b) indicates that the sign reversal of $E_{y}$ results in the sign reversal of $\kappa_{zx}$. This result suggests that the direction of the Hall current can be controlled by changing the direction of the electric field in the $x$-$y$ plane. This can be justified as follows. The sign reversal of $E_{x}$ and $E_{y}$ leads to the exchange between $\bm{D}^{\mathrm{ext},1}\leftrightarrow\bm{D}^{\mathrm{ext},2}$, and thus the ground-state wave functions of sublattice $1$ and $2$ [see Eq. (\ref{variationalGS})] are also swapped. Consequently, when we take complex conjugation of the BdG Hamiltonian (\ref{b2}) with the opposite signs of $E_{x}$ and $E_{y}$ and changing the sublattice index as $1\leftrightarrow 2$, the BdG Hamiltonian almost returns to the original one \footnote{The BdG Hamiltonian does not exactly return to
the original one if $\phi_{2}-\phi_{1}\neq\pm\pi$. However, the difference from $\pm\pi$ is usually small enough to be neglected.}. This implies that the sign change of $E_{x}$ and $E_{y}$ approximately corresponds to the following effective time reversal operation \cite{comment1}:
\begin{equation}\label{Eq:Effective_time_reversal}
    H_{\mathrm{BdG}}(\bm{k}) \rightarrow P^{\dag} H^{\ast}_{\mathrm{BdG}} (-\bm{k}) P, \quad P = I_{2\times 2}\otimes\sigma_{x},
\end{equation}
which leads to the reversal of the Hall current as shown in Fig. \ref{fig:thermal Hall conductivity} (b). The constant like behavior for $|E_{y}|\geq \SI{0.004}{MV/cm}$ reflects the fact that the ground state (\ref{variationalGS}) does not change much by varying $|E_{y}|$ due to $|C_{zy}^{1} E_{z}|\gg |C_{yy}^{1} E_{y}|$ \footnote{The ground state (\ref{variationalGS}) drastically changes around $|E_{y}| = 0$, where $\kappa_{zx}$ shows a sharp change.}.

We expect qualitatively similar results for KCuCl$_{3}$. However, it is more difficult to obtain reliable results in the KCuCl$_{3}$ case since there are more undetermined parameters than in the TlCuCl$_{3}$ case.

\medskip

{\it Discussion.}--- 
Here, we explain how the electric field induces and enhances the thermal Hall effect as in Fig. \ref{fig:thermal Hall conductivity} (a). Without an electric field, the difference between the variational parameters $\theta_{1}-\theta_{2}$ and $\phi_{1}-\phi_{2}$ in Eq.~(\ref{variationalGS}) are $0$ and $\pm\pi$, respectively \cite{Matsumoto2002, Oosawa2002, Matsumoto2004}. In this case, the Berry curvature vanishes due to the effective PT symmetry of the BdG Hamiltonian (\ref{b2}). However, the applied electric field gives rise to the difference between $\theta_{1}$ and $\theta_{2}$, which breaks the symmetry, resulting in the finite thermal Hall effect \footnotemark[1].

\begin{figure}[H]
 \centering
  \includegraphics[width=1.0\linewidth]{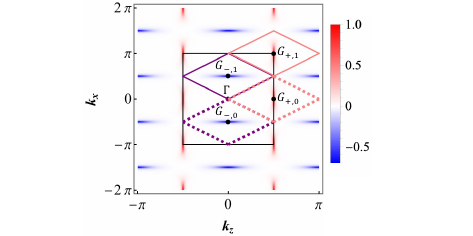}
  \caption{Distribution of the Berry curvature of the upper band ($n=1$) in the $k_{z}$-$k_{x}$ plane with $k_{y}=0$ under the applied electric field $\bm{E}=(0.0,0.016,1.6)~\si{MV/cm}$. The area enclosed by the solid black lines denotes the first Brillouin zone (${\rm BZ}$). The areas enclosed by the pink and purple rhombuses indicate the region $S_{+,j}(0)$ and $S_{-,j}(0)$, respectively (the solid ones for $j=1$ and the dashed ones for $j=0$).}
  \label{fig:Berrycurvature}
\end{figure}

We now argue that the electric field can open and control the band gap. Before applying the electric field, there are nodal lines $G_{\pm,j}$ in 
momentum space: $G_{+,j}=(j\pi,k_{y},\frac{\pi}{2})$, $G_{-,j}=(\frac{2j-1}{2}\pi,k_{y},0)$ (modulo reciprocal lattice vectors) for $j = 0, 1$, which are protected by the effective PT symmetry \footnotemark[1]. However, the applied electric field breaks the symmetry and opens the band gap at $G_{\pm,j}$ \footnotemark[1], each of which is a source of the Berry curvature as shown in Fig. \ref{fig:Berrycurvature} \footnote{As expected from Fig. \ref{fig:Berrycurvature}, the Chern numbers for these two bands are zero. Nontheless, the thermal Hall effect can manifest owing to the bosonic nature, i.e., thermally excited triplons contribute to the thermal Hall conductivity through the Bose distribution function $\rho(E_{n}(\bm{k})) = 1/(e^{\beta E_{n}(\bm{k})}-1)$ as seen in Eq. (\ref{thermalHalleffect}).}


For a fixed direction of the electric field, the band gap at $G_{\pm,j}$, which we denote by $E_{\mathrm{gap},\pm,j}(k_{y})$, behaves as \footnotemark[1] 
\begin{equation}\label{Edependenceofthebandgap}
    E_{\mathrm{gap},\pm,j}(k_{y}) \propto |\bm{E}|.
\end{equation}
Let us show that the electric field can increase $\kappa_{zx}$ by widening the band gap at $G_{\pm,j}$. For a rough estimation, we consider the high-temperature asymptotic form of $\kappa_{zx}$ in the following. In the temperature region $k_{\rm B}T \gg E_{1}(j\pi,k_{y},\frac{\pi}{2}), E_{1}(\frac{2j-1}{2}\pi,k_{y},0)$,
we can use the asymptotic form of $c_{2}(\rho)\sim \frac{\pi^{2}}{3}-\frac{1}{\rho}\ (\rho\rightarrow \infty)$ \cite{Samajdar2019}. By assuming that the Berry curvature is localized at $G_{\pm,j}$ and using $e^{\beta E_{n}(\bm{k})}\simeq 1+\beta E_{n}(\bm{k})$, $\Omega^{y}_{1}(\bm{k})=-\Omega^{y}_{2}(\bm{k})$, and Eq. (\ref{Edependenceofthebandgap}), we have
\begin{equation}
\label{Hightemperaturekappa}
\begin{split}
|\kappa_{zx}|&\simeq\frac{k_{\rm B}^{2}T}{\hbar}\left|\int_{\rm BZ} \frac{d^{3}k}{(2\pi)^{3}}\left[\frac{E_{1}(\bm{k})-E_{2}(\bm{k})}{k_{\rm B}T}\right]\Omega^{y}_{1}(\bm{k})\right|\\
&\simeq\frac{k_{\rm B}}{8\pi^{3}\hbar}\left|\sum_{\sigma,j}\int_{k_{y}}dk_{y} E_{\mathrm{gap},\sigma,j}(k_{y})\int_{S_{\sigma,j}(k_{y})} dk_{z}dk_{x} \Omega^{y}_{1}(\bm{k})\right|\\
&\simeq\frac{k_{\rm B}}{8\pi^{2}\hbar}|E_{\mathrm{gap},+}-E_{\mathrm{gap},-}|\propto |\bm{E}|,
\end{split}
\end{equation}
where the region $S_{\pm,j}(k_{y})$ is the area enclosed by the rhombus around $G_{\pm,j}$ in Fig. (\ref{fig:Berrycurvature}), and we have defined the average band gaps as $E_{\mathrm{gap},\pm} = \frac{1}{2}\sum_{j}\int_{k_y} dk_{y} E_{\mathrm{gap},\pm,j}(k_{y})$. In going from the second to the third line, we used $|\int_{S_{\pm,j}(k_{y})} dk_{z}dk_{x} \Omega^{y}_{1}(\bm{k})| \simeq \frac{\pi}{2}$. 
Clearly, Eq. (\ref{Hightemperaturekappa}) shows that $|\kappa_{zx}|$ increases with increasing electric field. Under $H = \SI{42}{T}$ and $\bm{E} = (0,0.016,1.6)~\si{MV/cm}$, $|\kappa_{zx}|$ in Eq. (\ref{Hightemperaturekappa}) is estimated as $\SI{0.048}{mW/K.m}$, whose order of magnitude is consistent with the numerical result in Fig. \ref{fig:thermal Hall conductivity} (a). 

\medskip

{\it Conclusion and outlook.}--- In this Letter, we have proposed the electric field induced thermal Hall effect of triplons in \textit{X}CuCl$_{3}$. We analyzed the isotropic Heisenberg model with symmetry-allowed interdimer and 
electric field-induced intradimer DM interactions. With this model, we showed that the electric field breaks the effective PT symmetry of the Hamiltonian and thus induces the thermal Hall effect, which can be observed experimentally in realistic electric and magnetic fields. 
Furthermore, we found that the electric field not only triggers the thermal Hall effect but also opens and enlarges the band gap at nodal lines, which are otherwise protected by the effective PT symmetry without an electric field, thereby enhancing the thermal Hall effect. 
We also showed that the sign change of $E_{x}$ and $E_{y}$ corresponds to the effective time reversal operation, which reverses the direction of the Hall current.

We anticipate that our proposal stimulates further experimental investigations and offers an approach to manipulating 
thermal Hall transport. We also expect that our theory should be applicable to a wide class of materials with magnetoelectric coupling. In particular, our approach may prove valuable for lattices whose symmetry properties are heretofore thought to preclude the thermal Hall effect, potentially broadening the research horizon in this field.
Finally, if the pressure induces the intradimer DM interaction \cite{Volkov2018, qi2020tunable, xu2021effect, Xing2022}, it can play the same role as the electric field.
\\


\begin{acknowledgments}

We thank Kosuke Fujiwara, Shojiro Kimura, and Karlo Penc for useful discussions. 
This work was supported by JSPS KAKENHI Grants No. JP18K03445, No. JP20K14411, No. JP23H01086, No. JP24K00546, MEXT KAKENHI Grant-in-Aid for Scientific Research on Innovative Areas “Quantum Liquid Crystals” (KAKENHI Grant No. JP22H04469), and for Transformative Research Areas A “Extreme Universe” (KAKENHI Grant No. JP21H05191). N.E. was supported by Forefront Physics and Mathematics Program to Drive Transformation (FoPM), a World-leading Innovative Graduate Study (WINGS) Program, the University of Tokyo and JSR Fellowship, the University of Tokyo. Y.A. was supported by JST PRESTO Grant No. JPMJPR2251. 

\end{acknowledgments}

\bibliography{Electric_field_induced_thermal_Hall_effect_arxiv_replace.bib}

\begin{thebibliography}{126}%
\makeatletter
\providecommand \@ifxundefined [1]{%
 \@ifx{#1\undefined}
}%
\providecommand \@ifnum [1]{%
 \ifnum #1\expandafter \@firstoftwo
 \else \expandafter \@secondoftwo
 \fi
}%
\providecommand \@ifx [1]{%
 \ifx #1\expandafter \@firstoftwo
 \else \expandafter \@secondoftwo
 \fi
}%
\providecommand \natexlab [1]{#1}%
\providecommand \enquote  [1]{``#1''}%
\providecommand \bibnamefont  [1]{#1}%
\providecommand \bibfnamefont [1]{#1}%
\providecommand \citenamefont [1]{#1}%
\providecommand \href@noop [0]{\@secondoftwo}%
\providecommand \href [0]{\begingroup \@sanitize@url \@href}%
\providecommand \@href[1]{\@@startlink{#1}\@@href}%
\providecommand \@@href[1]{\endgroup#1\@@endlink}%
\providecommand \@sanitize@url [0]{\catcode `\\12\catcode `\$12\catcode
  `\&12\catcode `\#12\catcode `\^12\catcode `\_12\catcode `\%12\relax}%
\providecommand \@@startlink[1]{}%
\providecommand \@@endlink[0]{}%
\providecommand \url  [0]{\begingroup\@sanitize@url \@url }%
\providecommand \@url [1]{\endgroup\@href {#1}{\urlprefix }}%
\providecommand \urlprefix  [0]{URL }%
\providecommand \Eprint [0]{\href }%
\providecommand \doibase [0]{http://dx.doi.org/}%
\providecommand \selectlanguage [0]{\@gobble}%
\providecommand \bibinfo  [0]{\@secondoftwo}%
\providecommand \bibfield  [0]{\@secondoftwo}%
\providecommand \translation [1]{[#1]}%
\providecommand \BibitemOpen [0]{}%
\providecommand \bibitemStop [0]{}%
\providecommand \bibitemNoStop [0]{.\EOS\space}%
\providecommand \EOS [0]{\spacefactor3000\relax}%
\providecommand \BibitemShut  [1]{\csname bibitem#1\endcsname}%
\let\auto@bib@innerbib\@empty
\bibitem [{\citenamefont {Nikuni}\ \emph {et~al.}(2000)\citenamefont {Nikuni},
  \citenamefont {Oshikawa}, \citenamefont {Oosawa},\ and\ \citenamefont
  {Tanaka}}]{Nikuni2000}%
  \BibitemOpen
  \bibfield  {author} {\bibinfo {author} {\bibfnamefont {T}~\bibnamefont
  {Nikuni}}, \bibinfo {author} {\bibfnamefont {M}~\bibnamefont {Oshikawa}},
  \bibinfo {author} {\bibfnamefont {A}~\bibnamefont {Oosawa}}, \ and\ \bibinfo
  {author} {\bibfnamefont {H}~\bibnamefont {Tanaka}},\ }\bibfield  {title}
  {\enquote {\bibinfo {title} {Bose-{E}instein {C}ondensation of {D}ilute
  {M}agnons in {T}l{C}u{C}l{$_{3}$}},}\ }\href@noop {} {\bibfield  {journal}
  {\bibinfo  {journal} {Phys. Rev. Lett.}\ }\textbf {\bibinfo {volume} {84}},\
  \bibinfo {pages} {5868} (\bibinfo {year} {2000})}\BibitemShut {NoStop}%
\bibitem [{\citenamefont {Tanaka}\ \emph {et~al.}(2001)\citenamefont {Tanaka},
  \citenamefont {Oosawa}, \citenamefont {Kato}, \citenamefont {Uekusa},
  \citenamefont {Ohashi}, \citenamefont {Kakurai},\ and\ \citenamefont
  {Hoser}}]{Tanaka2001}%
  \BibitemOpen
  \bibfield  {author} {\bibinfo {author} {\bibfnamefont {Hidekazu}\
  \bibnamefont {Tanaka}}, \bibinfo {author} {\bibfnamefont {Akira}\
  \bibnamefont {Oosawa}}, \bibinfo {author} {\bibfnamefont {Tetsuya}\
  \bibnamefont {Kato}}, \bibinfo {author} {\bibfnamefont {Hidehiro}\
  \bibnamefont {Uekusa}}, \bibinfo {author} {\bibfnamefont {Yuji}\ \bibnamefont
  {Ohashi}}, \bibinfo {author} {\bibfnamefont {Kazuhisa}\ \bibnamefont
  {Kakurai}}, \ and\ \bibinfo {author} {\bibfnamefont {Andreas}\ \bibnamefont
  {Hoser}},\ }\bibfield  {title} {\enquote {\bibinfo {title} {Observation of
  {F}ield-{I}nduced {T}ransverse {N}{\'e}el {O}rdering in the {S}pin {G}ap
  {S}ystem {T}l{C}u{C}l{$_{3}$}},}\ }\href@noop {} {\bibfield  {journal}
  {\bibinfo  {journal} {J. Phys. Soc. Jpn.}\ }\textbf {\bibinfo {volume}
  {70}},\ \bibinfo {pages} {939--942} (\bibinfo {year} {2001})}\BibitemShut
  {NoStop}%
\bibitem [{\citenamefont {Matsumoto}\ \emph {et~al.}(2002)\citenamefont
  {Matsumoto}, \citenamefont {Normand}, \citenamefont {Rice},\ and\
  \citenamefont {Sigrist}}]{Matsumoto2002}%
  \BibitemOpen
  \bibfield  {author} {\bibinfo {author} {\bibfnamefont {Masashige}\
  \bibnamefont {Matsumoto}}, \bibinfo {author} {\bibfnamefont {B}~\bibnamefont
  {Normand}}, \bibinfo {author} {\bibfnamefont {T.~M}\ \bibnamefont {Rice}}, \
  and\ \bibinfo {author} {\bibfnamefont {Manfred}\ \bibnamefont {Sigrist}},\
  }\bibfield  {title} {\enquote {\bibinfo {title} {Magnon {D}ispersion in the
  {F}ield-{I}nduced {M}agnetically {O}rdered {P}hase of
  {T}l{C}u{C}l{$_{3}$}},}\ }\href@noop {} {\bibfield  {journal} {\bibinfo
  {journal} {Phys. Rev. Lett.}\ }\textbf {\bibinfo {volume} {89}},\ \bibinfo
  {pages} {077203} (\bibinfo {year} {2002})}\BibitemShut {NoStop}%
\bibitem [{\citenamefont {Oosawa}\ \emph {et~al.}(2002)\citenamefont {Oosawa},
  \citenamefont {Takamasu}, \citenamefont {Tatani}, \citenamefont {Abe},
  \citenamefont {Tsujii}, \citenamefont {Suzuki}, \citenamefont {Tanaka},
  \citenamefont {Kido},\ and\ \citenamefont {Kindo}}]{Oosawa2002}%
  \BibitemOpen
  \bibfield  {author} {\bibinfo {author} {\bibfnamefont {A}~\bibnamefont
  {Oosawa}}, \bibinfo {author} {\bibfnamefont {T}~\bibnamefont {Takamasu}},
  \bibinfo {author} {\bibfnamefont {K}~\bibnamefont {Tatani}}, \bibinfo
  {author} {\bibfnamefont {H}~\bibnamefont {Abe}}, \bibinfo {author}
  {\bibfnamefont {N}~\bibnamefont {Tsujii}}, \bibinfo {author} {\bibfnamefont
  {O}~\bibnamefont {Suzuki}}, \bibinfo {author} {\bibfnamefont {H}~\bibnamefont
  {Tanaka}}, \bibinfo {author} {\bibfnamefont {G}~\bibnamefont {Kido}}, \ and\
  \bibinfo {author} {\bibfnamefont {K}~\bibnamefont {Kindo}},\ }\bibfield
  {title} {\enquote {\bibinfo {title} {Field-induced magnetic ordering in the
  quantum spin system {K}{C}u{C}l{$_{3}$}},}\ }\href@noop {} {\bibfield
  {journal} {\bibinfo  {journal} {Phys. Rev. B}\ }\textbf {\bibinfo {volume}
  {66}},\ \bibinfo {pages} {104405} (\bibinfo {year} {2002})}\BibitemShut
  {NoStop}%
\bibitem [{\citenamefont {Matsumoto}\ \emph {et~al.}(2004)\citenamefont
  {Matsumoto}, \citenamefont {Normand}, \citenamefont {Rice},\ and\
  \citenamefont {Sigrist}}]{Matsumoto2004}%
  \BibitemOpen
  \bibfield  {author} {\bibinfo {author} {\bibfnamefont {Masashige}\
  \bibnamefont {Matsumoto}}, \bibinfo {author} {\bibfnamefont {B}~\bibnamefont
  {Normand}}, \bibinfo {author} {\bibfnamefont {T.~M}\ \bibnamefont {Rice}}, \
  and\ \bibinfo {author} {\bibfnamefont {Manfred}\ \bibnamefont {Sigrist}},\
  }\bibfield  {title} {\enquote {\bibinfo {title} {Field-and pressure-induced
  magnetic quantum phase transitions in {T}l{C}u{C}l{$_{3}$}},}\ }\href@noop {}
  {\bibfield  {journal} {\bibinfo  {journal} {Phys. Rev. B}\ }\textbf {\bibinfo
  {volume} {69}},\ \bibinfo {pages} {054423} (\bibinfo {year}
  {2004})}\BibitemShut {NoStop}%
\bibitem [{\citenamefont {Yamada}\ \emph {et~al.}(2007)\citenamefont {Yamada},
  \citenamefont {Ono}, \citenamefont {Tanaka}, \citenamefont {Misguich},
  \citenamefont {Oshikawa},\ and\ \citenamefont {Sakakibara}}]{yamada2007}%
  \BibitemOpen
  \bibfield  {author} {\bibinfo {author} {\bibfnamefont {Fumiko}\ \bibnamefont
  {Yamada}}, \bibinfo {author} {\bibfnamefont {Toshio}\ \bibnamefont {Ono}},
  \bibinfo {author} {\bibfnamefont {Hidekazu}\ \bibnamefont {Tanaka}}, \bibinfo
  {author} {\bibfnamefont {Gr{\'e}goire}\ \bibnamefont {Misguich}}, \bibinfo
  {author} {\bibfnamefont {Masaki}\ \bibnamefont {Oshikawa}}, \ and\ \bibinfo
  {author} {\bibfnamefont {Toshiro}\ \bibnamefont {Sakakibara}},\ }\bibfield
  {title} {\enquote {\bibinfo {title} {Magnetic-{F}ield {I}nduced
  {B}ose--{E}instein {C}ondensation of {M}agnons and {C}ritical {B}ehavior in
  {I}nteracting {S}pin {D}imer {S}ystem {T}l{C}u{C}l{$_{3}$}},}\ }\href@noop {}
  {\bibfield  {journal} {\bibinfo  {journal} {J. Phys. Soc. Jpn.}\ }\textbf
  {\bibinfo {volume} {77}},\ \bibinfo {pages} {013701} (\bibinfo {year}
  {2007})}\BibitemShut {NoStop}%
\bibitem [{\citenamefont {Giamarchi}\ \emph {et~al.}(2008)\citenamefont
  {Giamarchi}, \citenamefont {R{\"u}egg},\ and\ \citenamefont
  {Tchernyshyov}}]{Giamarchi2008}%
  \BibitemOpen
  \bibfield  {author} {\bibinfo {author} {\bibfnamefont {Thierry}\ \bibnamefont
  {Giamarchi}}, \bibinfo {author} {\bibfnamefont {Christian}\ \bibnamefont
  {R{\"u}egg}}, \ and\ \bibinfo {author} {\bibfnamefont {Oleg}\ \bibnamefont
  {Tchernyshyov}},\ }\bibfield  {title} {\enquote {\bibinfo {title}
  {Bose-{E}instein condensation in magnetic insulators},}\ }\href@noop {}
  {\bibfield  {journal} {\bibinfo  {journal} {Nat. Phys.}\ }\textbf {\bibinfo
  {volume} {4}},\ \bibinfo {pages} {198--204} (\bibinfo {year}
  {2008})}\BibitemShut {NoStop}%
\bibitem [{\citenamefont {Cavadini}\ \emph {et~al.}(2001)\citenamefont
  {Cavadini}, \citenamefont {Heigold}, \citenamefont {Henggeler}, \citenamefont
  {Furrer}, \citenamefont {G{\"u}del}, \citenamefont {Kr{\"a}mer},\ and\
  \citenamefont {Mutka}}]{cavadini2001}%
  \BibitemOpen
  \bibfield  {author} {\bibinfo {author} {\bibfnamefont {N}~\bibnamefont
  {Cavadini}}, \bibinfo {author} {\bibfnamefont {G}~\bibnamefont {Heigold}},
  \bibinfo {author} {\bibfnamefont {W}~\bibnamefont {Henggeler}}, \bibinfo
  {author} {\bibfnamefont {A}~\bibnamefont {Furrer}}, \bibinfo {author}
  {\bibfnamefont {H-U}\ \bibnamefont {G{\"u}del}}, \bibinfo {author}
  {\bibfnamefont {Karl}\ \bibnamefont {Kr{\"a}mer}}, \ and\ \bibinfo {author}
  {\bibfnamefont {H}~\bibnamefont {Mutka}},\ }\bibfield  {title} {\enquote
  {\bibinfo {title} {Magnetic excitations in the quantum spin system
  {T}l{C}u{C}l{$_{3}$}},}\ }\href@noop {} {\bibfield  {journal} {\bibinfo
  {journal} {Phys. Rev. B}\ }\textbf {\bibinfo {volume} {63}},\ \bibinfo
  {pages} {172414} (\bibinfo {year} {2001})}\BibitemShut {NoStop}%
\bibitem [{\citenamefont {Cavadini}\ \emph {et~al.}(2002)\citenamefont
  {Cavadini}, \citenamefont {R{\"u}egg}, \citenamefont {Furrer}, \citenamefont
  {G{\"u}del}, \citenamefont {Kr{\"a}mer}, \citenamefont {Mutka},\ and\
  \citenamefont {Vorderwisch}}]{cavadini2002}%
  \BibitemOpen
  \bibfield  {author} {\bibinfo {author} {\bibfnamefont {N}~\bibnamefont
  {Cavadini}}, \bibinfo {author} {\bibfnamefont {Ch}~\bibnamefont {R{\"u}egg}},
  \bibinfo {author} {\bibfnamefont {A}~\bibnamefont {Furrer}}, \bibinfo
  {author} {\bibfnamefont {H-U}\ \bibnamefont {G{\"u}del}}, \bibinfo {author}
  {\bibfnamefont {Karl}\ \bibnamefont {Kr{\"a}mer}}, \bibinfo {author}
  {\bibfnamefont {H}~\bibnamefont {Mutka}}, \ and\ \bibinfo {author}
  {\bibfnamefont {P}~\bibnamefont {Vorderwisch}},\ }\bibfield  {title}
  {\enquote {\bibinfo {title} {Triplet excitations in low-{$H_{c}$} spin-gap
  systems {K}{C}u{C}l{$_{3}$} and {T}l{C}u{C}l{$_{3}$}: {A}n inelastic neutron
  scattering study},}\ }\href@noop {} {\bibfield  {journal} {\bibinfo
  {journal} {Phys. Rev. B}\ }\textbf {\bibinfo {volume} {65}},\ \bibinfo
  {pages} {132415} (\bibinfo {year} {2002})}\BibitemShut {NoStop}%
\bibitem [{\citenamefont {R{\"u}egg}\ \emph {et~al.}(2003)\citenamefont
  {R{\"u}egg}, \citenamefont {Cavadini}, \citenamefont {Furrer}, \citenamefont
  {G{\"u}del}, \citenamefont {Kr{\"a}mer}, \citenamefont {Mutka}, \citenamefont
  {Wildes}, \citenamefont {Habicht},\ and\ \citenamefont
  {Vorderwisch}}]{ruegg2003}%
  \BibitemOpen
  \bibfield  {author} {\bibinfo {author} {\bibfnamefont {Ch}~\bibnamefont
  {R{\"u}egg}}, \bibinfo {author} {\bibfnamefont {Nordal}\ \bibnamefont
  {Cavadini}}, \bibinfo {author} {\bibfnamefont {Albert}\ \bibnamefont
  {Furrer}}, \bibinfo {author} {\bibfnamefont {H-U}\ \bibnamefont {G{\"u}del}},
  \bibinfo {author} {\bibfnamefont {Karl}\ \bibnamefont {Kr{\"a}mer}}, \bibinfo
  {author} {\bibfnamefont {Hannu}\ \bibnamefont {Mutka}}, \bibinfo {author}
  {\bibfnamefont {Andrew}\ \bibnamefont {Wildes}}, \bibinfo {author}
  {\bibfnamefont {Klaus}\ \bibnamefont {Habicht}}, \ and\ \bibinfo {author}
  {\bibfnamefont {Peter}\ \bibnamefont {Vorderwisch}},\ }\bibfield  {title}
  {\enquote {\bibinfo {title} {Bose--{E}instein condensation of the triplet
  states in the magnetic insulator {T}l{C}u{C}l{$_{3}$}},}\ }\href@noop {}
  {\bibfield  {journal} {\bibinfo  {journal} {Nature}\ }\textbf {\bibinfo
  {volume} {423}},\ \bibinfo {pages} {62--65} (\bibinfo {year}
  {2003})}\BibitemShut {NoStop}%
\bibitem [{\citenamefont {Kimura}\ \emph {et~al.}(2016)\citenamefont {Kimura},
  \citenamefont {Kakihata}, \citenamefont {Sawada}, \citenamefont {Watanabe},
  \citenamefont {Matsumoto}, \citenamefont {Hagiwara},\ and\ \citenamefont
  {Tanaka}}]{Kimura2016}%
  \BibitemOpen
  \bibfield  {author} {\bibinfo {author} {\bibfnamefont {S}~\bibnamefont
  {Kimura}}, \bibinfo {author} {\bibfnamefont {K}~\bibnamefont {Kakihata}},
  \bibinfo {author} {\bibfnamefont {Y}~\bibnamefont {Sawada}}, \bibinfo
  {author} {\bibfnamefont {K}~\bibnamefont {Watanabe}}, \bibinfo {author}
  {\bibfnamefont {M}~\bibnamefont {Matsumoto}}, \bibinfo {author}
  {\bibfnamefont {M}~\bibnamefont {Hagiwara}}, \ and\ \bibinfo {author}
  {\bibfnamefont {H}~\bibnamefont {Tanaka}},\ }\bibfield  {title} {\enquote
  {\bibinfo {title} {Ferroelectricity by {B}ose-{E}instein condensation in a
  quantum magnet},}\ }\href@noop {} {\bibfield  {journal} {\bibinfo  {journal}
  {Nat. Commun.}\ }\textbf {\bibinfo {volume} {7}},\ \bibinfo {pages} {12822}
  (\bibinfo {year} {2016})}\BibitemShut {NoStop}%
\bibitem [{\citenamefont {Kimura}\ \emph {et~al.}(2017)\citenamefont {Kimura},
  \citenamefont {Kakihata}, \citenamefont {Sawada}, \citenamefont {Watanabe},
  \citenamefont {Matsumoto}, \citenamefont {Hagiwara},\ and\ \citenamefont
  {Tanaka}}]{Kimura2017}%
  \BibitemOpen
  \bibfield  {author} {\bibinfo {author} {\bibfnamefont {Shojiro}\ \bibnamefont
  {Kimura}}, \bibinfo {author} {\bibfnamefont {Kento}\ \bibnamefont
  {Kakihata}}, \bibinfo {author} {\bibfnamefont {Yuya}\ \bibnamefont {Sawada}},
  \bibinfo {author} {\bibfnamefont {Kazuo}\ \bibnamefont {Watanabe}}, \bibinfo
  {author} {\bibfnamefont {Masashige}\ \bibnamefont {Matsumoto}}, \bibinfo
  {author} {\bibfnamefont {Masayuki}\ \bibnamefont {Hagiwara}}, \ and\ \bibinfo
  {author} {\bibfnamefont {Hidekazu}\ \bibnamefont {Tanaka}},\ }\bibfield
  {title} {\enquote {\bibinfo {title} {Magnetoelectric effect in the quantum
  spin gap system {T}l{C}u{C}l{$_{3}$}},}\ }\href@noop {} {\bibfield  {journal}
  {\bibinfo  {journal} {Phys. Rev. B}\ }\textbf {\bibinfo {volume} {95}},\
  \bibinfo {pages} {184420} (\bibinfo {year} {2017})}\BibitemShut {NoStop}%
\bibitem [{\citenamefont {Kimura}\ \emph {et~al.}(2018)\citenamefont {Kimura},
  \citenamefont {Matsumoto}, \citenamefont {Akaki}, \citenamefont {Hagiwara},
  \citenamefont {Kindo},\ and\ \citenamefont {Tanaka}}]{Kimura2018}%
  \BibitemOpen
  \bibfield  {author} {\bibinfo {author} {\bibfnamefont {Shojiro}\ \bibnamefont
  {Kimura}}, \bibinfo {author} {\bibfnamefont {Masashige}\ \bibnamefont
  {Matsumoto}}, \bibinfo {author} {\bibfnamefont {Mitsuru}\ \bibnamefont
  {Akaki}}, \bibinfo {author} {\bibfnamefont {Masayuki}\ \bibnamefont
  {Hagiwara}}, \bibinfo {author} {\bibfnamefont {Koichi}\ \bibnamefont
  {Kindo}}, \ and\ \bibinfo {author} {\bibfnamefont {Hidekazu}\ \bibnamefont
  {Tanaka}},\ }\bibfield  {title} {\enquote {\bibinfo {title} {Electric dipole
  spin resonance in a quantum spin dimer system driven by magnetoelectric
  coupling},}\ }\href@noop {} {\bibfield  {journal} {\bibinfo  {journal} {Phys.
  Rev. B}\ }\textbf {\bibinfo {volume} {97}},\ \bibinfo {pages} {140406(R)}
  (\bibinfo {year} {2018})}\BibitemShut {NoStop}%
\bibitem [{\citenamefont {Kimura}\ \emph {et~al.}(2020)\citenamefont {Kimura},
  \citenamefont {Matsumoto},\ and\ \citenamefont {Tanaka}}]{Kimura2020}%
  \BibitemOpen
  \bibfield  {author} {\bibinfo {author} {\bibfnamefont {Shojiro}\ \bibnamefont
  {Kimura}}, \bibinfo {author} {\bibfnamefont {Masashige}\ \bibnamefont
  {Matsumoto}}, \ and\ \bibinfo {author} {\bibfnamefont {Hidekazu}\
  \bibnamefont {Tanaka}},\ }\bibfield  {title} {\enquote {\bibinfo {title}
  {Electrical {S}witching of the {N}onreciprocal {D}irectional {M}icrowave
  {R}esponse in a {T}riplon {B}ose-{E}instein {C}ondensate},}\ }\href@noop {}
  {\bibfield  {journal} {\bibinfo  {journal} {Phys. Rev. Lett.}\ }\textbf
  {\bibinfo {volume} {124}},\ \bibinfo {pages} {217401} (\bibinfo {year}
  {2020})}\BibitemShut {NoStop}%
\bibitem [{\citenamefont {Ramesh}\ and\ \citenamefont
  {Spaldin}(2007)}]{ramesh2007multiferroics}%
  \BibitemOpen
  \bibfield  {author} {\bibinfo {author} {\bibfnamefont {Ramaroorthy}\
  \bibnamefont {Ramesh}}\ and\ \bibinfo {author} {\bibfnamefont {Nicola~A}\
  \bibnamefont {Spaldin}},\ }\bibfield  {title} {\enquote {\bibinfo {title}
  {Multiferroics: progress and prospects in thin films},}\ }\href@noop {}
  {\bibfield  {journal} {\bibinfo  {journal} {Nat. Mater.}\ }\textbf {\bibinfo
  {volume} {6}},\ \bibinfo {pages} {21--29} (\bibinfo {year}
  {2007})}\BibitemShut {NoStop}%
\bibitem [{\citenamefont {Cheong}\ and\ \citenamefont
  {Mostovoy}(2007)}]{cheong2007multiferroics}%
  \BibitemOpen
  \bibfield  {author} {\bibinfo {author} {\bibfnamefont {Sang-Wook}\
  \bibnamefont {Cheong}}\ and\ \bibinfo {author} {\bibfnamefont {Maxim}\
  \bibnamefont {Mostovoy}},\ }\bibfield  {title} {\enquote {\bibinfo {title}
  {Multiferroics: a magnetic twist for ferroelectricity},}\ }\href@noop {}
  {\bibfield  {journal} {\bibinfo  {journal} {Nat. Mater.}\ }\textbf {\bibinfo
  {volume} {6}},\ \bibinfo {pages} {13--20} (\bibinfo {year}
  {2007})}\BibitemShut {NoStop}%
\bibitem [{\citenamefont {Bibes}\ and\ \citenamefont
  {Barth{\'e}l{\'e}my}(2008)}]{bibes2008towards}%
  \BibitemOpen
  \bibfield  {author} {\bibinfo {author} {\bibfnamefont {Manuel}\ \bibnamefont
  {Bibes}}\ and\ \bibinfo {author} {\bibfnamefont {Agn{\`e}s}\ \bibnamefont
  {Barth{\'e}l{\'e}my}},\ }\bibfield  {title} {\enquote {\bibinfo {title}
  {Towards a magnetoelectric memory},}\ }\href@noop {} {\bibfield  {journal}
  {\bibinfo  {journal} {Nat. Mater.}\ }\textbf {\bibinfo {volume} {7}},\
  \bibinfo {pages} {425--426} (\bibinfo {year} {2008})}\BibitemShut {NoStop}%
\bibitem [{\citenamefont {Liu}\ and\ \citenamefont
  {Vignale}(2011)}]{liu2011electric}%
  \BibitemOpen
  \bibfield  {author} {\bibinfo {author} {\bibfnamefont {Tianyu}\ \bibnamefont
  {Liu}}\ and\ \bibinfo {author} {\bibfnamefont {G}~\bibnamefont {Vignale}},\
  }\bibfield  {title} {\enquote {\bibinfo {title} {Electric {C}ontrol of {S}pin
  {C}urrents and {S}pin-{W}ave {L}ogic},}\ }\href@noop {} {\bibfield  {journal}
  {\bibinfo  {journal} {Phys. Rev. Lett.}\ }\textbf {\bibinfo {volume} {106}},\
  \bibinfo {pages} {247203} (\bibinfo {year} {2011})}\BibitemShut {NoStop}%
\bibitem [{\citenamefont {Tokura}\ \emph {et~al.}(2014)\citenamefont {Tokura},
  \citenamefont {Seki},\ and\ \citenamefont
  {Nagaosa}}]{tokura2014multiferroics}%
  \BibitemOpen
  \bibfield  {author} {\bibinfo {author} {\bibfnamefont {Yoshinori}\
  \bibnamefont {Tokura}}, \bibinfo {author} {\bibfnamefont {Shinichiro}\
  \bibnamefont {Seki}}, \ and\ \bibinfo {author} {\bibfnamefont {Naoto}\
  \bibnamefont {Nagaosa}},\ }\bibfield  {title} {\enquote {\bibinfo {title}
  {Multiferroics of spin origin},}\ }\href@noop {} {\bibfield  {journal}
  {\bibinfo  {journal} {Rep. Prog. Phys.}\ }\textbf {\bibinfo {volume} {77}},\
  \bibinfo {pages} {076501} (\bibinfo {year} {2014})}\BibitemShut {NoStop}%
\bibitem [{\citenamefont {Yang}\ \emph {et~al.}(2018)\citenamefont {Yang},
  \citenamefont {Boulle}, \citenamefont {Cros}, \citenamefont {Fert},\ and\
  \citenamefont {Chshiev}}]{yang2018controlling}%
  \BibitemOpen
  \bibfield  {author} {\bibinfo {author} {\bibfnamefont {Hongxin}\ \bibnamefont
  {Yang}}, \bibinfo {author} {\bibfnamefont {Olivier}\ \bibnamefont {Boulle}},
  \bibinfo {author} {\bibfnamefont {Vincent}\ \bibnamefont {Cros}}, \bibinfo
  {author} {\bibfnamefont {Albert}\ \bibnamefont {Fert}}, \ and\ \bibinfo
  {author} {\bibfnamefont {Mairbek}\ \bibnamefont {Chshiev}},\ }\bibfield
  {title} {\enquote {\bibinfo {title} {Controlling {D}zyaloshinskii-{M}oriya
  {I}nteraction via {C}hirality {D}ependent {A}tomic-{L}ayer {S}tacking,
  {I}nsulator {C}apping and {E}lectric {F}ield},}\ }\href@noop {} {\bibfield
  {journal} {\bibinfo  {journal} {Sci. Rep.}\ }\textbf {\bibinfo {volume}
  {8}},\ \bibinfo {pages} {12356} (\bibinfo {year} {2018})}\BibitemShut
  {NoStop}%
\bibitem [{\citenamefont {Zhang}\ \emph {et~al.}(2018)\citenamefont {Zhang},
  \citenamefont {Zhong}, \citenamefont {Zang}, \citenamefont {Zhang},
  \citenamefont {Yu}, \citenamefont {Han}, \citenamefont {Liu}, \citenamefont
  {Yan}, \citenamefont {Kang},\ and\ \citenamefont
  {Mei}}]{zhang2018electrical}%
  \BibitemOpen
  \bibfield  {author} {\bibinfo {author} {\bibfnamefont {W}~\bibnamefont
  {Zhang}}, \bibinfo {author} {\bibfnamefont {H}~\bibnamefont {Zhong}},
  \bibinfo {author} {\bibfnamefont {R}~\bibnamefont {Zang}}, \bibinfo {author}
  {\bibfnamefont {Y}~\bibnamefont {Zhang}}, \bibinfo {author} {\bibfnamefont
  {S}~\bibnamefont {Yu}}, \bibinfo {author} {\bibfnamefont {G}~\bibnamefont
  {Han}}, \bibinfo {author} {\bibfnamefont {GL}~\bibnamefont {Liu}}, \bibinfo
  {author} {\bibfnamefont {SS}~\bibnamefont {Yan}}, \bibinfo {author}
  {\bibfnamefont {S}~\bibnamefont {Kang}}, \ and\ \bibinfo {author}
  {\bibfnamefont {LM}~\bibnamefont {Mei}},\ }\bibfield  {title} {\enquote
  {\bibinfo {title} {Electrical field enhanced interfacial
  {D}zyaloshinskii-{M}oriya interaction in {M}g{O}/{F}e/{P}t system},}\
  }\href@noop {} {\bibfield  {journal} {\bibinfo  {journal} {App. Phys. Lett.}\
  }\textbf {\bibinfo {volume} {113}} (\bibinfo {year} {2018})}\BibitemShut
  {NoStop}%
\bibitem [{\citenamefont {Srivastava}\ \emph {et~al.}(2018)\citenamefont
  {Srivastava}, \citenamefont {Schott}, \citenamefont {Juge}, \citenamefont
  {Krizakova}, \citenamefont {Belmeguenai}, \citenamefont {Roussign{\'e}},
  \citenamefont {Bernand-Mantel}, \citenamefont {Ranno}, \citenamefont
  {Pizzini}, \citenamefont {Ch{\'e}rif} \emph {et~al.}}]{srivastava2018large}%
  \BibitemOpen
  \bibfield  {author} {\bibinfo {author} {\bibfnamefont {Titiksha}\
  \bibnamefont {Srivastava}}, \bibinfo {author} {\bibfnamefont {Marine}\
  \bibnamefont {Schott}}, \bibinfo {author} {\bibfnamefont {Rom{\'e}o}\
  \bibnamefont {Juge}}, \bibinfo {author} {\bibfnamefont {Viola}\ \bibnamefont
  {Krizakova}}, \bibinfo {author} {\bibfnamefont {Mohamed}\ \bibnamefont
  {Belmeguenai}}, \bibinfo {author} {\bibfnamefont {Yves}\ \bibnamefont
  {Roussign{\'e}}}, \bibinfo {author} {\bibfnamefont {Anne}\ \bibnamefont
  {Bernand-Mantel}}, \bibinfo {author} {\bibfnamefont {Laurent}\ \bibnamefont
  {Ranno}}, \bibinfo {author} {\bibfnamefont {Stefania}\ \bibnamefont
  {Pizzini}}, \bibinfo {author} {\bibfnamefont {Salim-Mourad}\ \bibnamefont
  {Ch{\'e}rif}},  \emph {et~al.},\ }\bibfield  {title} {\enquote {\bibinfo
  {title} {Large-{V}oltage {T}uning of {D}zyaloshinskii--{M}oriya
  {I}nteractions: {A} {R}oute toward {D}ynamic {C}ontrol of {S}kyrmion
  {C}hirality},}\ }\href@noop {} {\bibfield  {journal} {\bibinfo  {journal}
  {Nano Lett.}\ }\textbf {\bibinfo {volume} {18}},\ \bibinfo {pages}
  {4871--4877} (\bibinfo {year} {2018})}\BibitemShut {NoStop}%
\bibitem [{\citenamefont {Rana}\ and\ \citenamefont {Otani}(2019)}]{rana2019}%
  \BibitemOpen
  \bibfield  {author} {\bibinfo {author} {\bibfnamefont {Bivas}\ \bibnamefont
  {Rana}}\ and\ \bibinfo {author} {\bibfnamefont {YoshiChika}\ \bibnamefont
  {Otani}},\ }\bibfield  {title} {\enquote {\bibinfo {title} {Towards magnonic
  devices based on voltage-controlled magnetic anisotropy},}\ }\href@noop {}
  {\bibfield  {journal} {\bibinfo  {journal} {Commun. Phys.}\ }\textbf
  {\bibinfo {volume} {2}},\ \bibinfo {pages} {90} (\bibinfo {year}
  {2019})}\BibitemShut {NoStop}%
\bibitem [{\citenamefont {Li}\ \emph {et~al.}(2021)\citenamefont {Li},
  \citenamefont {Yao},\ and\ \citenamefont {Chen}}]{li2021}%
  \BibitemOpen
  \bibfield  {author} {\bibinfo {author} {\bibfnamefont {Chao-Kai}\
  \bibnamefont {Li}}, \bibinfo {author} {\bibfnamefont {Xu-Ping}\ \bibnamefont
  {Yao}}, \ and\ \bibinfo {author} {\bibfnamefont {Gang}\ \bibnamefont
  {Chen}},\ }\bibfield  {title} {\enquote {\bibinfo {title} {Writing and
  deleting skyrmions with electric fields in a multiferroic heterostructure},}\
  }\href@noop {} {\bibfield  {journal} {\bibinfo  {journal} {Phys. Rev. Res.}\
  }\textbf {\bibinfo {volume} {3}},\ \bibinfo {pages} {L012026} (\bibinfo
  {year} {2021})}\BibitemShut {NoStop}%
\bibitem [{\citenamefont {Mankovsky}\ \emph {et~al.}(2021)\citenamefont
  {Mankovsky}, \citenamefont {Simon}, \citenamefont {Polesya}, \citenamefont
  {Marmodoro},\ and\ \citenamefont {Ebert}}]{mankovsky2021electric}%
  \BibitemOpen
  \bibfield  {author} {\bibinfo {author} {\bibfnamefont {S}~\bibnamefont
  {Mankovsky}}, \bibinfo {author} {\bibfnamefont {E}~\bibnamefont {Simon}},
  \bibinfo {author} {\bibfnamefont {S}~\bibnamefont {Polesya}}, \bibinfo
  {author} {\bibfnamefont {A}~\bibnamefont {Marmodoro}}, \ and\ \bibinfo
  {author} {\bibfnamefont {H}~\bibnamefont {Ebert}},\ }\bibfield  {title}
  {\enquote {\bibinfo {title} {Electric-field control of exchange
  interactions},}\ }\href@noop {} {\bibfield  {journal} {\bibinfo  {journal}
  {Phys. Rev. B}\ }\textbf {\bibinfo {volume} {104}},\ \bibinfo {pages}
  {174443} (\bibinfo {year} {2021})}\BibitemShut {NoStop}%
\bibitem [{\citenamefont {Huang}\ \emph {et~al.}(2021)\citenamefont {Huang},
  \citenamefont {Jiang}, \citenamefont {Zhu}, \citenamefont {Pan},
  \citenamefont {Fan}, \citenamefont {Ma}, \citenamefont {Hu},\ and\
  \citenamefont {Shi}}]{huang2021tuning}%
  \BibitemOpen
  \bibfield  {author} {\bibinfo {author} {\bibfnamefont {C}~\bibnamefont
  {Huang}}, \bibinfo {author} {\bibfnamefont {LZ}~\bibnamefont {Jiang}},
  \bibinfo {author} {\bibfnamefont {Y}~\bibnamefont {Zhu}}, \bibinfo {author}
  {\bibfnamefont {YF}~\bibnamefont {Pan}}, \bibinfo {author} {\bibfnamefont
  {JY}~\bibnamefont {Fan}}, \bibinfo {author} {\bibfnamefont {CL}~\bibnamefont
  {Ma}}, \bibinfo {author} {\bibfnamefont {J}~\bibnamefont {Hu}}, \ and\
  \bibinfo {author} {\bibfnamefont {DN}~\bibnamefont {Shi}},\ }\bibfield
  {title} {\enquote {\bibinfo {title} {Tuning {D}zyaloshinskii--{M}oriya
  interaction via an electric field at the {C}o/h-{B}{N} interface},}\
  }\href@noop {} {\bibfield  {journal} {\bibinfo  {journal} {Phys. Chem. Chem.
  Phys.}\ }\textbf {\bibinfo {volume} {23}},\ \bibinfo {pages} {22246--22250}
  (\bibinfo {year} {2021})}\BibitemShut {NoStop}%
\bibitem [{\citenamefont {Richter}\ \emph {et~al.}(2022)\citenamefont
  {Richter}, \citenamefont {Ohanyan}, \citenamefont {Schulenburg},\ and\
  \citenamefont {Schnack}}]{richter2022}%
  \BibitemOpen
  \bibfield  {author} {\bibinfo {author} {\bibfnamefont {Johannes}\
  \bibnamefont {Richter}}, \bibinfo {author} {\bibfnamefont {Vadim}\
  \bibnamefont {Ohanyan}}, \bibinfo {author} {\bibfnamefont {J{\"o}rg}\
  \bibnamefont {Schulenburg}}, \ and\ \bibinfo {author} {\bibfnamefont
  {J{\"u}rgen}\ \bibnamefont {Schnack}},\ }\bibfield  {title} {\enquote
  {\bibinfo {title} {Electric field driven flat bands: {E}nhanced
  magnetoelectric and electrocaloric effects in frustrated quantum magnets},}\
  }\href@noop {} {\bibfield  {journal} {\bibinfo  {journal} {Phys. Rev. B}\
  }\textbf {\bibinfo {volume} {105}},\ \bibinfo {pages} {054420} (\bibinfo
  {year} {2022})}\BibitemShut {NoStop}%
\bibitem [{\citenamefont {Katsura}\ \emph {et~al.}(2010)\citenamefont
  {Katsura}, \citenamefont {Nagaosa},\ and\ \citenamefont {Lee}}]{Katsura2010}%
  \BibitemOpen
  \bibfield  {author} {\bibinfo {author} {\bibfnamefont {Hosho}\ \bibnamefont
  {Katsura}}, \bibinfo {author} {\bibfnamefont {Naoto}\ \bibnamefont
  {Nagaosa}}, \ and\ \bibinfo {author} {\bibfnamefont {Patrick~A}\ \bibnamefont
  {Lee}},\ }\bibfield  {title} {\enquote {\bibinfo {title} {Theory of the
  {T}hermal {H}all {E}ffect in {Q}uantum {M}agnets},}\ }\href@noop {}
  {\bibfield  {journal} {\bibinfo  {journal} {Phys. Rev. Lett.}\ }\textbf
  {\bibinfo {volume} {104}},\ \bibinfo {pages} {066403} (\bibinfo {year}
  {2010})}\BibitemShut {NoStop}%
\bibitem [{\citenamefont {Onose}\ \emph {et~al.}(2010)\citenamefont {Onose},
  \citenamefont {Ideue}, \citenamefont {Katsura}, \citenamefont {Shiomi},
  \citenamefont {Nagaosa},\ and\ \citenamefont {Tokura}}]{Onose2010}%
  \BibitemOpen
  \bibfield  {author} {\bibinfo {author} {\bibfnamefont {Y}~\bibnamefont
  {Onose}}, \bibinfo {author} {\bibfnamefont {T}~\bibnamefont {Ideue}},
  \bibinfo {author} {\bibfnamefont {H}~\bibnamefont {Katsura}}, \bibinfo
  {author} {\bibfnamefont {Y}~\bibnamefont {Shiomi}}, \bibinfo {author}
  {\bibfnamefont {N}~\bibnamefont {Nagaosa}}, \ and\ \bibinfo {author}
  {\bibfnamefont {Y}~\bibnamefont {Tokura}},\ }\bibfield  {title} {\enquote
  {\bibinfo {title} {Observation of the {M}agnon {H}all {E}ffect},}\
  }\href@noop {} {\bibfield  {journal} {\bibinfo  {journal} {Science}\ }\textbf
  {\bibinfo {volume} {329}},\ \bibinfo {pages} {297--299} (\bibinfo {year}
  {2010})}\BibitemShut {NoStop}%
\bibitem [{\citenamefont {Matsumoto}\ and\ \citenamefont
  {Murakami}(2011)}]{Matsumoto2011}%
  \BibitemOpen
  \bibfield  {author} {\bibinfo {author} {\bibfnamefont {Ryo}\ \bibnamefont
  {Matsumoto}}\ and\ \bibinfo {author} {\bibfnamefont {Shuichi}\ \bibnamefont
  {Murakami}},\ }\bibfield  {title} {\enquote {\bibinfo {title} {Theoretical
  {P}rediction of a {R}otating {M}agnon {W}ave {P}acket in {F}erromagnets},}\
  }\href@noop {} {\bibfield  {journal} {\bibinfo  {journal} {Phys. Rev. Lett.}\
  }\textbf {\bibinfo {volume} {106}},\ \bibinfo {pages} {197202} (\bibinfo
  {year} {2011})}\BibitemShut {NoStop}%
\bibitem [{\citenamefont {Ideue}\ \emph {et~al.}(2012)\citenamefont {Ideue},
  \citenamefont {Onose}, \citenamefont {Katsura}, \citenamefont {Shiomi},
  \citenamefont {Ishiwata}, \citenamefont {Nagaosa},\ and\ \citenamefont
  {Tokura}}]{Ideue2012}%
  \BibitemOpen
  \bibfield  {author} {\bibinfo {author} {\bibfnamefont {T}~\bibnamefont
  {Ideue}}, \bibinfo {author} {\bibfnamefont {Y}~\bibnamefont {Onose}},
  \bibinfo {author} {\bibfnamefont {H}~\bibnamefont {Katsura}}, \bibinfo
  {author} {\bibfnamefont {Y}~\bibnamefont {Shiomi}}, \bibinfo {author}
  {\bibfnamefont {S}~\bibnamefont {Ishiwata}}, \bibinfo {author} {\bibfnamefont
  {N}~\bibnamefont {Nagaosa}}, \ and\ \bibinfo {author} {\bibfnamefont
  {Y}~\bibnamefont {Tokura}},\ }\bibfield  {title} {\enquote {\bibinfo {title}
  {Effect of lattice geometry on magnon {H}all effect in ferromagnetic
  insulators},}\ }\href@noop {} {\bibfield  {journal} {\bibinfo  {journal}
  {Phys. Rev. B}\ }\textbf {\bibinfo {volume} {85}},\ \bibinfo {pages} {134411}
  (\bibinfo {year} {2012})}\BibitemShut {NoStop}%
\bibitem [{\citenamefont {Shindou}\ \emph {et~al.}(2013)\citenamefont
  {Shindou}, \citenamefont {Matsumoto}, \citenamefont {Murakami},\ and\
  \citenamefont {Ohe}}]{Shindou2013}%
  \BibitemOpen
  \bibfield  {author} {\bibinfo {author} {\bibfnamefont {Ryuichi}\ \bibnamefont
  {Shindou}}, \bibinfo {author} {\bibfnamefont {Ryo}\ \bibnamefont
  {Matsumoto}}, \bibinfo {author} {\bibfnamefont {Shuichi}\ \bibnamefont
  {Murakami}}, \ and\ \bibinfo {author} {\bibfnamefont {J.~I}\ \bibnamefont
  {Ohe}},\ }\bibfield  {title} {\enquote {\bibinfo {title} {Topological chiral
  magnonic edge mode in a magnonic crystal},}\ }\href@noop {} {\bibfield
  {journal} {\bibinfo  {journal} {Phys. Rev. B}\ }\textbf {\bibinfo {volume}
  {87}},\ \bibinfo {pages} {174427} (\bibinfo {year} {2013})}\BibitemShut
  {NoStop}%
\bibitem [{\citenamefont {Zhang}\ \emph {et~al.}(2013)\citenamefont {Zhang},
  \citenamefont {Ren}, \citenamefont {Wang},\ and\ \citenamefont
  {Li}}]{Zhang2013}%
  \BibitemOpen
  \bibfield  {author} {\bibinfo {author} {\bibfnamefont {Lifa}\ \bibnamefont
  {Zhang}}, \bibinfo {author} {\bibfnamefont {Jie}\ \bibnamefont {Ren}},
  \bibinfo {author} {\bibfnamefont {Jian-Sheng}\ \bibnamefont {Wang}}, \ and\
  \bibinfo {author} {\bibfnamefont {Baowen}\ \bibnamefont {Li}},\ }\bibfield
  {title} {\enquote {\bibinfo {title} {Topological magnon insulator in
  insulating ferromagnet},}\ }\href@noop {} {\bibfield  {journal} {\bibinfo
  {journal} {Phys. Rev. B}\ }\textbf {\bibinfo {volume} {87}},\ \bibinfo
  {pages} {144101} (\bibinfo {year} {2013})}\BibitemShut {NoStop}%
\bibitem [{\citenamefont {Matsumoto}\ \emph {et~al.}(2014)\citenamefont
  {Matsumoto}, \citenamefont {Shindou},\ and\ \citenamefont
  {Murakami}}]{Matsumoto2014}%
  \BibitemOpen
  \bibfield  {author} {\bibinfo {author} {\bibfnamefont {Ryo}\ \bibnamefont
  {Matsumoto}}, \bibinfo {author} {\bibfnamefont {Ryuichi}\ \bibnamefont
  {Shindou}}, \ and\ \bibinfo {author} {\bibfnamefont {Shuichi}\ \bibnamefont
  {Murakami}},\ }\bibfield  {title} {\enquote {\bibinfo {title} {Thermal hall
  effect of magnons in magnets with dipolar interaction},}\ }\href@noop {}
  {\bibfield  {journal} {\bibinfo  {journal} {Phys. Rev. B}\ }\textbf {\bibinfo
  {volume} {89}},\ \bibinfo {pages} {054420} (\bibinfo {year}
  {2014})}\BibitemShut {NoStop}%
\bibitem [{\citenamefont {Mook}\ \emph
  {et~al.}(2014{\natexlab{a}})\citenamefont {Mook}, \citenamefont {Henk},\ and\
  \citenamefont {Mertig}}]{Mook2014edge}%
  \BibitemOpen
  \bibfield  {author} {\bibinfo {author} {\bibfnamefont {Alexander}\
  \bibnamefont {Mook}}, \bibinfo {author} {\bibfnamefont {J{\"u}rgen}\
  \bibnamefont {Henk}}, \ and\ \bibinfo {author} {\bibfnamefont {Ingrid}\
  \bibnamefont {Mertig}},\ }\bibfield  {title} {\enquote {\bibinfo {title}
  {Edge states in topological magnon insulators},}\ }\href@noop {} {\bibfield
  {journal} {\bibinfo  {journal} {Phys. Rev. B}\ }\textbf {\bibinfo {volume}
  {90}},\ \bibinfo {pages} {024412} (\bibinfo {year}
  {2014}{\natexlab{a}})}\BibitemShut {NoStop}%
\bibitem [{\citenamefont {Mook}\ \emph
  {et~al.}(2014{\natexlab{b}})\citenamefont {Mook}, \citenamefont {Henk},\ and\
  \citenamefont {Mertig}}]{Mook2014}%
  \BibitemOpen
  \bibfield  {author} {\bibinfo {author} {\bibfnamefont {Alexander}\
  \bibnamefont {Mook}}, \bibinfo {author} {\bibfnamefont {J{\"u}rgen}\
  \bibnamefont {Henk}}, \ and\ \bibinfo {author} {\bibfnamefont {Ingrid}\
  \bibnamefont {Mertig}},\ }\bibfield  {title} {\enquote {\bibinfo {title}
  {Magnon {H}all effect and topology in kagome lattices: {A} theoretical
  investigation},}\ }\href@noop {} {\bibfield  {journal} {\bibinfo  {journal}
  {Phys. Rev. B}\ }\textbf {\bibinfo {volume} {89}},\ \bibinfo {pages} {134409}
  (\bibinfo {year} {2014}{\natexlab{b}})}\BibitemShut {NoStop}%
\bibitem [{\citenamefont {Hirschberger}\ \emph {et~al.}(2015)\citenamefont
  {Hirschberger}, \citenamefont {Chisnell}, \citenamefont {Lee},\ and\
  \citenamefont {Ong}}]{hirschberger2015}%
  \BibitemOpen
  \bibfield  {author} {\bibinfo {author} {\bibfnamefont {Max}\ \bibnamefont
  {Hirschberger}}, \bibinfo {author} {\bibfnamefont {Robin}\ \bibnamefont
  {Chisnell}}, \bibinfo {author} {\bibfnamefont {Young~S}\ \bibnamefont {Lee}},
  \ and\ \bibinfo {author} {\bibfnamefont {Nai~Phuan}\ \bibnamefont {Ong}},\
  }\bibfield  {title} {\enquote {\bibinfo {title} {Thermal {H}all {E}ffect of
  {S}pin {E}xcitations in a {K}agome {M}agnet},}\ }\href@noop {} {\bibfield
  {journal} {\bibinfo  {journal} {Phys. Rev. Lett.}\ }\textbf {\bibinfo
  {volume} {115}},\ \bibinfo {pages} {106603} (\bibinfo {year}
  {2015})}\BibitemShut {NoStop}%
\bibitem [{\citenamefont {Chisnell}\ \emph {et~al.}(2015)\citenamefont
  {Chisnell}, \citenamefont {Helton}, \citenamefont {Freedman}, \citenamefont
  {Singh}, \citenamefont {Bewley}, \citenamefont {Nocera},\ and\ \citenamefont
  {Lee}}]{chisnell2015}%
  \BibitemOpen
  \bibfield  {author} {\bibinfo {author} {\bibfnamefont {R}~\bibnamefont
  {Chisnell}}, \bibinfo {author} {\bibfnamefont {J.~S}\ \bibnamefont {Helton}},
  \bibinfo {author} {\bibfnamefont {D.~E}\ \bibnamefont {Freedman}}, \bibinfo
  {author} {\bibfnamefont {D.~K}\ \bibnamefont {Singh}}, \bibinfo {author}
  {\bibfnamefont {R.~I}\ \bibnamefont {Bewley}}, \bibinfo {author}
  {\bibfnamefont {D.~G}\ \bibnamefont {Nocera}}, \ and\ \bibinfo {author}
  {\bibfnamefont {Y.~S}\ \bibnamefont {Lee}},\ }\bibfield  {title} {\enquote
  {\bibinfo {title} {Topological {M}agnon {B}ands in a {K}agome {L}attice
  {F}erromagnet},}\ }\href@noop {} {\bibfield  {journal} {\bibinfo  {journal}
  {Phys. Rev. Lett.}\ }\textbf {\bibinfo {volume} {115}},\ \bibinfo {pages}
  {147201} (\bibinfo {year} {2015})}\BibitemShut {NoStop}%
\bibitem [{\citenamefont {Xu}\ \emph {et~al.}(2016)\citenamefont {Xu},
  \citenamefont {Ohtsuki},\ and\ \citenamefont {Shindou}}]{Xu2016}%
  \BibitemOpen
  \bibfield  {author} {\bibinfo {author} {\bibfnamefont {Baolong}\ \bibnamefont
  {Xu}}, \bibinfo {author} {\bibfnamefont {Tomi}\ \bibnamefont {Ohtsuki}}, \
  and\ \bibinfo {author} {\bibfnamefont {Ryuichi}\ \bibnamefont {Shindou}},\
  }\bibfield  {title} {\enquote {\bibinfo {title} {Integer quantum magnon
  {H}all plateau-plateau transition in a spin-ice model},}\ }\href@noop {}
  {\bibfield  {journal} {\bibinfo  {journal} {Phys. Rev. B}\ }\textbf {\bibinfo
  {volume} {94}},\ \bibinfo {pages} {220403(R)} (\bibinfo {year}
  {2016})}\BibitemShut {NoStop}%
\bibitem [{\citenamefont {Cheng}\ \emph {et~al.}(2016)\citenamefont {Cheng},
  \citenamefont {Okamoto},\ and\ \citenamefont {Xiao}}]{cheng2016}%
  \BibitemOpen
  \bibfield  {author} {\bibinfo {author} {\bibfnamefont {Ran}\ \bibnamefont
  {Cheng}}, \bibinfo {author} {\bibfnamefont {Satoshi}\ \bibnamefont
  {Okamoto}}, \ and\ \bibinfo {author} {\bibfnamefont {Di}~\bibnamefont
  {Xiao}},\ }\bibfield  {title} {\enquote {\bibinfo {title} {Spin {N}ernst
  {E}ffect of {M}agnons in {C}ollinear {A}ntiferromagnets},}\ }\href@noop {}
  {\bibfield  {journal} {\bibinfo  {journal} {Phys. Rev. Lett.}\ }\textbf
  {\bibinfo {volume} {117}},\ \bibinfo {pages} {217202} (\bibinfo {year}
  {2016})}\BibitemShut {NoStop}%
\bibitem [{\citenamefont {Zyuzin}\ and\ \citenamefont
  {Kovalev}(2016)}]{zyuzin2016}%
  \BibitemOpen
  \bibfield  {author} {\bibinfo {author} {\bibfnamefont {Vladimir~A}\
  \bibnamefont {Zyuzin}}\ and\ \bibinfo {author} {\bibfnamefont {Alexey~A}\
  \bibnamefont {Kovalev}},\ }\bibfield  {title} {\enquote {\bibinfo {title}
  {Magnon {S}pin {N}ernst {E}ffect in {A}ntiferromagnets},}\ }\href@noop {}
  {\bibfield  {journal} {\bibinfo  {journal} {Phys. Rev. Lett.}\ }\textbf
  {\bibinfo {volume} {117}},\ \bibinfo {pages} {217203} (\bibinfo {year}
  {2016})}\BibitemShut {NoStop}%
\bibitem [{\citenamefont {Kim}\ \emph {et~al.}(2016)\citenamefont {Kim},
  \citenamefont {Ochoa}, \citenamefont {Zarzuela},\ and\ \citenamefont
  {Tserkovnyak}}]{kim2016realization}%
  \BibitemOpen
  \bibfield  {author} {\bibinfo {author} {\bibfnamefont {Se~Kwon}\ \bibnamefont
  {Kim}}, \bibinfo {author} {\bibfnamefont {H{\'e}ctor}\ \bibnamefont {Ochoa}},
  \bibinfo {author} {\bibfnamefont {Ricardo}\ \bibnamefont {Zarzuela}}, \ and\
  \bibinfo {author} {\bibfnamefont {Yaroslav}\ \bibnamefont {Tserkovnyak}},\
  }\bibfield  {title} {\enquote {\bibinfo {title} {Realization of the
  {H}aldane-{K}ane-{M}ele {M}odel in a {S}ystem of {L}ocalized {S}pins},}\
  }\href@noop {} {\bibfield  {journal} {\bibinfo  {journal} {Phys. Rev. Lett.}\
  }\textbf {\bibinfo {volume} {117}},\ \bibinfo {pages} {227201} (\bibinfo
  {year} {2016})}\BibitemShut {NoStop}%
\bibitem [{\citenamefont {Shiomi}\ \emph {et~al.}(2017)\citenamefont {Shiomi},
  \citenamefont {Takashima},\ and\ \citenamefont {Saitoh}}]{shiomi2017}%
  \BibitemOpen
  \bibfield  {author} {\bibinfo {author} {\bibfnamefont {Y}~\bibnamefont
  {Shiomi}}, \bibinfo {author} {\bibfnamefont {R}~\bibnamefont {Takashima}}, \
  and\ \bibinfo {author} {\bibfnamefont {E}~\bibnamefont {Saitoh}},\ }\bibfield
   {title} {\enquote {\bibinfo {title} {Experimental evidence consistent with a
  magnon {N}ernst effect in the antiferromagnetic insulator
  {M}n{P}{S}{$_{3}$}},}\ }\href@noop {} {\bibfield  {journal} {\bibinfo
  {journal} {Phys. Rev. B}\ }\textbf {\bibinfo {volume} {96}},\ \bibinfo
  {pages} {134425} (\bibinfo {year} {2017})}\BibitemShut {NoStop}%
\bibitem [{\citenamefont {Laurell}\ and\ \citenamefont
  {Fiete}(2017)}]{laurell2017}%
  \BibitemOpen
  \bibfield  {author} {\bibinfo {author} {\bibfnamefont {Pontus}\ \bibnamefont
  {Laurell}}\ and\ \bibinfo {author} {\bibfnamefont {Gregory~A}\ \bibnamefont
  {Fiete}},\ }\bibfield  {title} {\enquote {\bibinfo {title} {Topological
  {M}agnon {B}ands and {U}nconventional {S}uperconductivity in {P}yrochlore
  {I}ridate {T}hin {F}ilms},}\ }\href@noop {} {\bibfield  {journal} {\bibinfo
  {journal} {Phys. Rev. Lett.}\ }\textbf {\bibinfo {volume} {118}},\ \bibinfo
  {pages} {177201} (\bibinfo {year} {2017})}\BibitemShut {NoStop}%
\bibitem [{\citenamefont {Nakata}\ \emph
  {et~al.}(2017{\natexlab{a}})\citenamefont {Nakata}, \citenamefont
  {Klinovaja},\ and\ \citenamefont {Loss}}]{nakata2017magnonic}%
  \BibitemOpen
  \bibfield  {author} {\bibinfo {author} {\bibfnamefont {Kouki}\ \bibnamefont
  {Nakata}}, \bibinfo {author} {\bibfnamefont {Jelena}\ \bibnamefont
  {Klinovaja}}, \ and\ \bibinfo {author} {\bibfnamefont {Daniel}\ \bibnamefont
  {Loss}},\ }\bibfield  {title} {\enquote {\bibinfo {title} {Magnonic quantum
  {H}all effect and {W}iedemann-{F}ranz law},}\ }\href@noop {} {\bibfield
  {journal} {\bibinfo  {journal} {Phys. Rev. B}\ }\textbf {\bibinfo {volume}
  {95}},\ \bibinfo {pages} {125429} (\bibinfo {year}
  {2017}{\natexlab{a}})}\BibitemShut {NoStop}%
\bibitem [{\citenamefont {Nakata}\ \emph
  {et~al.}(2017{\natexlab{b}})\citenamefont {Nakata}, \citenamefont {Kim},
  \citenamefont {Klinovaja},\ and\ \citenamefont {Loss}}]{nakata2017}%
  \BibitemOpen
  \bibfield  {author} {\bibinfo {author} {\bibfnamefont {Kouki}\ \bibnamefont
  {Nakata}}, \bibinfo {author} {\bibfnamefont {Se~Kwon}\ \bibnamefont {Kim}},
  \bibinfo {author} {\bibfnamefont {Jelena}\ \bibnamefont {Klinovaja}}, \ and\
  \bibinfo {author} {\bibfnamefont {Daniel}\ \bibnamefont {Loss}},\ }\bibfield
  {title} {\enquote {\bibinfo {title} {Magnonic topological insulators in
  antiferromagnets},}\ }\href@noop {} {\bibfield  {journal} {\bibinfo
  {journal} {Phys. Rev. B}\ }\textbf {\bibinfo {volume} {96}},\ \bibinfo
  {pages} {224414} (\bibinfo {year} {2017}{\natexlab{b}})}\BibitemShut
  {NoStop}%
\bibitem [{\citenamefont {Owerre}(2017)}]{owerre2017}%
  \BibitemOpen
  \bibfield  {author} {\bibinfo {author} {\bibfnamefont {S.~A}\ \bibnamefont
  {Owerre}},\ }\bibfield  {title} {\enquote {\bibinfo {title} {Topological
  thermal {H}all effect in frustrated kagome antiferromagnets},}\ }\href@noop
  {} {\bibfield  {journal} {\bibinfo  {journal} {Phys. Rev. B}\ }\textbf
  {\bibinfo {volume} {95}},\ \bibinfo {pages} {014422} (\bibinfo {year}
  {2017})}\BibitemShut {NoStop}%
\bibitem [{\citenamefont {Murakami}\ and\ \citenamefont
  {Okamoto}(2017)}]{murakami2017thermal}%
  \BibitemOpen
  \bibfield  {author} {\bibinfo {author} {\bibfnamefont {Shuichi}\ \bibnamefont
  {Murakami}}\ and\ \bibinfo {author} {\bibfnamefont {Akihiro}\ \bibnamefont
  {Okamoto}},\ }\bibfield  {title} {\enquote {\bibinfo {title} {Thermal {H}all
  {E}ffect of {M}agnons},}\ }\href@noop {} {\bibfield  {journal} {\bibinfo
  {journal} {J. Phys. Soc. Jpn.}\ }\textbf {\bibinfo {volume} {86}},\ \bibinfo
  {pages} {011010} (\bibinfo {year} {2017})}\BibitemShut {NoStop}%
\bibitem [{\citenamefont {Laurell}\ and\ \citenamefont
  {Fiete}(2018)}]{laurell2018magnon}%
  \BibitemOpen
  \bibfield  {author} {\bibinfo {author} {\bibfnamefont {Pontus}\ \bibnamefont
  {Laurell}}\ and\ \bibinfo {author} {\bibfnamefont {Gregory~A}\ \bibnamefont
  {Fiete}},\ }\bibfield  {title} {\enquote {\bibinfo {title} {Magnon thermal
  {H}all effect in kagome antiferromagnets with {D}zyaloshinskii-{M}oriya
  interactions},}\ }\href@noop {} {\bibfield  {journal} {\bibinfo  {journal}
  {Phys. Rev. B}\ }\textbf {\bibinfo {volume} {98}},\ \bibinfo {pages} {094419}
  (\bibinfo {year} {2018})}\BibitemShut {NoStop}%
\bibitem [{\citenamefont {Cookmeyer}\ and\ \citenamefont
  {Moore}(2018)}]{cookmeyer2018}%
  \BibitemOpen
  \bibfield  {author} {\bibinfo {author} {\bibfnamefont {Tessa}\ \bibnamefont
  {Cookmeyer}}\ and\ \bibinfo {author} {\bibfnamefont {Joel~E}\ \bibnamefont
  {Moore}},\ }\bibfield  {title} {\enquote {\bibinfo {title} {Spin-wave
  analysis of the low-temperature thermal {H}all effect in the candidate
  {K}itaev spin liquid {$\alpha$}-{R}u{C}l{$_{3}$}},}\ }\href@noop {}
  {\bibfield  {journal} {\bibinfo  {journal} {Phys. Rev. B}\ }\textbf {\bibinfo
  {volume} {98}},\ \bibinfo {pages} {060412(R)} (\bibinfo {year}
  {2018})}\BibitemShut {NoStop}%
\bibitem [{\citenamefont {Doki}\ \emph {et~al.}(2018)\citenamefont {Doki},
  \citenamefont {Akazawa}, \citenamefont {Lee}, \citenamefont {Han},
  \citenamefont {Sugii}, \citenamefont {Shimozawa}, \citenamefont {Kawashima},
  \citenamefont {Oda}, \citenamefont {Yoshida},\ and\ \citenamefont
  {Yamashita}}]{doki2018}%
  \BibitemOpen
  \bibfield  {author} {\bibinfo {author} {\bibfnamefont {Hayato}\ \bibnamefont
  {Doki}}, \bibinfo {author} {\bibfnamefont {Masatoshi}\ \bibnamefont
  {Akazawa}}, \bibinfo {author} {\bibfnamefont {Hyun-Yong}\ \bibnamefont
  {Lee}}, \bibinfo {author} {\bibfnamefont {Jung~Hoon}\ \bibnamefont {Han}},
  \bibinfo {author} {\bibfnamefont {Kaori}\ \bibnamefont {Sugii}}, \bibinfo
  {author} {\bibfnamefont {Masaaki}\ \bibnamefont {Shimozawa}}, \bibinfo
  {author} {\bibfnamefont {Naoki}\ \bibnamefont {Kawashima}}, \bibinfo {author}
  {\bibfnamefont {Migaku}\ \bibnamefont {Oda}}, \bibinfo {author}
  {\bibfnamefont {Hiroyuki}\ \bibnamefont {Yoshida}}, \ and\ \bibinfo {author}
  {\bibfnamefont {Minoru}\ \bibnamefont {Yamashita}},\ }\bibfield  {title}
  {\enquote {\bibinfo {title} {Spin {T}hermal {H}all {C}onductivity of a
  {K}agome {A}ntiferromagnet},}\ }\href@noop {} {\bibfield  {journal} {\bibinfo
   {journal} {Phys. Rev. Lett.}\ }\textbf {\bibinfo {volume} {121}},\ \bibinfo
  {pages} {097203} (\bibinfo {year} {2018})}\BibitemShut {NoStop}%
\bibitem [{\citenamefont {McClarty}\ \emph {et~al.}(2018)\citenamefont
  {McClarty}, \citenamefont {Dong}, \citenamefont {Gohlke}, \citenamefont
  {Rau}, \citenamefont {Pollmann}, \citenamefont {Moessner},\ and\
  \citenamefont {Penc}}]{mcclarty2018}%
  \BibitemOpen
  \bibfield  {author} {\bibinfo {author} {\bibfnamefont {P.~A}\ \bibnamefont
  {McClarty}}, \bibinfo {author} {\bibfnamefont {X-Y}\ \bibnamefont {Dong}},
  \bibinfo {author} {\bibfnamefont {M}~\bibnamefont {Gohlke}}, \bibinfo
  {author} {\bibfnamefont {J.~G}\ \bibnamefont {Rau}}, \bibinfo {author}
  {\bibfnamefont {F}~\bibnamefont {Pollmann}}, \bibinfo {author} {\bibfnamefont
  {R}~\bibnamefont {Moessner}}, \ and\ \bibinfo {author} {\bibfnamefont
  {K}~\bibnamefont {Penc}},\ }\bibfield  {title} {\enquote {\bibinfo {title}
  {Topological magnons in {K}itaev magnets at high fields},}\ }\href@noop {}
  {\bibfield  {journal} {\bibinfo  {journal} {Phys. Rev. B}\ }\textbf {\bibinfo
  {volume} {98}},\ \bibinfo {pages} {060404(R)} (\bibinfo {year}
  {2018})}\BibitemShut {NoStop}%
\bibitem [{\citenamefont {Joshi}(2018)}]{joshi2018}%
  \BibitemOpen
  \bibfield  {author} {\bibinfo {author} {\bibfnamefont {Darshan~G}\
  \bibnamefont {Joshi}},\ }\bibfield  {title} {\enquote {\bibinfo {title}
  {Topological excitations in the ferromagnetic {K}itaev-{H}eisenberg model},}\
  }\href@noop {} {\bibfield  {journal} {\bibinfo  {journal} {Phys. Rev. B}\
  }\textbf {\bibinfo {volume} {98}},\ \bibinfo {pages} {060405(R)} (\bibinfo
  {year} {2018})}\BibitemShut {NoStop}%
\bibitem [{\citenamefont {Zyuzin}\ and\ \citenamefont
  {Kovalev}(2018)}]{zyuzin2018}%
  \BibitemOpen
  \bibfield  {author} {\bibinfo {author} {\bibfnamefont {Vladimir~A}\
  \bibnamefont {Zyuzin}}\ and\ \bibinfo {author} {\bibfnamefont {Alexey~A}\
  \bibnamefont {Kovalev}},\ }\bibfield  {title} {\enquote {\bibinfo {title}
  {Spin {H}all and {N}ernst effects of {W}eyl magnons},}\ }\href@noop {}
  {\bibfield  {journal} {\bibinfo  {journal} {Phys. Rev. B}\ }\textbf {\bibinfo
  {volume} {97}},\ \bibinfo {pages} {174407} (\bibinfo {year}
  {2018})}\BibitemShut {NoStop}%
\bibitem [{\citenamefont {Mook}\ \emph {et~al.}(2019)\citenamefont {Mook},
  \citenamefont {Henk},\ and\ \citenamefont {Mertig}}]{Mook2019}%
  \BibitemOpen
  \bibfield  {author} {\bibinfo {author} {\bibfnamefont {Alexander}\
  \bibnamefont {Mook}}, \bibinfo {author} {\bibfnamefont {J{\"u}rgen}\
  \bibnamefont {Henk}}, \ and\ \bibinfo {author} {\bibfnamefont {Ingrid}\
  \bibnamefont {Mertig}},\ }\bibfield  {title} {\enquote {\bibinfo {title}
  {Thermal {H}all effect in noncollinear coplanar insulating
  antiferromagnets},}\ }\href@noop {} {\bibfield  {journal} {\bibinfo
  {journal} {Phys. Rev. B}\ }\textbf {\bibinfo {volume} {99}},\ \bibinfo
  {pages} {014427} (\bibinfo {year} {2019})}\BibitemShut {NoStop}%
\bibitem [{\citenamefont {Kondo}\ \emph
  {et~al.}(2019{\natexlab{a}})\citenamefont {Kondo}, \citenamefont {Akagi},\
  and\ \citenamefont {Katsura}}]{kondoz2}%
  \BibitemOpen
  \bibfield  {author} {\bibinfo {author} {\bibfnamefont {Hiroki}\ \bibnamefont
  {Kondo}}, \bibinfo {author} {\bibfnamefont {Yutaka}\ \bibnamefont {Akagi}}, \
  and\ \bibinfo {author} {\bibfnamefont {Hosho}\ \bibnamefont {Katsura}},\
  }\bibfield  {title} {\enquote {\bibinfo {title} {{$\mathbb Z_{2}$}
  topological invariant for magnon spin {H}all systems},}\ }\href@noop {}
  {\bibfield  {journal} {\bibinfo  {journal} {Phys. Rev. B}\ }\textbf {\bibinfo
  {volume} {99}},\ \bibinfo {pages} {041110(R)} (\bibinfo {year}
  {2019}{\natexlab{a}})}\BibitemShut {NoStop}%
\bibitem [{\citenamefont {Kawano}\ and\ \citenamefont
  {Hotta}(2019)}]{Kawano2019}%
  \BibitemOpen
  \bibfield  {author} {\bibinfo {author} {\bibfnamefont {Masataka}\
  \bibnamefont {Kawano}}\ and\ \bibinfo {author} {\bibfnamefont {Chisa}\
  \bibnamefont {Hotta}},\ }\bibfield  {title} {\enquote {\bibinfo {title}
  {Thermal {H}all effect and topological edge states in a square-lattice
  antiferromagnet},}\ }\href@noop {} {\bibfield  {journal} {\bibinfo  {journal}
  {Phys. Rev. B}\ }\textbf {\bibinfo {volume} {99}},\ \bibinfo {pages} {054422}
  (\bibinfo {year} {2019})}\BibitemShut {NoStop}%
\bibitem [{\citenamefont {Kondo}\ \emph
  {et~al.}(2019{\natexlab{b}})\citenamefont {Kondo}, \citenamefont {Akagi},\
  and\ \citenamefont {Katsura}}]{kondo3D}%
  \BibitemOpen
  \bibfield  {author} {\bibinfo {author} {\bibfnamefont {Hiroki}\ \bibnamefont
  {Kondo}}, \bibinfo {author} {\bibfnamefont {Yutaka}\ \bibnamefont {Akagi}}, \
  and\ \bibinfo {author} {\bibfnamefont {Hosho}\ \bibnamefont {Katsura}},\
  }\bibfield  {title} {\enquote {\bibinfo {title} {Three-dimensional
  topological magnon systems},}\ }\href@noop {} {\bibfield  {journal} {\bibinfo
   {journal} {Phys. Rev. B}\ }\textbf {\bibinfo {volume} {100}},\ \bibinfo
  {pages} {144401} (\bibinfo {year} {2019}{\natexlab{b}})}\BibitemShut
  {NoStop}%
\bibitem [{\citenamefont {Kim}\ \emph {et~al.}(2019)\citenamefont {Kim},
  \citenamefont {Nakata}, \citenamefont {Loss},\ and\ \citenamefont
  {Tserkovnyak}}]{Kim2019}%
  \BibitemOpen
  \bibfield  {author} {\bibinfo {author} {\bibfnamefont {Se~Kwon}\ \bibnamefont
  {Kim}}, \bibinfo {author} {\bibfnamefont {Kouki}\ \bibnamefont {Nakata}},
  \bibinfo {author} {\bibfnamefont {Daniel}\ \bibnamefont {Loss}}, \ and\
  \bibinfo {author} {\bibfnamefont {Yaroslav}\ \bibnamefont {Tserkovnyak}},\
  }\bibfield  {title} {\enquote {\bibinfo {title} {Tunable {M}agnonic {T}hermal
  {H}all {E}ffect in {S}kyrmion {C}rystal {P}hases of {F}errimagnets},}\
  }\href@noop {} {\bibfield  {journal} {\bibinfo  {journal} {Phys. Rev. Lett.}\
  }\textbf {\bibinfo {volume} {122}},\ \bibinfo {pages} {057204} (\bibinfo
  {year} {2019})}\BibitemShut {NoStop}%
\bibitem [{\citenamefont {Hwang}\ \emph {et~al.}(2020)\citenamefont {Hwang},
  \citenamefont {Trivedi},\ and\ \citenamefont {Randeria}}]{hwang2020}%
  \BibitemOpen
  \bibfield  {author} {\bibinfo {author} {\bibfnamefont {Kyusung}\ \bibnamefont
  {Hwang}}, \bibinfo {author} {\bibfnamefont {Nandini}\ \bibnamefont
  {Trivedi}}, \ and\ \bibinfo {author} {\bibfnamefont {Mohit}\ \bibnamefont
  {Randeria}},\ }\bibfield  {title} {\enquote {\bibinfo {title} {Topological
  {M}agnons with {N}odal-{L}ine and {T}riple-{P}oint {D}egeneracies:
  {I}mplications for {T}hermal {H}all {E}ffect in {P}yrochlore {I}ridates},}\
  }\href@noop {} {\bibfield  {journal} {\bibinfo  {journal} {Phys. Rev. Lett.}\
  }\textbf {\bibinfo {volume} {125}},\ \bibinfo {pages} {047203} (\bibinfo
  {year} {2020})}\BibitemShut {NoStop}%
\bibitem [{\citenamefont {Akagi}(2020)}]{Akagi2020}%
  \BibitemOpen
  \bibfield  {author} {\bibinfo {author} {\bibfnamefont {Yutaka}\ \bibnamefont
  {Akagi}},\ }\bibfield  {title} {\enquote {\bibinfo {title} {Topological
  {I}nvariant for {B}osonic {B}ogoliubov--de {G}ennes {S}ystems with
  {D}isorder},}\ }\href@noop {} {\bibfield  {journal} {\bibinfo  {journal} {J.
  Phys. Soc. Jpn.}\ }\textbf {\bibinfo {volume} {89}},\ \bibinfo {pages}
  {123601} (\bibinfo {year} {2020})}\BibitemShut {NoStop}%
\bibitem [{\citenamefont {Kondo}\ \emph {et~al.}(2020)\citenamefont {Kondo},
  \citenamefont {Akagi},\ and\ \citenamefont {Katsura}}]{kondo2020}%
  \BibitemOpen
  \bibfield  {author} {\bibinfo {author} {\bibfnamefont {Hiroki}\ \bibnamefont
  {Kondo}}, \bibinfo {author} {\bibfnamefont {Yutaka}\ \bibnamefont {Akagi}}, \
  and\ \bibinfo {author} {\bibfnamefont {Hosho}\ \bibnamefont {Katsura}},\
  }\bibfield  {title} {\enquote {\bibinfo {title} {Non-hermiticity and
  topological invariants of magnon {B}ogoliubov--de {G}ennes systems},}\
  }\href@noop {} {\bibfield  {journal} {\bibinfo  {journal} {Prog Theor Exp
  Phys.}\ }\textbf {\bibinfo {volume} {2020}},\ \bibinfo {pages} {12A104}
  (\bibinfo {year} {2020})}\BibitemShut {NoStop}%
\bibitem [{\citenamefont {Nakata}\ and\ \citenamefont
  {Kim}(2021)}]{nakata2021}%
  \BibitemOpen
  \bibfield  {author} {\bibinfo {author} {\bibfnamefont {Kouki}\ \bibnamefont
  {Nakata}}\ and\ \bibinfo {author} {\bibfnamefont {Se~Kwon}\ \bibnamefont
  {Kim}},\ }\bibfield  {title} {\enquote {\bibinfo {title} {Topological {H}all
  {E}ffects of {M}agnons in {F}errimagnets},}\ }\href@noop {} {\bibfield
  {journal} {\bibinfo  {journal} {J. Phys. Soc. Jpn.}\ }\textbf {\bibinfo
  {volume} {90}},\ \bibinfo {pages} {081004} (\bibinfo {year}
  {2021})}\BibitemShut {NoStop}%
\bibitem [{\citenamefont {Kondo}\ and\ \citenamefont
  {Akagi}(2021)}]{kondodirac}%
  \BibitemOpen
  \bibfield  {author} {\bibinfo {author} {\bibfnamefont {Hiroki}\ \bibnamefont
  {Kondo}}\ and\ \bibinfo {author} {\bibfnamefont {Yutaka}\ \bibnamefont
  {Akagi}},\ }\bibfield  {title} {\enquote {\bibinfo {title} {Dirac {S}urface
  {S}tates in {M}agnonic {A}nalogs of {T}opological {C}rystalline
  {I}nsulators},}\ }\href@noop {} {\bibfield  {journal} {\bibinfo  {journal}
  {Phys. Rev. Lett.}\ }\textbf {\bibinfo {volume} {127}},\ \bibinfo {pages}
  {177201} (\bibinfo {year} {2021})}\BibitemShut {NoStop}%
\bibitem [{\citenamefont {Fujiwara}\ \emph {et~al.}(2022)\citenamefont
  {Fujiwara}, \citenamefont {Kitamura},\ and\ \citenamefont
  {Morimoto}}]{Fujiwara2022}%
  \BibitemOpen
  \bibfield  {author} {\bibinfo {author} {\bibfnamefont {Kosuke}\ \bibnamefont
  {Fujiwara}}, \bibinfo {author} {\bibfnamefont {Sota}\ \bibnamefont
  {Kitamura}}, \ and\ \bibinfo {author} {\bibfnamefont {Takahiro}\ \bibnamefont
  {Morimoto}},\ }\bibfield  {title} {\enquote {\bibinfo {title} {Thermal {H}all
  responses in frustrated honeycomb spin systems},}\ }\href@noop {} {\bibfield
  {journal} {\bibinfo  {journal} {Phys. Rev. B}\ }\textbf {\bibinfo {volume}
  {106}},\ \bibinfo {pages} {035113} (\bibinfo {year} {2022})}\BibitemShut
  {NoStop}%
\bibitem [{\citenamefont {Kim}\ and\ \citenamefont {Kim}(2022)}]{kim2022}%
  \BibitemOpen
  \bibfield  {author} {\bibinfo {author} {\bibfnamefont {Hongseok}\
  \bibnamefont {Kim}}\ and\ \bibinfo {author} {\bibfnamefont {Se~Kwon}\
  \bibnamefont {Kim}},\ }\bibfield  {title} {\enquote {\bibinfo {title}
  {Topological phase transition in magnon bands in a honeycomb ferromagnet
  driven by sublattice symmetry breaking},}\ }\href@noop {} {\bibfield
  {journal} {\bibinfo  {journal} {Phys. Rev. B}\ }\textbf {\bibinfo {volume}
  {106}},\ \bibinfo {pages} {104430} (\bibinfo {year} {2022})}\BibitemShut
  {NoStop}%
\bibitem [{\citenamefont {McClarty}(2022)}]{mcclarty2022}%
  \BibitemOpen
  \bibfield  {author} {\bibinfo {author} {\bibfnamefont {Paul~A}\ \bibnamefont
  {McClarty}},\ }\bibfield  {title} {\enquote {\bibinfo {title} {Topological
  {M}agnons: {A} {R}eview},}\ }\href@noop {} {\bibfield  {journal} {\bibinfo
  {journal} {Annu. Rev. Condens. Matter Phys.}\ }\textbf {\bibinfo {volume}
  {13}},\ \bibinfo {pages} {171--190} (\bibinfo {year} {2022})}\BibitemShut
  {NoStop}%
\bibitem [{\citenamefont {Neumann}\ \emph {et~al.}(2022)\citenamefont
  {Neumann}, \citenamefont {Mook}, \citenamefont {Henk},\ and\ \citenamefont
  {Mertig}}]{Neumann2022}%
  \BibitemOpen
  \bibfield  {author} {\bibinfo {author} {\bibfnamefont {Robin~R}\ \bibnamefont
  {Neumann}}, \bibinfo {author} {\bibfnamefont {Alexander}\ \bibnamefont
  {Mook}}, \bibinfo {author} {\bibfnamefont {J{\"u}rgen}\ \bibnamefont {Henk}},
  \ and\ \bibinfo {author} {\bibfnamefont {Ingrid}\ \bibnamefont {Mertig}},\
  }\bibfield  {title} {\enquote {\bibinfo {title} {Thermal {H}all {E}ffect of
  {M}agnons in {C}ollinear {A}ntiferromagnetic {I}nsulators: {S}ignatures of
  {M}agnetic and {T}opological {P}hase {T}ransitions},}\ }\href@noop {}
  {\bibfield  {journal} {\bibinfo  {journal} {Phys. Rev. Lett.}\ }\textbf
  {\bibinfo {volume} {128}},\ \bibinfo {pages} {117201} (\bibinfo {year}
  {2022})}\BibitemShut {NoStop}%
\bibitem [{\citenamefont {Go}\ \emph {et~al.}(2024)\citenamefont {Go},
  \citenamefont {An}, \citenamefont {Lee},\ and\ \citenamefont {Kim}}]{go2023}%
  \BibitemOpen
  \bibfield  {author} {\bibinfo {author} {\bibfnamefont {Gyungchoon}\
  \bibnamefont {Go}}, \bibinfo {author} {\bibfnamefont {Daehyeon}\ \bibnamefont
  {An}}, \bibinfo {author} {\bibfnamefont {Hyun-Woo}\ \bibnamefont {Lee}}, \
  and\ \bibinfo {author} {\bibfnamefont {Se~Kwon}\ \bibnamefont {Kim}},\
  }\bibfield  {title} {\enquote {\bibinfo {title} {Magnon {O}rbital {N}ernst
  {E}ffect in {H}oneycomb {A}ntiferromagnets without {S}pin--{O}rbit
  {C}oupling},}\ }\href@noop {} {\bibfield  {journal} {\bibinfo  {journal}
  {Nano Lett.}\ }\textbf {\bibinfo {volume} {24}},\ \bibinfo {pages}
  {5968--5974} (\bibinfo {year} {2024})}\BibitemShut {NoStop}%
\bibitem [{\citenamefont {Zhuo}\ \emph {et~al.}(2023)\citenamefont {Zhuo},
  \citenamefont {Kang}, \citenamefont {Manchon},\ and\ \citenamefont
  {Cheng}}]{zhuo2023topological}%
  \BibitemOpen
  \bibfield  {author} {\bibinfo {author} {\bibfnamefont {Fengjun}\ \bibnamefont
  {Zhuo}}, \bibinfo {author} {\bibfnamefont {Jian}\ \bibnamefont {Kang}},
  \bibinfo {author} {\bibfnamefont {Aur{\'e}lien}\ \bibnamefont {Manchon}}, \
  and\ \bibinfo {author} {\bibfnamefont {Zhenxiang}\ \bibnamefont {Cheng}},\
  }\bibfield  {title} {\enquote {\bibinfo {title} {Topological {P}hases in
  {M}agnonics},}\ }\href@noop {} {\bibfield  {journal} {\bibinfo  {journal}
  {Advanced Physics Research}\ ,\ \bibinfo {pages} {2300054}} (\bibinfo {year}
  {2023})}\BibitemShut {NoStop}%
\bibitem [{\citenamefont {Zhang}\ \emph {et~al.}(2024)\citenamefont {Zhang},
  \citenamefont {Gao},\ and\ \citenamefont {Chen}}]{zhang2023}%
  \BibitemOpen
  \bibfield  {author} {\bibinfo {author} {\bibfnamefont {Xiao-Tian}\
  \bibnamefont {Zhang}}, \bibinfo {author} {\bibfnamefont {Yong~Hao}\
  \bibnamefont {Gao}}, \ and\ \bibinfo {author} {\bibfnamefont {Gang}\
  \bibnamefont {Chen}},\ }\bibfield  {title} {\enquote {\bibinfo {title}
  {Thermal {H}all effects in quantum magnets},}\ }\href@noop {} {\bibfield
  {journal} {\bibinfo  {journal} {Physics Reports}\ }\textbf {\bibinfo {volume}
  {1070}},\ \bibinfo {pages} {1--59} (\bibinfo {year} {2024})}\BibitemShut
  {NoStop}%
\bibitem [{\citenamefont {Raghu}\ and\ \citenamefont
  {Haldane}(2008)}]{Raghu2008}%
  \BibitemOpen
  \bibfield  {author} {\bibinfo {author} {\bibfnamefont {Srinivas}\
  \bibnamefont {Raghu}}\ and\ \bibinfo {author} {\bibfnamefont {Frederick
  Duncan~Michael}\ \bibnamefont {Haldane}},\ }\bibfield  {title} {\enquote
  {\bibinfo {title} {Analogs of quantum-{H}all-effect edge states in photonic
  crystals},}\ }\href@noop {} {\bibfield  {journal} {\bibinfo  {journal} {Phys.
  Rev. A}\ }\textbf {\bibinfo {volume} {78}},\ \bibinfo {pages} {033834}
  (\bibinfo {year} {2008})}\BibitemShut {NoStop}%
\bibitem [{\citenamefont {Petrescu}\ \emph {et~al.}(2012)\citenamefont
  {Petrescu}, \citenamefont {Houck},\ and\ \citenamefont
  {Le~Hur}}]{Petrescu2012}%
  \BibitemOpen
  \bibfield  {author} {\bibinfo {author} {\bibfnamefont {Alexandru}\
  \bibnamefont {Petrescu}}, \bibinfo {author} {\bibfnamefont {Andrew~A}\
  \bibnamefont {Houck}}, \ and\ \bibinfo {author} {\bibfnamefont {Karyn}\
  \bibnamefont {Le~Hur}},\ }\bibfield  {title} {\enquote {\bibinfo {title}
  {Anomalous {H}all effects of light and chiral edge modes on the {K}agom{\'e}
  lattice},}\ }\href@noop {} {\bibfield  {journal} {\bibinfo  {journal} {Phys.
  Rev. A}\ }\textbf {\bibinfo {volume} {86}},\ \bibinfo {pages} {053804}
  (\bibinfo {year} {2012})}\BibitemShut {NoStop}%
\bibitem [{\citenamefont {Rechtsman}\ \emph {et~al.}(2013)\citenamefont
  {Rechtsman}, \citenamefont {Zeuner}, \citenamefont {Plotnik}, \citenamefont
  {Lumer}, \citenamefont {Podolsky}, \citenamefont {Dreisow}, \citenamefont
  {Nolte}, \citenamefont {Segev},\ and\ \citenamefont
  {Szameit}}]{Rechtsman2013}%
  \BibitemOpen
  \bibfield  {author} {\bibinfo {author} {\bibfnamefont {Mikael~C}\
  \bibnamefont {Rechtsman}}, \bibinfo {author} {\bibfnamefont {Julia~M}\
  \bibnamefont {Zeuner}}, \bibinfo {author} {\bibfnamefont {Yonatan}\
  \bibnamefont {Plotnik}}, \bibinfo {author} {\bibfnamefont {Yaakov}\
  \bibnamefont {Lumer}}, \bibinfo {author} {\bibfnamefont {Daniel}\
  \bibnamefont {Podolsky}}, \bibinfo {author} {\bibfnamefont {Felix}\
  \bibnamefont {Dreisow}}, \bibinfo {author} {\bibfnamefont {Stefan}\
  \bibnamefont {Nolte}}, \bibinfo {author} {\bibfnamefont {Mordechai}\
  \bibnamefont {Segev}}, \ and\ \bibinfo {author} {\bibfnamefont {Alexander}\
  \bibnamefont {Szameit}},\ }\bibfield  {title} {\enquote {\bibinfo {title}
  {Photonic {F}loquet topological insulators},}\ }\href@noop {} {\bibfield
  {journal} {\bibinfo  {journal} {Nature}\ }\textbf {\bibinfo {volume} {496}},\
  \bibinfo {pages} {196--200} (\bibinfo {year} {2013})}\BibitemShut {NoStop}%
\bibitem [{\citenamefont {Hafezi}\ \emph {et~al.}(2013)\citenamefont {Hafezi},
  \citenamefont {Mittal}, \citenamefont {Fan}, \citenamefont {Migdall},\ and\
  \citenamefont {Taylor}}]{Hafezi2013}%
  \BibitemOpen
  \bibfield  {author} {\bibinfo {author} {\bibfnamefont {Mohammad}\
  \bibnamefont {Hafezi}}, \bibinfo {author} {\bibfnamefont {Sunil}\
  \bibnamefont {Mittal}}, \bibinfo {author} {\bibfnamefont {J}~\bibnamefont
  {Fan}}, \bibinfo {author} {\bibfnamefont {A}~\bibnamefont {Migdall}}, \ and\
  \bibinfo {author} {\bibfnamefont {JM}~\bibnamefont {Taylor}},\ }\bibfield
  {title} {\enquote {\bibinfo {title} {Imaging topological edge states in
  silicon photonics},}\ }\href@noop {} {\bibfield  {journal} {\bibinfo
  {journal} {Nat. Photonics}\ }\textbf {\bibinfo {volume} {7}},\ \bibinfo
  {pages} {1001--1005} (\bibinfo {year} {2013})}\BibitemShut {NoStop}%
\bibitem [{\citenamefont {Ben-Abdallah}(2016)}]{Ben-Abdallah2016}%
  \BibitemOpen
  \bibfield  {author} {\bibinfo {author} {\bibfnamefont {Philippe}\
  \bibnamefont {Ben-Abdallah}},\ }\bibfield  {title} {\enquote {\bibinfo
  {title} {Photon {T}hermal {H}all {E}ffect},}\ }\href@noop {} {\bibfield
  {journal} {\bibinfo  {journal} {Phys. Rev. Lett.}\ }\textbf {\bibinfo
  {volume} {116}},\ \bibinfo {pages} {084301} (\bibinfo {year}
  {2016})}\BibitemShut {NoStop}%
\bibitem [{\citenamefont {Strohm}\ \emph {et~al.}(2005)\citenamefont {Strohm},
  \citenamefont {Rikken},\ and\ \citenamefont {Wyder}}]{Strohm2005}%
  \BibitemOpen
  \bibfield  {author} {\bibinfo {author} {\bibfnamefont {C}~\bibnamefont
  {Strohm}}, \bibinfo {author} {\bibfnamefont {G.~L. J.~A}\ \bibnamefont
  {Rikken}}, \ and\ \bibinfo {author} {\bibfnamefont {P}~\bibnamefont
  {Wyder}},\ }\bibfield  {title} {\enquote {\bibinfo {title} {Phenomenological
  {E}vidence for the {P}honon {H}all {E}ffect},}\ }\href@noop {} {\bibfield
  {journal} {\bibinfo  {journal} {Phys. Rev. Lett.}\ }\textbf {\bibinfo
  {volume} {95}},\ \bibinfo {pages} {155901} (\bibinfo {year}
  {2005})}\BibitemShut {NoStop}%
\bibitem [{\citenamefont {Sheng}\ \emph {et~al.}(2006)\citenamefont {Sheng},
  \citenamefont {Sheng},\ and\ \citenamefont {Ting}}]{Sheng2006}%
  \BibitemOpen
  \bibfield  {author} {\bibinfo {author} {\bibfnamefont {L}~\bibnamefont
  {Sheng}}, \bibinfo {author} {\bibfnamefont {D.~N}\ \bibnamefont {Sheng}}, \
  and\ \bibinfo {author} {\bibfnamefont {C.~S}\ \bibnamefont {Ting}},\
  }\bibfield  {title} {\enquote {\bibinfo {title} {Theory of the {P}honon
  {H}all {E}ffect in {P}aramagnetic {D}ielectrics},}\ }\href@noop {} {\bibfield
   {journal} {\bibinfo  {journal} {Phys. Rev. Lett.}\ }\textbf {\bibinfo
  {volume} {96}},\ \bibinfo {pages} {155901} (\bibinfo {year}
  {2006})}\BibitemShut {NoStop}%
\bibitem [{\citenamefont {Kagan}\ and\ \citenamefont
  {Maksimov}(2008)}]{Kagan2008}%
  \BibitemOpen
  \bibfield  {author} {\bibinfo {author} {\bibfnamefont {Yu}~\bibnamefont
  {Kagan}}\ and\ \bibinfo {author} {\bibfnamefont {L.~A}\ \bibnamefont
  {Maksimov}},\ }\bibfield  {title} {\enquote {\bibinfo {title} {Anomalous
  {H}all {E}ffect for the {P}honon {H}eat {C}onductivity in {P}aramagnetic
  {D}ielectrics},}\ }\href@noop {} {\bibfield  {journal} {\bibinfo  {journal}
  {Phys. Rev. Lett.}\ }\textbf {\bibinfo {volume} {100}},\ \bibinfo {pages}
  {145902} (\bibinfo {year} {2008})}\BibitemShut {NoStop}%
\bibitem [{\citenamefont {Zhang}\ \emph {et~al.}(2010)\citenamefont {Zhang},
  \citenamefont {Ren}, \citenamefont {Wang},\ and\ \citenamefont
  {Li}}]{Zhang2010}%
  \BibitemOpen
  \bibfield  {author} {\bibinfo {author} {\bibfnamefont {Lifa}\ \bibnamefont
  {Zhang}}, \bibinfo {author} {\bibfnamefont {Jie}\ \bibnamefont {Ren}},
  \bibinfo {author} {\bibfnamefont {Jian-Sheng}\ \bibnamefont {Wang}}, \ and\
  \bibinfo {author} {\bibfnamefont {Baowen}\ \bibnamefont {Li}},\ }\bibfield
  {title} {\enquote {\bibinfo {title} {Topological {N}ature of the {P}honon
  {H}all {E}ffect},}\ }\href@noop {} {\bibfield  {journal} {\bibinfo  {journal}
  {Phys. Rev. Lett.}\ }\textbf {\bibinfo {volume} {105}},\ \bibinfo {pages}
  {225901} (\bibinfo {year} {2010})}\BibitemShut {NoStop}%
\bibitem [{\citenamefont {Zhang}\ \emph {et~al.}(2011)\citenamefont {Zhang},
  \citenamefont {Ren}, \citenamefont {Wang},\ and\ \citenamefont
  {Li}}]{Zhang2011}%
  \BibitemOpen
  \bibfield  {author} {\bibinfo {author} {\bibfnamefont {Lifa}\ \bibnamefont
  {Zhang}}, \bibinfo {author} {\bibfnamefont {Jie}\ \bibnamefont {Ren}},
  \bibinfo {author} {\bibfnamefont {Jian-Sheng}\ \bibnamefont {Wang}}, \ and\
  \bibinfo {author} {\bibfnamefont {Baowen}\ \bibnamefont {Li}},\ }\bibfield
  {title} {\enquote {\bibinfo {title} {The phonon {H}all effect: theory and
  application},}\ }\href@noop {} {\bibfield  {journal} {\bibinfo  {journal} {J.
  Phys. Condens.}\ }\textbf {\bibinfo {volume} {23}},\ \bibinfo {pages}
  {305402} (\bibinfo {year} {2011})}\BibitemShut {NoStop}%
\bibitem [{\citenamefont {Qin}\ \emph {et~al.}(2012)\citenamefont {Qin},
  \citenamefont {Zhou},\ and\ \citenamefont {Shi}}]{Qin2012}%
  \BibitemOpen
  \bibfield  {author} {\bibinfo {author} {\bibfnamefont {Tao}\ \bibnamefont
  {Qin}}, \bibinfo {author} {\bibfnamefont {Jianhui}\ \bibnamefont {Zhou}}, \
  and\ \bibinfo {author} {\bibfnamefont {Junren}\ \bibnamefont {Shi}},\
  }\bibfield  {title} {\enquote {\bibinfo {title} {Berry curvature and the
  phonon {H}all effect},}\ }\href@noop {} {\bibfield  {journal} {\bibinfo
  {journal} {Phys. Rev. B}\ }\textbf {\bibinfo {volume} {86}},\ \bibinfo
  {pages} {104305} (\bibinfo {year} {2012})}\BibitemShut {NoStop}%
\bibitem [{\citenamefont {Romh{\'a}nyi}\ \emph {et~al.}(2015)\citenamefont
  {Romh{\'a}nyi}, \citenamefont {Penc},\ and\ \citenamefont
  {Ganesh}}]{Romhanyi2015}%
  \BibitemOpen
  \bibfield  {author} {\bibinfo {author} {\bibfnamefont {Judit}\ \bibnamefont
  {Romh{\'a}nyi}}, \bibinfo {author} {\bibfnamefont {Karlo}\ \bibnamefont
  {Penc}}, \ and\ \bibinfo {author} {\bibfnamefont {Ramachandran}\ \bibnamefont
  {Ganesh}},\ }\bibfield  {title} {\enquote {\bibinfo {title} {Hall effect of
  triplons in a dimerized quantum magnet},}\ }\href@noop {} {\bibfield
  {journal} {\bibinfo  {journal} {Nat. Commun.}\ }\textbf {\bibinfo {volume}
  {6}},\ \bibinfo {pages} {6805} (\bibinfo {year} {2015})}\BibitemShut
  {NoStop}%
\bibitem [{\citenamefont {Malki}\ and\ \citenamefont
  {Schmidt}(2017)}]{Malki2017}%
  \BibitemOpen
  \bibfield  {author} {\bibinfo {author} {\bibfnamefont {M}~\bibnamefont
  {Malki}}\ and\ \bibinfo {author} {\bibfnamefont {K.~P}\ \bibnamefont
  {Schmidt}},\ }\bibfield  {title} {\enquote {\bibinfo {title} {Magnetic
  {C}hern bands and triplon {H}all effect in an extended {S}hastry-{S}utherland
  model},}\ }\href@noop {} {\bibfield  {journal} {\bibinfo  {journal} {Phys.
  Rev. B}\ }\textbf {\bibinfo {volume} {95}},\ \bibinfo {pages} {195137}
  (\bibinfo {year} {2017})}\BibitemShut {NoStop}%
\bibitem [{\citenamefont {McClarty}\ \emph {et~al.}(2017)\citenamefont
  {McClarty}, \citenamefont {Kr{\"u}ger}, \citenamefont {Guidi}, \citenamefont
  {Parker}, \citenamefont {Refson}, \citenamefont {Parker}, \citenamefont
  {Prabhakaran},\ and\ \citenamefont {Coldea}}]{McClarty2017}%
  \BibitemOpen
  \bibfield  {author} {\bibinfo {author} {\bibfnamefont {Paul~A}\ \bibnamefont
  {McClarty}}, \bibinfo {author} {\bibfnamefont {F}~\bibnamefont {Kr{\"u}ger}},
  \bibinfo {author} {\bibfnamefont {Tatiana}\ \bibnamefont {Guidi}}, \bibinfo
  {author} {\bibfnamefont {SF}~\bibnamefont {Parker}}, \bibinfo {author}
  {\bibfnamefont {Keith}\ \bibnamefont {Refson}}, \bibinfo {author}
  {\bibfnamefont {AW}~\bibnamefont {Parker}}, \bibinfo {author} {\bibfnamefont
  {Dharmalingam}\ \bibnamefont {Prabhakaran}}, \ and\ \bibinfo {author}
  {\bibfnamefont {Radu}\ \bibnamefont {Coldea}},\ }\bibfield  {title} {\enquote
  {\bibinfo {title} {Topological triplon modes and bound states in a
  {S}hastry--{S}utherland magnet},}\ }\href@noop {} {\bibfield  {journal}
  {\bibinfo  {journal} {Nat. Phys.}\ }\textbf {\bibinfo {volume} {13}},\
  \bibinfo {pages} {736--741} (\bibinfo {year} {2017})}\BibitemShut {NoStop}%
\bibitem [{\citenamefont {Joshi}\ and\ \citenamefont
  {Schnyder}(2019)}]{joshi2019}%
  \BibitemOpen
  \bibfield  {author} {\bibinfo {author} {\bibfnamefont {Darshan~G}\
  \bibnamefont {Joshi}}\ and\ \bibinfo {author} {\bibfnamefont {Andreas~P}\
  \bibnamefont {Schnyder}},\ }\bibfield  {title} {\enquote {\bibinfo {title}
  {{$\mathbb Z_{2}$} topological quantum paramagnet on a honeycomb bilayer},}\
  }\href@noop {} {\bibfield  {journal} {\bibinfo  {journal} {Phys. Rev. B}\
  }\textbf {\bibinfo {volume} {100}},\ \bibinfo {pages} {020407(R)} (\bibinfo
  {year} {2019})}\BibitemShut {NoStop}%
\bibitem [{\citenamefont {Sun}\ \emph {et~al.}(2021)\citenamefont {Sun},
  \citenamefont {Sengupta}, \citenamefont {Nam},\ and\ \citenamefont
  {Yang}}]{Sun2021}%
  \BibitemOpen
  \bibfield  {author} {\bibinfo {author} {\bibfnamefont {Hao}\ \bibnamefont
  {Sun}}, \bibinfo {author} {\bibfnamefont {Pinaki}\ \bibnamefont {Sengupta}},
  \bibinfo {author} {\bibfnamefont {Donguk}\ \bibnamefont {Nam}}, \ and\
  \bibinfo {author} {\bibfnamefont {Bo}~\bibnamefont {Yang}},\ }\bibfield
  {title} {\enquote {\bibinfo {title} {Negative thermal {H}all conductance in a
  two-dimer {S}hastry-{S}utherland model with a {$\pi$}-flux {D}irac
  triplon},}\ }\href@noop {} {\bibfield  {journal} {\bibinfo  {journal} {Phys.
  Rev. B}\ }\textbf {\bibinfo {volume} {103}},\ \bibinfo {pages} {L140404}
  (\bibinfo {year} {2021})}\BibitemShut {NoStop}%
\bibitem [{\citenamefont {Bhowmick}\ and\ \citenamefont
  {Sengupta}(2021)}]{Bhowmick2021}%
  \BibitemOpen
  \bibfield  {author} {\bibinfo {author} {\bibfnamefont {Dhiman}\ \bibnamefont
  {Bhowmick}}\ and\ \bibinfo {author} {\bibfnamefont {Pinaki}\ \bibnamefont
  {Sengupta}},\ }\bibfield  {title} {\enquote {\bibinfo {title} {Weyl triplons
  in {S}r{C}u{$_{2}$}({B}{O}{$_{3}$}){$_{2}$}},}\ }\href@noop {} {\bibfield
  {journal} {\bibinfo  {journal} {Phys. Rev. B}\ }\textbf {\bibinfo {volume}
  {104}},\ \bibinfo {pages} {085121} (\bibinfo {year} {2021})}\BibitemShut
  {NoStop}%
\bibitem [{\citenamefont {Suetsugu}\ \emph {et~al.}(2022)\citenamefont
  {Suetsugu}, \citenamefont {Yokoi}, \citenamefont {Totsuka}, \citenamefont
  {Ono}, \citenamefont {Tanaka}, \citenamefont {Kasahara}, \citenamefont
  {Kasahara}, \citenamefont {Chengchao}, \citenamefont {Kageyama},\ and\
  \citenamefont {Matsuda}}]{Suetsugu2022}%
  \BibitemOpen
  \bibfield  {author} {\bibinfo {author} {\bibfnamefont {S}~\bibnamefont
  {Suetsugu}}, \bibinfo {author} {\bibfnamefont {T}~\bibnamefont {Yokoi}},
  \bibinfo {author} {\bibfnamefont {K}~\bibnamefont {Totsuka}}, \bibinfo
  {author} {\bibfnamefont {T}~\bibnamefont {Ono}}, \bibinfo {author}
  {\bibfnamefont {I}~\bibnamefont {Tanaka}}, \bibinfo {author} {\bibfnamefont
  {S}~\bibnamefont {Kasahara}}, \bibinfo {author} {\bibfnamefont
  {Y}~\bibnamefont {Kasahara}}, \bibinfo {author} {\bibfnamefont
  {Z}~\bibnamefont {Chengchao}}, \bibinfo {author} {\bibfnamefont
  {H}~\bibnamefont {Kageyama}}, \ and\ \bibinfo {author} {\bibfnamefont
  {Y}~\bibnamefont {Matsuda}},\ }\bibfield  {title} {\enquote {\bibinfo {title}
  {Intrinsic suppression of the topological thermal {H}all effect in an exactly
  solvable quantum magnet},}\ }\href@noop {} {\bibfield  {journal} {\bibinfo
  {journal} {Phys. Rev. B}\ }\textbf {\bibinfo {volume} {105}},\ \bibinfo
  {pages} {024415} (\bibinfo {year} {2022})}\BibitemShut {NoStop}%
\bibitem [{\citenamefont {Thomasen}\ \emph {et~al.}(2021)\citenamefont
  {Thomasen}, \citenamefont {Penc}, \citenamefont {Shannon},\ and\
  \citenamefont {Romh{\'a}nyi}}]{thomasen2021}%
  \BibitemOpen
  \bibfield  {author} {\bibinfo {author} {\bibfnamefont {Andreas}\ \bibnamefont
  {Thomasen}}, \bibinfo {author} {\bibfnamefont {Karlo}\ \bibnamefont {Penc}},
  \bibinfo {author} {\bibfnamefont {Nic}\ \bibnamefont {Shannon}}, \ and\
  \bibinfo {author} {\bibfnamefont {Judit}\ \bibnamefont {Romh{\'a}nyi}},\
  }\bibfield  {title} {\enquote {\bibinfo {title} {Fragility of {$\mathbb
  Z_{2}$} topological invariant characterizing triplet excitations in a bilayer
  kagome magnet},}\ }\href@noop {} {\bibfield  {journal} {\bibinfo  {journal}
  {Phys. Rev. B}\ }\textbf {\bibinfo {volume} {104}},\ \bibinfo {pages}
  {104412} (\bibinfo {year} {2021})}\BibitemShut {NoStop}%
\bibitem [{\citenamefont {Saha-Dasgupta}\ and\ \citenamefont
  {Valenti}(2002)}]{saha2002comparative}%
  \BibitemOpen
  \bibfield  {author} {\bibinfo {author} {\bibfnamefont {T}~\bibnamefont
  {Saha-Dasgupta}}\ and\ \bibinfo {author} {\bibfnamefont {R}~\bibnamefont
  {Valenti}},\ }\bibfield  {title} {\enquote {\bibinfo {title} {Comparative
  study between two quantum spin systems {K}{C}u{C}l{$_{3}$} and
  {T}l{C}u{C}l{$_{3}$}},}\ }\href@noop {} {\bibfield  {journal} {\bibinfo
  {journal} {EPL}\ }\textbf {\bibinfo {volume} {60}},\ \bibinfo {pages} {309}
  (\bibinfo {year} {2002})}\BibitemShut {NoStop}%
\bibitem [{Note1()}]{Note1}%
  \BibitemOpen
  \bibinfo {note} {See Supplemental Material for details}\BibitemShut {NoStop}%
\bibitem [{\citenamefont {Kaplan}\ and\ \citenamefont
  {Mahanti}(2011)}]{Kaplan2011}%
  \BibitemOpen
  \bibfield  {author} {\bibinfo {author} {\bibfnamefont {Thomas~A}\
  \bibnamefont {Kaplan}}\ and\ \bibinfo {author} {\bibfnamefont {Subhendra~D}\
  \bibnamefont {Mahanti}},\ }\bibfield  {title} {\enquote {\bibinfo {title}
  {Canted-spin-caused electric dipoles: {A} local symmetry theory},}\
  }\href@noop {} {\bibfield  {journal} {\bibinfo  {journal} {Phys. Rev. B}\
  }\textbf {\bibinfo {volume} {83}},\ \bibinfo {pages} {174432} (\bibinfo
  {year} {2011})}\BibitemShut {NoStop}%
\bibitem [{Note2()}]{Note2}%
  \BibitemOpen
  \bibinfo {note} {In the later analysis, we only focus on the lowest triplon
  mode. Since the $z$-component of the electric field-induced intradimer DM
  interaction term barely affects the lowest mode, we can ignore the $z$
  component in the present analysis.}\BibitemShut {Stop}%
\bibitem [{Note4()}]{Note4}%
  \BibitemOpen
  \bibinfo {note} {The values in Table \ref {polarizationtensor} are taken from
  \cite {Kimura2020}. Although the values in Table \ref {polarizationtensor}
  look different from the values in \cite {Kimura2020}, they are consistent if
  we use the same coordinate system as in \cite {Kimura2020}.}\BibitemShut
  {Stop}%
\bibitem [{\citenamefont {Sachdev}\ and\ \citenamefont
  {Bhatt}(1990)}]{Sachdev1990}%
  \BibitemOpen
  \bibfield  {author} {\bibinfo {author} {\bibfnamefont {Subir}\ \bibnamefont
  {Sachdev}}\ and\ \bibinfo {author} {\bibfnamefont {R.~N}\ \bibnamefont
  {Bhatt}},\ }\bibfield  {title} {\enquote {\bibinfo {title} {Bond-operator
  representation of quantum spins: {M}ean-field theory of frustrated quantum
  {H}eisenberg antiferromagnets},}\ }\href@noop {} {\bibfield  {journal}
  {\bibinfo  {journal} {Phys. Rev. B}\ }\textbf {\bibinfo {volume} {41}},\
  \bibinfo {pages} {9323} (\bibinfo {year} {1990})}\BibitemShut {NoStop}%
\bibitem [{Note7()}]{Note7}%
  \BibitemOpen
  \bibinfo {note} {We used the method of Ref. \cite {fukui2005}.}\BibitemShut
  {Stop}%
\bibitem [{\citenamefont {Lottermoser}\ \emph {et~al.}(2004)\citenamefont
  {Lottermoser}, \citenamefont {Lonkai}, \citenamefont {Amann}, \citenamefont
  {Hohlwein}, \citenamefont {Ihringer},\ and\ \citenamefont
  {Fiebig}}]{Lottermoser2004}%
  \BibitemOpen
  \bibfield  {author} {\bibinfo {author} {\bibfnamefont {Thomas}\ \bibnamefont
  {Lottermoser}}, \bibinfo {author} {\bibfnamefont {Thomas}\ \bibnamefont
  {Lonkai}}, \bibinfo {author} {\bibfnamefont {Uwe}\ \bibnamefont {Amann}},
  \bibinfo {author} {\bibfnamefont {Dietmar}\ \bibnamefont {Hohlwein}},
  \bibinfo {author} {\bibfnamefont {J{\"o}rg}\ \bibnamefont {Ihringer}}, \ and\
  \bibinfo {author} {\bibfnamefont {Manfred}\ \bibnamefont {Fiebig}},\
  }\bibfield  {title} {\enquote {\bibinfo {title} {Magnetic phase control by an
  electric field},}\ }\href@noop {} {\bibfield  {journal} {\bibinfo  {journal}
  {Nature}\ }\textbf {\bibinfo {volume} {430}},\ \bibinfo {pages} {541--544}
  (\bibinfo {year} {2004})}\BibitemShut {NoStop}%
\bibitem [{Note8()}]{Note8}%
  \BibitemOpen
  \bibinfo {note} {The BdG Hamiltonian does not exactly return to the original
  one if $\phi _{2}-\phi _{1}\protect \neq \pm \pi $. However, the difference
  from $\pm \pi $ is usually small enough to be neglected.}\BibitemShut {Stop}%
\bibitem [{com()}]{comment1}%
  \BibitemOpen
  \href@noop {} {}\bibinfo {note} {The qualitative explanation is as follows.
  The swapping between the ground state of sublattice 1 and 2 effectively
  corresponds to the sign change of the symmetry-allowed interdimer DM
  interaction due to its antisymmetry. This leads to the reversal of fictitious
  fluxes induced by the interdimer DM interaction (see also the Supplemental
  Material for the explanation of fictitious fluxes), resulting in the time
  reversal operation.}\BibitemShut {Stop}%
\bibitem [{Note9()}]{Note9}%
  \BibitemOpen
  \bibinfo {note} {The ground state (\ref {variationalGS}) drastically changes
  around $|E_{y}| = 0$, where $\kappa _{zx}$ shows a sharp change.}\BibitemShut
  {Stop}%
\bibitem [{Note10()}]{Note10}%
  \BibitemOpen
  \bibinfo {note} {As expected from Fig. \ref {fig:Berrycurvature}, the Chern
  numbers for these two bands are zero. Nontheless, the thermal Hall effect can
  manifest owing to the bosonic nature, i.e., thermally excited triplons
  contribute to the thermal Hall conductivity through the Bose distribution
  function $\rho (E_{n}(\protect \bm {k})) = 1/(e^{\beta E_{n}(\protect \bm
  {k})}-1)$ as seen in Eq. (\ref {thermalHalleffect}).}\BibitemShut {Stop}%
\bibitem [{\citenamefont {Samajdar}\ \emph {et~al.}(2019)\citenamefont
  {Samajdar}, \citenamefont {Chatterjee}, \citenamefont {Sachdev},\ and\
  \citenamefont {Scheurer}}]{Samajdar2019}%
  \BibitemOpen
  \bibfield  {author} {\bibinfo {author} {\bibfnamefont {Rhine}\ \bibnamefont
  {Samajdar}}, \bibinfo {author} {\bibfnamefont {Shubhayu}\ \bibnamefont
  {Chatterjee}}, \bibinfo {author} {\bibfnamefont {Subir}\ \bibnamefont
  {Sachdev}}, \ and\ \bibinfo {author} {\bibfnamefont {Mathias~S}\ \bibnamefont
  {Scheurer}},\ }\bibfield  {title} {\enquote {\bibinfo {title} {Thermal {H}all
  effect in square-lattice spin liquids: {A} {S}chwinger boson mean-field
  study},}\ }\href@noop {} {\bibfield  {journal} {\bibinfo  {journal} {Phys.
  Rev. B}\ }\textbf {\bibinfo {volume} {99}},\ \bibinfo {pages} {165126}
  (\bibinfo {year} {2019})}\BibitemShut {NoStop}%
\bibitem [{\citenamefont {Volkov}\ \emph {et~al.}(2018)\citenamefont {Volkov},
  \citenamefont {Sheka}, \citenamefont {Gaididei}, \citenamefont {Kravchuk},
  \citenamefont {R{\"o}{\ss}ler}, \citenamefont {Fassbender},\ and\
  \citenamefont {Makarov}}]{Volkov2018}%
  \BibitemOpen
  \bibfield  {author} {\bibinfo {author} {\bibfnamefont {Oleksii~M}\
  \bibnamefont {Volkov}}, \bibinfo {author} {\bibfnamefont {Denis~D}\
  \bibnamefont {Sheka}}, \bibinfo {author} {\bibfnamefont {Yuri}\ \bibnamefont
  {Gaididei}}, \bibinfo {author} {\bibfnamefont {Volodymyr~P}\ \bibnamefont
  {Kravchuk}}, \bibinfo {author} {\bibfnamefont {Ulrich~K}\ \bibnamefont
  {R{\"o}{\ss}ler}}, \bibinfo {author} {\bibfnamefont {J{\"u}rgen}\
  \bibnamefont {Fassbender}}, \ and\ \bibinfo {author} {\bibfnamefont {Denys}\
  \bibnamefont {Makarov}},\ }\bibfield  {title} {\enquote {\bibinfo {title}
  {Mesoscale {D}zyaloshinskii-{M}oriya interaction: geometrical tailoring of
  the magnetochirality},}\ }\href@noop {} {\bibfield  {journal} {\bibinfo
  {journal} {Sci. Rep.}\ }\textbf {\bibinfo {volume} {8}},\ \bibinfo {pages}
  {866} (\bibinfo {year} {2018})}\BibitemShut {NoStop}%
\bibitem [{\citenamefont {Qi}\ \emph {et~al.}(2020)\citenamefont {Qi},
  \citenamefont {Jiang},\ and\ \citenamefont {Mi}}]{qi2020tunable}%
  \BibitemOpen
  \bibfield  {author} {\bibinfo {author} {\bibfnamefont {Shengmei}\
  \bibnamefont {Qi}}, \bibinfo {author} {\bibfnamefont {Jiawei}\ \bibnamefont
  {Jiang}}, \ and\ \bibinfo {author} {\bibfnamefont {Wenbo}\ \bibnamefont
  {Mi}},\ }\bibfield  {title} {\enquote {\bibinfo {title} {Tunable valley
  polarization, magnetic anisotropy and {D}zyaloshinskii--{M}oriya interaction
  in two-dimensional intrinsic ferromagnetic {J}anus 2{H}-{V}{S}e{X} ({X} {$=$}
  {S}, {T}e) monolayers},}\ }\href@noop {} {\bibfield  {journal} {\bibinfo
  {journal} {Phys. Chem. Chem. Phys.}\ }\textbf {\bibinfo {volume} {22}},\
  \bibinfo {pages} {23597--23608} (\bibinfo {year} {2020})}\BibitemShut
  {NoStop}%
\bibitem [{\citenamefont {Xu}\ \emph {et~al.}(2021)\citenamefont {Xu},
  \citenamefont {Wang}, \citenamefont {Xu},\ and\ \citenamefont
  {Hu}}]{xu2021effect}%
  \BibitemOpen
  \bibfield  {author} {\bibinfo {author} {\bibfnamefont {Chong}\ \bibnamefont
  {Xu}}, \bibinfo {author} {\bibfnamefont {Qian-Jun}\ \bibnamefont {Wang}},
  \bibinfo {author} {\bibfnamefont {Bin}\ \bibnamefont {Xu}}, \ and\ \bibinfo
  {author} {\bibfnamefont {Jun}\ \bibnamefont {Hu}},\ }\bibfield  {title}
  {\enquote {\bibinfo {title} {Effect of biaxial strain and hydrostatic
  pressure on the magnetic properties of bilayer {C}r{I}{$_{3}$}},}\
  }\href@noop {} {\bibfield  {journal} {\bibinfo  {journal} {Front. Phys.}\
  }\textbf {\bibinfo {volume} {16}},\ \bibinfo {pages} {53502} (\bibinfo {year}
  {2021})}\BibitemShut {NoStop}%
\bibitem [{\citenamefont {Xing}\ \emph {et~al.}(2022)\citenamefont {Xing},
  \citenamefont {Chen}, \citenamefont {Xu}, \citenamefont {Li},\ and\
  \citenamefont {Zhang}}]{Xing2022}%
  \BibitemOpen
  \bibfield  {author} {\bibinfo {author} {\bibfnamefont {Yuheng}\ \bibnamefont
  {Xing}}, \bibinfo {author} {\bibfnamefont {Hao}\ \bibnamefont {Chen}},
  \bibinfo {author} {\bibfnamefont {Ning}\ \bibnamefont {Xu}}, \bibinfo
  {author} {\bibfnamefont {Xiao}\ \bibnamefont {Li}}, \ and\ \bibinfo {author}
  {\bibfnamefont {Lifa}\ \bibnamefont {Zhang}},\ }\bibfield  {title} {\enquote
  {\bibinfo {title} {Valley modulation and single-edge transport of magnons in
  breathing kagome ferromagnets},}\ }\href@noop {} {\bibfield  {journal}
  {\bibinfo  {journal} {Phys. Rev. B}\ }\textbf {\bibinfo {volume} {105}},\
  \bibinfo {pages} {104409} (\bibinfo {year} {2022})}\BibitemShut {NoStop}%
\bibitem [{\citenamefont {Fukui}\ \emph {et~al.}(2005)\citenamefont {Fukui},
  \citenamefont {Hatsugai},\ and\ \citenamefont {Suzuki}}]{fukui2005}%
  \BibitemOpen
  \bibfield  {author} {\bibinfo {author} {\bibfnamefont {Takahiro}\
  \bibnamefont {Fukui}}, \bibinfo {author} {\bibfnamefont {Yasuhiro}\
  \bibnamefont {Hatsugai}}, \ and\ \bibinfo {author} {\bibfnamefont {Hiroshi}\
  \bibnamefont {Suzuki}},\ }\bibfield  {title} {\enquote {\bibinfo {title}
  {Chern {N}umbers in {D}iscretized {B}rillouin {Z}one: {E}fficient {M}ethod of
  {C}omputing ({S}pin) {H}all {C}onductances},}\ }\href@noop {} {\bibfield
  {journal} {\bibinfo  {journal} {J. Phys. Soc. Jpn.}\ }\textbf {\bibinfo
  {volume} {74}},\ \bibinfo {pages} {1674--1677} (\bibinfo {year}
  {2005})}\BibitemShut {NoStop}%
\bibitem [{Note11()}]{Note11}%
  \BibitemOpen
  \bibinfo {note} {See Sec. \ref {chap:Validity_P} for more
  details.}\BibitemShut {Stop}%
\bibitem [{\citenamefont {Solovyev}\ \emph {et~al.}(2021)\citenamefont
  {Solovyev}, \citenamefont {Ono},\ and\ \citenamefont
  {Nikolaev}}]{solovyev2021magnetically}%
  \BibitemOpen
  \bibfield  {author} {\bibinfo {author} {\bibfnamefont {Igor}\ \bibnamefont
  {Solovyev}}, \bibinfo {author} {\bibfnamefont {Ryota}\ \bibnamefont {Ono}}, \
  and\ \bibinfo {author} {\bibfnamefont {Sergey}\ \bibnamefont {Nikolaev}},\
  }\bibfield  {title} {\enquote {\bibinfo {title} {Magnetically {I}nduced
  {P}olarization in {C}entrosymmetric {B}onds},}\ }\href@noop {} {\bibfield
  {journal} {\bibinfo  {journal} {Phys. Rev. Lett.}\ }\textbf {\bibinfo
  {volume} {127}},\ \bibinfo {pages} {187601} (\bibinfo {year}
  {2021})}\BibitemShut {NoStop}%
\bibitem [{\citenamefont {Lee}\ \emph {et~al.}(2015)\citenamefont {Lee},
  \citenamefont {Han},\ and\ \citenamefont {Lee}}]{lee2015thermal}%
  \BibitemOpen
  \bibfield  {author} {\bibinfo {author} {\bibfnamefont {Hyunyong}\
  \bibnamefont {Lee}}, \bibinfo {author} {\bibfnamefont {Jung~Hoon}\
  \bibnamefont {Han}}, \ and\ \bibinfo {author} {\bibfnamefont {Patrick~A}\
  \bibnamefont {Lee}},\ }\bibfield  {title} {\enquote {\bibinfo {title}
  {Thermal {H}all effect of spins in a paramagnet},}\ }\href@noop {} {\bibfield
   {journal} {\bibinfo  {journal} {Phys. Rev. B}\ }\textbf {\bibinfo {volume}
  {91}},\ \bibinfo {pages} {125413} (\bibinfo {year} {2015})}\BibitemShut
  {NoStop}%
\bibitem [{\citenamefont {Owerre}(2016)}]{owerre2016magnon}%
  \BibitemOpen
  \bibfield  {author} {\bibinfo {author} {\bibfnamefont {SA}~\bibnamefont
  {Owerre}},\ }\bibfield  {title} {\enquote {\bibinfo {title} {Magnon {H}all
  effect in {A}{B}-stacked bilayer honeycomb quantum magnets},}\ }\href@noop {}
  {\bibfield  {journal} {\bibinfo  {journal} {Phys. Rev. B}\ }\textbf {\bibinfo
  {volume} {94}},\ \bibinfo {pages} {094405} (\bibinfo {year}
  {2016})}\BibitemShut {NoStop}%
\bibitem [{\citenamefont {Buzo}\ and\ \citenamefont
  {Doretto}(2024)}]{buzo2023thermal}%
  \BibitemOpen
  \bibfield  {author} {\bibinfo {author} {\bibfnamefont {Lucas~S}\ \bibnamefont
  {Buzo}}\ and\ \bibinfo {author} {\bibfnamefont {RL}~\bibnamefont {Doretto}},\
  }\bibfield  {title} {\enquote {\bibinfo {title} {Thermal {H}all conductivity
  of a valence bond solid phase in the square lattice {$J_{1}$}-{$J_{2}$}
  antiferromagnet {H}eisenberg model with a {D}zyaloshinskii-{M}oriya
  interaction},}\ }\href@noop {} {\bibfield  {journal} {\bibinfo  {journal}
  {Physical Review B}\ }\textbf {\bibinfo {volume} {109}},\ \bibinfo {pages}
  {134405} (\bibinfo {year} {2024})}\BibitemShut {NoStop}%
\bibitem [{\citenamefont {Zak}(1989)}]{zak1989berry}%
  \BibitemOpen
  \bibfield  {author} {\bibinfo {author} {\bibfnamefont {J}~\bibnamefont
  {Zak}},\ }\bibfield  {title} {\enquote {\bibinfo {title} {Berry’s phase for
  energy bands in solids},}\ }\href@noop {} {\bibfield  {journal} {\bibinfo
  {journal} {Phys. Rev. Lett.}\ }\textbf {\bibinfo {volume} {62}},\ \bibinfo
  {pages} {2747} (\bibinfo {year} {1989})}\BibitemShut {NoStop}%
\bibitem [{\citenamefont {Hatsugai}(2006)}]{hatsugai2006quantized}%
  \BibitemOpen
  \bibfield  {author} {\bibinfo {author} {\bibfnamefont {Yasuhiro}\
  \bibnamefont {Hatsugai}},\ }\bibfield  {title} {\enquote {\bibinfo {title}
  {Quantized {B}erry {P}hases as a {L}ocal {O}rder {P}arameter of a {Q}uantum
  {L}iquid},}\ }\href@noop {} {\bibfield  {journal} {\bibinfo  {journal} {J.
  Phys. Soc. Jpn.}\ }\textbf {\bibinfo {volume} {75}},\ \bibinfo {pages}
  {123601} (\bibinfo {year} {2006})}\BibitemShut {NoStop}%
\bibitem [{\citenamefont {Hasan}\ and\ \citenamefont
  {Kane}(2010)}]{hasan2010colloquium}%
  \BibitemOpen
  \bibfield  {author} {\bibinfo {author} {\bibfnamefont {M~Zahid}\ \bibnamefont
  {Hasan}}\ and\ \bibinfo {author} {\bibfnamefont {Charles~L}\ \bibnamefont
  {Kane}},\ }\bibfield  {title} {\enquote {\bibinfo {title} {Colloquium:
  topological insulators},}\ }\href@noop {} {\bibfield  {journal} {\bibinfo
  {journal} {Rev. Mod. Phys.}\ }\textbf {\bibinfo {volume} {82}},\ \bibinfo
  {pages} {3045} (\bibinfo {year} {2010})}\BibitemShut {NoStop}%
\bibitem [{\citenamefont {Sun}\ \emph {et~al.}(2012)\citenamefont {Sun},
  \citenamefont {Liu}, \citenamefont {Hemmerich},\ and\ \citenamefont
  {Das~Sarma}}]{sun2012topological}%
  \BibitemOpen
  \bibfield  {author} {\bibinfo {author} {\bibfnamefont {Kai}\ \bibnamefont
  {Sun}}, \bibinfo {author} {\bibfnamefont {W~Vincent}\ \bibnamefont {Liu}},
  \bibinfo {author} {\bibfnamefont {Andreas}\ \bibnamefont {Hemmerich}}, \ and\
  \bibinfo {author} {\bibfnamefont {S}~\bibnamefont {Das~Sarma}},\ }\bibfield
  {title} {\enquote {\bibinfo {title} {Topological semimetal in a fermionic
  optical lattice},}\ }\href@noop {} {\bibfield  {journal} {\bibinfo  {journal}
  {Nat. Phys.}\ }\textbf {\bibinfo {volume} {8}},\ \bibinfo {pages} {67--70}
  (\bibinfo {year} {2012})}\BibitemShut {NoStop}%
\bibitem [{\citenamefont {Hou}(2013)}]{hou2013hidden}%
  \BibitemOpen
  \bibfield  {author} {\bibinfo {author} {\bibfnamefont {Jing-Min}\
  \bibnamefont {Hou}},\ }\bibfield  {title} {\enquote {\bibinfo {title}
  {Hidden-{S}ymmetry-{P}rotected {T}opological {S}emimetals on a {S}quare
  {L}attice},}\ }\href@noop {} {\bibfield  {journal} {\bibinfo  {journal}
  {Phys. Rev. Lett.}\ }\textbf {\bibinfo {volume} {111}},\ \bibinfo {pages}
  {130403} (\bibinfo {year} {2013})}\BibitemShut {NoStop}%
\bibitem [{\citenamefont {Wagner}\ \emph {et~al.}(2017)\citenamefont {Wagner},
  \citenamefont {Dangel}, \citenamefont {Cartarius}, \citenamefont {Main},\
  and\ \citenamefont {Wunner}}]{wagner2017numerical}%
  \BibitemOpen
  \bibfield  {author} {\bibinfo {author} {\bibfnamefont {Marcel}\ \bibnamefont
  {Wagner}}, \bibinfo {author} {\bibfnamefont {Felix}\ \bibnamefont {Dangel}},
  \bibinfo {author} {\bibfnamefont {Holger}\ \bibnamefont {Cartarius}},
  \bibinfo {author} {\bibfnamefont {J{\"o}rg}\ \bibnamefont {Main}}, \ and\
  \bibinfo {author} {\bibfnamefont {G{\"u}nter}\ \bibnamefont {Wunner}},\
  }\bibfield  {title} {\enquote {\bibinfo {title} {Numerical calculation of the
  complex berry phase in non-{H}ermitian systems},}\ }\href@noop {} {\bibfield
  {journal} {\bibinfo  {journal} {arXiv:1708.03230}\ } (\bibinfo {year}
  {2017})}\BibitemShut {NoStop}%
\bibitem [{\citenamefont {Dangel}\ \emph {et~al.}(2018)\citenamefont {Dangel},
  \citenamefont {Wagner}, \citenamefont {Cartarius}, \citenamefont {Main},\
  and\ \citenamefont {Wunner}}]{dangel2018topological}%
  \BibitemOpen
  \bibfield  {author} {\bibinfo {author} {\bibfnamefont {Felix}\ \bibnamefont
  {Dangel}}, \bibinfo {author} {\bibfnamefont {Marcel}\ \bibnamefont {Wagner}},
  \bibinfo {author} {\bibfnamefont {Holger}\ \bibnamefont {Cartarius}},
  \bibinfo {author} {\bibfnamefont {J{\"o}rg}\ \bibnamefont {Main}}, \ and\
  \bibinfo {author} {\bibfnamefont {G{\"u}nter}\ \bibnamefont {Wunner}},\
  }\bibfield  {title} {\enquote {\bibinfo {title} {Topological invariants in
  dissipative extensions of the {S}u-{S}chrieffer-{H}eeger model},}\
  }\href@noop {} {\bibfield  {journal} {\bibinfo  {journal} {Phys. Rev. A}\
  }\textbf {\bibinfo {volume} {98}},\ \bibinfo {pages} {013628} (\bibinfo
  {year} {2018})}\BibitemShut {NoStop}%
\bibitem [{\citenamefont {Hirayama}\ \emph {et~al.}(2018)\citenamefont
  {Hirayama}, \citenamefont {Okugawa},\ and\ \citenamefont
  {Murakami}}]{hirayama2018topological}%
  \BibitemOpen
  \bibfield  {author} {\bibinfo {author} {\bibfnamefont {Motoaki}\ \bibnamefont
  {Hirayama}}, \bibinfo {author} {\bibfnamefont {Ryo}\ \bibnamefont {Okugawa}},
  \ and\ \bibinfo {author} {\bibfnamefont {Shuichi}\ \bibnamefont {Murakami}},\
  }\bibfield  {title} {\enquote {\bibinfo {title} {Topological {S}emimetals
  {S}tudied by {A}b {I}nitio {C}alculations},}\ }\href@noop {} {\bibfield
  {journal} {\bibinfo  {journal} {J. Phys. Soc. Jpn.}\ }\textbf {\bibinfo
  {volume} {87}},\ \bibinfo {pages} {041002} (\bibinfo {year}
  {2018})}\BibitemShut {NoStop}%
\bibitem [{\citenamefont {Kawabata}\ \emph {et~al.}(2019)\citenamefont
  {Kawabata}, \citenamefont {Shiozaki}, \citenamefont {Ueda},\ and\
  \citenamefont {Sato}}]{kawabata2019symmetry}%
  \BibitemOpen
  \bibfield  {author} {\bibinfo {author} {\bibfnamefont {Kohei}\ \bibnamefont
  {Kawabata}}, \bibinfo {author} {\bibfnamefont {Ken}\ \bibnamefont
  {Shiozaki}}, \bibinfo {author} {\bibfnamefont {Masahito}\ \bibnamefont
  {Ueda}}, \ and\ \bibinfo {author} {\bibfnamefont {Masatoshi}\ \bibnamefont
  {Sato}},\ }\bibfield  {title} {\enquote {\bibinfo {title} {Symmetry and
  {T}opology in {N}on-{H}ermitian {P}hysics},}\ }\href@noop {} {\bibfield
  {journal} {\bibinfo  {journal} {Phys. Rev. X}\ }\textbf {\bibinfo {volume}
  {9}},\ \bibinfo {pages} {041015} (\bibinfo {year} {2019})}\BibitemShut
  {NoStop}%
\bibitem [{\citenamefont {Tsubota}\ \emph {et~al.}(2022)\citenamefont
  {Tsubota}, \citenamefont {Yang}, \citenamefont {Akagi},\ and\ \citenamefont
  {Katsura}}]{tsubota2022symmetry}%
  \BibitemOpen
  \bibfield  {author} {\bibinfo {author} {\bibfnamefont {Shoichi}\ \bibnamefont
  {Tsubota}}, \bibinfo {author} {\bibfnamefont {Hong}\ \bibnamefont {Yang}},
  \bibinfo {author} {\bibfnamefont {Yutaka}\ \bibnamefont {Akagi}}, \ and\
  \bibinfo {author} {\bibfnamefont {Hosho}\ \bibnamefont {Katsura}},\
  }\bibfield  {title} {\enquote {\bibinfo {title} {Symmetry-protected
  quantization of complex {B}erry phases in non-{H}ermitian many-body
  systems},}\ }\href@noop {} {\bibfield  {journal} {\bibinfo  {journal} {Phys.
  Rev. B}\ }\textbf {\bibinfo {volume} {105}},\ \bibinfo {pages} {L201113}
  (\bibinfo {year} {2022})}\BibitemShut {NoStop}%
\bibitem [{\citenamefont {Okuma}(2023)}]{okuma2023bosonic}%
  \BibitemOpen
  \bibfield  {author} {\bibinfo {author} {\bibfnamefont {Nobuyuki}\
  \bibnamefont {Okuma}},\ }\bibfield  {title} {\enquote {\bibinfo {title}
  {Bosonic {A}ndreev bound state},}\ }\href@noop {} {\bibfield  {journal}
  {\bibinfo  {journal} {arXiv:2310.09197}\ } (\bibinfo {year}
  {2023})}\BibitemShut {NoStop}%
\bibitem [{Note12()}]{Note12}%
  \BibitemOpen
  \bibinfo {note} {We used the method in Ref. \cite {fukui2005}.}\BibitemShut
  {Stop}%
\bibitem [{Note13()}]{Note13}%
  \BibitemOpen
  \bibinfo {note} {We can usually ignore the effect of $|\protect \bm {E}|$ on
  the phase factor $\phi _{m}+\Phi _{m}$ for a fixed direction of the electric
  field.}\BibitemShut {Stop}%
\end{thebibliography}%

\widetext
\renewcommand{\theequation}{S\arabic{equation}}
\renewcommand{\thefigure}{S\arabic{figure}}
\setcounter{equation}{0}
\setcounter{figure}{0}

\begin{center}
	\textbf{\large Supplemental Material for: “Electric field induced thermal Hall effect of triplons in the quantum dimer magnets \textit{X}CuCl$_{3}$ (\textit{X} = Tl,\ K)”}
\end{center}

\widetext
\renewcommand{\theequation}{S\arabic{equation}}
\setcounter{equation}{0}
\renewcommand{\thefigure}{S\arabic{figure}}
\setcounter{figure}{0}

\section{Symmetry analysis}
\subsection{Crystal structure and transformation property of spins under the symmetry}\label{sec:DMsymmetry}
\begin{figure}[H]
    \centering
    \includegraphics[width=0.9\linewidth]{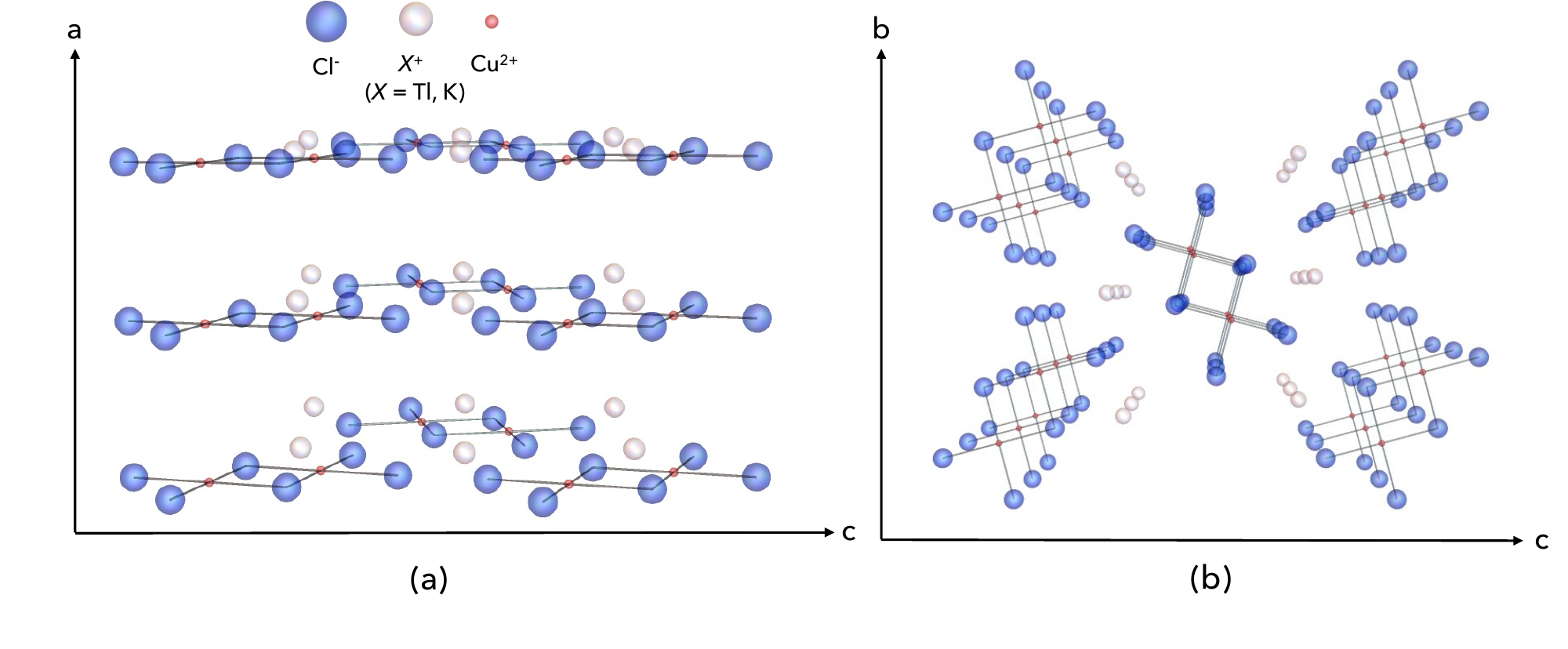}
    \caption{Crystal sructure of $X$CuCl$_{3}$: (a) $ac$ plane; (b) $bc$ plane. The blue, white, and red balls represent Cl$^{-}$, $X^{+}$ ($X=$ Tl, K), and Cu$^{2+}$ ions, respectively. The $S=1/2$ spins are carried by Cu$^{2+}$ ions.}
    \label{fig:Crystal_structure}
\end{figure}

\begin{figure}[H]
 \centering
  \includegraphics[width=0.5\linewidth]{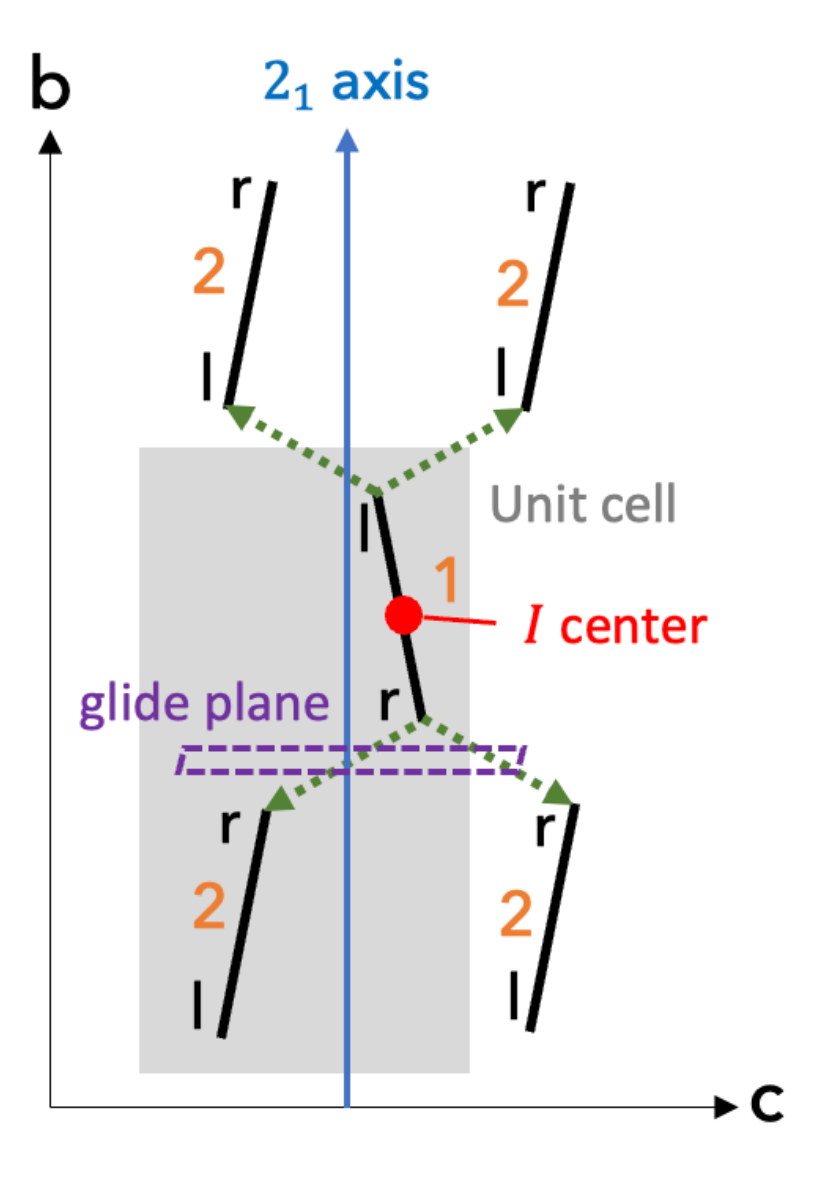}
  \caption{A schematic picture of symmetries in $X$CuCl$_{3}$ in the $bc$ plane. The gray rectangle indicates the structural unit cell. The dotted green arrows indicate the sign convention for $\bm{D}^{\mathrm{int}}$. The red point represents the inversion center, which is centered at each dimer. The solid blue arrow indicates the $2_{1}$ spiral axis parallel to the $\hat{b}(z)$ axis. The purple dotted rhombus indicates the glide plane perpendicular to the $\hat{b}(z)$ axis.
  }
  \label{fig:S1}
\end{figure}

\indent \textit{X}CuCl$_{3}$ (\textit{X} = Tl,\ K) has a monoclinic crystal structure [see Fig. \ref{fig:Crystal_structure}] with space group $P2_{1}/c$ \cite{Tanaka2001, cavadini2001, cavadini2002, ruegg2003}. This space group has four symmetry operations, $E$, $I(0,0,0)$, $2_{1}(b:0,*,-1/4)$, and $c(-b/4)$ apart from Bravais lattice translations [see Fig. \ref{fig:S1}]. We consider the transformation property of spins under $I(0,0,0)$, $2_{1}(b:0,*,-1/4)$, and $c(-b/4)$. Let us denote by $S^{m}_{\mu,l/r}(\bm{R})$ the $\mu(=x,y,z)$ component of the left/right spin of the dimer in the unit cell at the position $\bm{R}$ on the sublattice $m$ ($\{\bm{R},m\}$). The spins are transformed under $I(0,0,0)$ as follows:
\begin{equation}\label{inversion}
   I(0,0,0):\begin{cases}
    S^{1}_{\mu,l/r}(\bm{R}) \rightarrow S^{1}_{\mu,r/l}(\bm{R})\\
    S^{2}_{\mu,l/r}(\bm{R}) \rightarrow S^{2}_{\mu,r/l}(\bm{R}+2\hat{a}+\hat{b}+\hat{c})
   \end{cases}.
\end{equation}
Since $I(0,0,0)$ is the inversion operation, the spins do not change sign. The symmetry operation $2_{1}(b:0,*,-1/4)$ corresponds to a $\pi$ rotation about an axis parallel to $\hat{b}(z)$ and a half translation along the axis. Therefore, two of the three components of the spins change sign:
\begin{equation}\label{spiral}
   2_{1}(b:0,*,-1/4):\begin{cases}
    S^{1}_{\mu,l/r}(\bm{R}) \rightarrow \delta_{\mu}S^{2}_{\mu,r/l}(\bm{R}+\hat{a}+\hat{b})\\
    S^{2}_{\mu,l/r}(\bm{R}) \rightarrow \delta_{\mu}S^{1}_{\mu,r/l}(\bm{R}+\hat{a})
   \end{cases},
\end{equation}
where $\delta_{x} = \delta_{y} = -1$ and $\delta_{z} = 1$. The symmetry operation $c(-b/4)$ represents a reflection through a glide plane perpendicular to the $\hat{b}(z)$ axis and a half translation along the $\hat{c}$ axis. Hence, two of the three components of the spins change sign:
\begin{equation}\label{glide}
   c(-b/4):\begin{cases}
    S^{1}_{\mu,l/r}(\bm{R}) \rightarrow \delta_{\mu}S^{2}_{\mu,l/r}(\bm{R}+\hat{a}+\hat{c})\\
    S^{2}_{\mu,l/r}(\bm{R}) \rightarrow \delta_{\mu}S^{1}_{\mu,l/r}(\bm{R}-\hat{a})
   \end{cases}.
\end{equation}

\subsection{The relative signs of the inter-dimer DM vectors}
\indent We can determine the relative signs of the inter-dimer DM vectors using Eqs. (\ref{inversion}), (\ref{spiral}), and (\ref{glide}). The inter-dimer DM interaction term [Eq. (3) in the main text] can be written as
\begin{equation}
\sum_{\eta=0,1}
D^{\mathrm{int}}_{r}(\eta(2\hat{a}+\hat{c}))\hat{z}\cdot [\bm{S}^{1}_{r}(\bm{R})\times \bm{S}^{2}_{r}(\bm{R}+\eta(2\hat{a}+\hat{c}))]+D^{\mathrm{int}}_{l}(\hat{b}+\eta(2\hat{a}+\hat{c}))\hat{z}\cdot [\bm{S}^{1}_{l}(\bm{R})\times \bm{S}^{2}_{l}(\bm{R}+\hat{b}+\eta(2\hat{a}+\hat{c}))],
\end{equation}
where $D^{\mathrm{int}}_{\alpha}(\bm{d}) = |\bm{D}^{\mathrm{int}}_{\alpha}(\bm{d})|$ ($\alpha=l, r$). Here, we note that the Hamiltonian is unchanged under the symmetry operation. Applying $I(0,0,0)$, we find
\begin{equation}
    D^{\mathrm{int}}_{r}(\bm{0}) = D^{\mathrm{int}}_{l}(2\hat{a}+\hat{b}+\hat{c}).
\end{equation}
while applying $2_{1}(b:0,*,-1/4)$, $c(-b/4)$, and translations, we find
\begin{equation}
    D^{\mathrm{int}}_{r}(\bm{0}) = -D^{\mathrm{int}}_{l}(\hat{b}),
\end{equation}
and
\begin{equation}
    D^{\mathrm{int}}_{r}(\bm{0}) = -D^{\mathrm{int}}_{r}(2\hat{a}+\hat{c}).
\end{equation}
The obtained results are summarized in Fig. 1 (b) in the main text.

\subsection{Derivation of Eqs. (6) and (7)}
\indent Here, we derive Eqs. (6) and (7) of the main text. The local polarization on dimers $\{\bm{R},m\}$ is expressed using the polarization tensor and the left and right spins of the dimer \footnote{See Sec. \ref{chap:Validity_P} for more details.}
\begin{equation}\label{polarizationdef}
\bm{P}^{m}(\bm{R})=\tilde{C}^{m}(\bm{S}^{m}_{l}(\bm{R})\times\bm{S}^{m}_{r}(\bm{R})).
\end{equation}
Also, the local polarization on dimers $\{\bm{R},m\}$ are transformed under $I(0,0,0)$, $2_{1}(b:0,*,-1/4)$, and $c(-b/4)$ as follows:
\begin{equation}\label{pinversion}
   I(0,0,0):\begin{cases}
   P^{1}_{\mu}(\bm{R})\rightarrow -P^{1}_{\mu}(\bm{R})\\
   P^{2}_{\mu}(\bm{R})\rightarrow -P^{2}_{\mu}(\bm{R}+2\hat{a}+\hat{b}+\hat{c})
   \end{cases},
\end{equation}
\begin{equation}\label{pspiral}
   2_{1}(b:0,*,-1/4):\begin{cases}
   P^{1}_{\mu}(\bm{R})\rightarrow \delta_{\mu}P^{2}_{\mu}(\bm{R}+\hat{a}+\hat{b})\\
   P^{2}_{\mu}(\bm{R})\rightarrow \delta_{\mu}P^{1}_{\mu}(\bm{R}+\hat{a})
   \end{cases},
\end{equation}
\begin{equation}\label{pglide}
   c(-b/4):\begin{cases}
   P^{1}_{\mu}(\bm{R})\rightarrow -\delta_{\mu}P^{2}_{\mu}(\bm{R}+\hat{a}+\hat{c})\\
   P^{2}_{\mu}(\bm{R})\rightarrow -\delta_{\mu}P^{1}_{\mu}(\bm{R}-\hat{a})
   \end{cases},
\end{equation}
where $\delta_{x}=\delta_{y}=-1$, and $\delta_{z}=1$. We can determine how the polarization tensor $\tilde{C}^{m}$ is transformed under these symmetry operations by the above equations.

Using Eqs. (\ref{inversion}), (\ref{polarizationdef}), and (\ref{pinversion}), we find that the symmetry under $I(0,0,0)$ does not impose any constraints on the elements of the polarization tensor $\tilde{C}^{m}$. Thus, we can express the polarization tensor of the sublattice $1$ by nine independent components $C_{\mu\nu}^{1} (\mu, \nu=x,y,z)$ \cite{Kaplan2011, Kimura2016, Kimura2017, Kimura2018, Kimura2020}
\begin{equation}\label{C1matrix}
  \tilde{C}^{1}=
    \begin{pmatrix}
    C_{xx}^{1} & C_{xy}^{1} & C_{xz}^{1}\\
    C_{yx}^{1} & C_{yy}^{1} & C_{yz}^{1}\\
    C_{zx}^{1} & C_{zy}^{1} & C_{zz}^{1}\\
    \end{pmatrix}.
\end{equation}
Applying $2_{1}(b:0,*,-1/4)$, $c(-b/4)$, and translations, we can obtain the matrix expression of the polarization tensor $\tilde{C}^{2}$ in terms of $C_{\mu\nu}^{1}$ in (\ref{C1matrix}) as
\begin{equation}\label{C2matrix}
  \tilde{C}^{2} =
    \begin{pmatrix}
    -C_{xx}^{1} & -C_{xy}^{1} & C_{xz}^{1}\\
    -C_{yx}^{1} & -C_{yy}^{1} & C_{yz}^{1}\\
    C_{zx}^{1} & C_{zy}^{1} & -C_{zz}^{1}\\
    \end{pmatrix}.
\end{equation}
Here, we used Eqs. (\ref{spiral}), (\ref{glide}), (\ref{polarizationdef}), (\ref{pspiral}), and (\ref{pglide}).
By substituting Eqs. (\ref{C1matrix}), and (\ref{C2matrix}) into Eq. (5) in the main text, we obtain Eq. (6) in the main text.\\

\section{The validity of the representation of the polarization operator}\label{chap:Validity_P}
\indent Here we 
provide a justification for introducing the polarization operator (\ref{polarizationdef}) or the electric field-induced intra-dimer DM interaction [Eq. (5) in the main text]. From the viewpoint of time-reversal symmetry, spin-dependent electric polarization is most commonly described by quadratic terms in spin operators, particularly the vector spin chirality term $\bm{S}_{l}(\bm{R})\times \bm{S}_{r}(\bm{R})$, which is capable of generating electric polarization irrespective of the local symmetry between two spins \cite{tokura2014multiferroics, Kimura2017}. Although the vector spin chirality $\bm{S}_{l}(\bm{R})\times \bm{S}_{r}(\bm{R})$ is zero in a spin singlet state, it has a non-zero matrix element between the spin singlet and triplet states \cite{Kimura2016, Kimura2017, Kimura2018, Kimura2020}. Therefore, from the point of view of symmetry, it is generally expected that the magnetic order due to the BEC of triplons would induce vector spin chirality-type electric polarization of the form $\tilde{C}\cdot (\bm{S}_{l}(\bm{R})\times \bm{S}_{r}(\bm{R}))$, where $\tilde C$ is the $3\times 3$ tensor appearing in Eq. \ref{polarizationdef} \cite{Kaplan2011, Kimura2016, Kimura2017}. 
Theoretical models assuming this form of polarization operator not only reproduce the experimentally observed spontaneous polarization in $X$CuCl$_3$ \cite{Kimura2016, Kimura2017} but also well explain various experimental results in these materials, including ESR spectrum \cite{Kimura2018} and nonreciprocal directional dichroism \cite{Kimura2020}. For these reasons, the assumption of $\tilde{C}\cdot (\bm{S}_{l}(\bm{R})\times \bm{S}_{r}(\bm{R}))$ type electric polarization (\ref{polarizationdef}) in this model is phenomenological but eminently justifiable.

\section{Undetermined parameters of the polarization tensor}
\label{chap:Net_polarization}
\indent Here, we delve into why certain components of the polarization tensor were left undetermined in previous studies. To explain this, we first express the net electric polarization induced in the system employing the polarization tensor $C^{1}_{\mu\nu}$ in Eq. (7) in the main text. By calculating the expectation value of Eq. (\ref{polarizationdef}) with respect to the ground state [Eq. (9) in the main text], we arrive at the following expression
\begin{equation}\label{Eq:net_polarization}
    \begin{split}
        (\bm{P}^{1}(\bm{R})+\bm{P}^{2}(\bm{R}))_{x} &= \frac{1}{\sqrt{8}}[-\sin2\theta_{1}(\sin\phi_{1}C^{1}_{xx}+\cos\phi_{1}C^{1}_{xy})+\sin2\theta_{2}(\sin\phi_{2}C^{1}_{xx}+\cos\phi_{2}C^{1}_{xy})],\\
        (\bm{P}^{1}(\bm{R})+\bm{P}^{2}(\bm{R}))_{y} &= \frac{1}{\sqrt{8}}[-\sin2\theta_{1}(\sin\phi_{1}C^{1}_{yx}+\cos\phi_{1}C^{1}_{yy})+\sin2\theta_{2}(\sin\phi_{2}C^{1}_{yx}+\cos\phi_{2}C^{1}_{yy})],\\
        (\bm{P}^{1}(\bm{R})+\bm{P}^{2}(\bm{R}))_{z} &= \frac{1}{\sqrt{8}}[-\sin2\theta_{1}(\sin\phi_{1}C^{1}_{zx}+\cos\phi_{1}C^{1}_{zy})+\sin2\theta_{2}(\sin\phi_{2}C^{1}_{zx}+\cos\phi_{2}C^{1}_{zy})].
    \end{split}
\end{equation}
Assuming that $\theta_{1} = \theta_{2} = \theta$ and $\phi_{2} - \phi_{1} = \pm \pi$, which are satisfied in the absence of an electric field \cite{Matsumoto2002, Oosawa2002, Matsumoto2004} [see also Sec. \ref{sec:variational_eq}], we can calculate the spontaneous polarization as
\begin{equation}\label{Eq:net_spontaneous_polarization}
    \begin{split}
        (\bm{P}^{1}(\bm{R})+\bm{P}^{2}(\bm{R}))_x &= -\frac{1}{\sqrt{2}} \sin2\theta (\sin\phi_{1} C^{1}_{xx} + \cos\phi_{1} C^{1}_{xy}),\\
        (\bm{P}^{1}(\bm{R})+\bm{P}^{2}(\bm{R}))_y &= -\frac{1}{\sqrt{2}} \sin2\theta (\sin\phi_{1} C^{1}_{yx} + \cos\phi_{1} C^{1}_{yy}),\\
        (\bm{P}^{1}(\bm{R})+\bm{P}^{2}(\bm{R}))_z &= 0,
    \end{split} 
\end{equation}
which implies the presence of glide symmetry $c(b/4)$ in the BEC phase of \textit{X}CuCl$_3$ \cite{Kimura2017}. The absence of $C^{1}_{zx}$ and $C^{1}_{zy}$ in the equation above results in the indeterminacy of them
in previous experiments \cite{Kimura2016, Kimura2017, Kimura2018, Kimura2020}. 
To determine them, it is necessary to observe the polarization induced by a strong electric field, where $\theta_{1} = \theta_{2} = \theta$ and $\phi_{2} - \phi_{1} = \pm \pi$ do not hold. 
We anticipate that measuring the thermal Hall conductivity induced by an electric field will provide constraints on the hitherto unknown elements of the polarization tensor $C_{zx}^{1}$ and $C_{zy}^{1}$ in \textit{X}CuCl$_{3}$, offering a valuable direction for future studies. An alternative method for determining the polarization tensor involves computing the polarization induced by an electric field using the atomic orbital model derived from density functional theory calculations, following a similar approach as in the previous study \cite{solovyev2021magnetically}.

\section{Qualitative picture of the electric field induced thermal Hall effect}
\begin{figure}[H]
    \centering
    \includegraphics[width=0.8\linewidth]{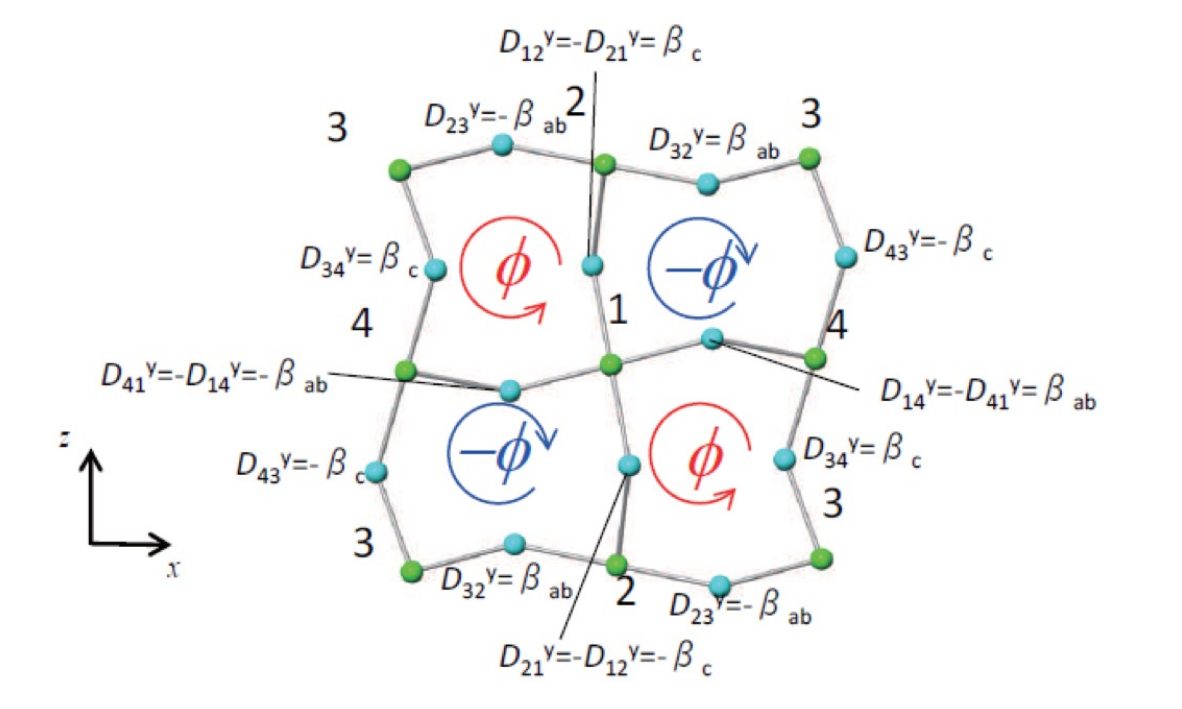}
    \caption{Fictitious fluxes induced by the DM interactions in the pseudocubic $zx$ plane of the GdFeO$_{3}$-type distorted perovskite structure, with a magnetic field applied along the pseudocubic $y$ axis. The positive direction of fluxes is taken to be counterclockwise. This figure is taken from Ref. \cite{Ideue2012}.}
    \label{fig:Fluxes_distoted_perovskite}
\end{figure}

\indent Here, we provide a qualitative understanding of the electric field induced thermal Hall effect, particularly in the context of the \textit{no-go} condition for magnons in ferromagnets. Previous studies \cite{Onose2010, Ideue2012, Kawano2019} have demonstrated that the thermal Hall effect of magnons is anticipated when the DM interaction introduces complex hopping terms in the free-magnon Hamiltonian, expressed as $iDSb^{\dag}_{\bm{R}_i}b_{\bm{R}_j} + h.c.$ (where $D$ represents the magnitude of the DM vector $\bm{D}(\bm{R}_j - \bm{R}_i)$ and $S$ is the magnitude of the spin). The complex phase factor $\phi = D/J$ ($J$ is the magnitude of the exchange interaction) acts as an effective vector potential or fictitious flux for magnons. However, previous studies \cite{Ideue2012, Kawano2019} have also shown that the thermal Hall effect of magnons is absent in ferromagnets with edge-sharing geometry such as square, honeycomb, and cubic lattices, despite the presence of the DM interactions inducing fictitious fluxes. This limitation is known as the \textit{no-go} condition for magnons in ferromagnets. To illustrate this, we examine the GdFeO$_{3}$-type distorted perovskite structure as an example \cite{Ideue2012}.
Fig. \ref{fig:Fluxes_distoted_perovskite} shows the fictitious fluxes caused by the DM interactions in this structure.

In this pseudocubic $zx$ plane, magnons gain the phase of $\phi$ (red character) when moving on the loop $1\rightarrow 2\rightarrow 3\rightarrow 4\rightarrow 1$ in Fig. \ref{fig:Fluxes_distoted_perovskite}, whereas the phase of $-\phi$ (blue character) when going around on the adjacent loop $1\rightarrow 4\rightarrow 3\rightarrow 2\rightarrow 1$ due to $\bm{D}(\bm{R}_{j}-\bm{R}_{i})=-\bm{D}(\bm{R}_{i}-\bm{R}_{j})$. The absence of the thermal Hall effect in the system with the staggered flux pattern can be roughly understood as follows: 
if we translate the system by half a unit lattice vector in the $x$ direction and rotate the plane around the $x$ axis by $\pi$, the original flux pattern is restored. 
This implies $\kappa_{zx}=-\kappa_{zx}$, and thus the thermal Hall conductivity 
vanishes identically. As this example suggests, the crystal symmetry combined with 
time-reversal symmetry cancels out the effect of the flux pattern induced by the DM interactions in 
systems with edge-sharing lattices. Consequently, the thermal Hall effect of magnons has mainly been proposed in the corner-sharing kagome \cite{Katsura2010, Ideue2012, Mook2014, hirschberger2015, lee2015thermal, owerre2017, laurell2018magnon} and pyrochlore lattices \cite{Onose2010, Ideue2012, laurell2017}, and also the Haldane-type honeycomb lattice \cite{kim2016realization, owerre2016magnon}.

\begin{figure}[H]
    \centering
    \includegraphics[width=1.0\linewidth]{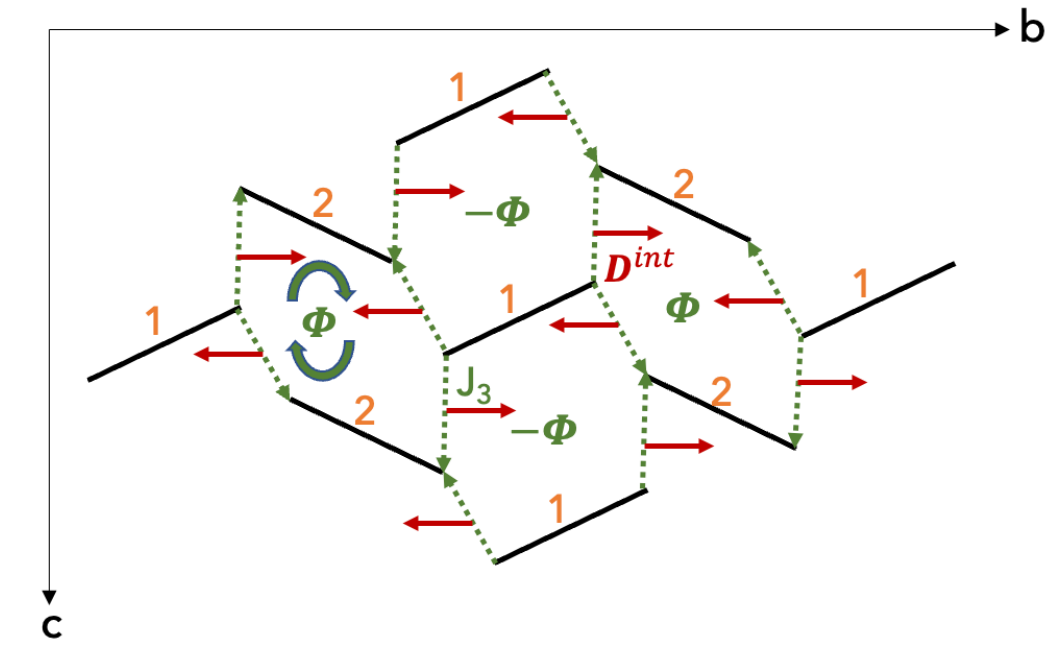}
    \caption{Fictitious fluxes $\Phi$ induced by the inter-dimer DM interaction in the $bc$ plane of \textit{X}CuCl$_3$. The thick black lines indicate the dimers and the solid brown (dotted green) arrows denote the direction of (sign convention for) inter-dimer DM interaction. The positive direction of fluxes is taken to be clockwise.}
    \label{fig:Fictious_fluxes_XCuCl3}
\end{figure}

Although the above \textit{no-go} condition does not directly apply to antiferromagnets \cite{Kawano2019} nor quantum dimer magnets \cite{buzo2023thermal}, similar restrictions exist regarding the lattice geometries. Indeed, \textit{X}CuCl$_3$ do not exhibit the thermal Hall effect without an electric field, even if the symmetry-allowed inter-dimer DM interaction [Eq. (3) in the main text] generates the complex phase factor or fictitious flux for triplons as depicted in Fig. \ref{fig:Fictious_fluxes_XCuCl3}. From 
Fig. (\ref{fig:Fictious_fluxes_XCuCl3}), we can find that the \textit{X}CuCl$_3$ have a similar flux pattern as 
the GdFeO$_3$-type distorted perovskite structure in Fig. \ref{fig:Fluxes_distoted_perovskite}. Indeed, the absence of the thermal Hall effect in \textit{X}CuCl$_{3}$ without an electric field can be qualitatively understood as follows. If we apply the symmetry operation that reflects the system through a glide plane perpendicular to the $\hat{b}(z)$ axis and translate the system by half a 
unit lattice vector in the direction of $\hat{c}$, the flux pattern returns to the original one. This implies that $\kappa_{zx} = - \kappa_{zx}$, and thus the thermal Hall conductivity is zero. Therefore, in the absence of an electric field, the crystal symmetry, in combination with 
time-reversal symmetry, cancels out the effect of fictitious fluxes induced by the inter-dimer DM interaction. However, the electric field-induced DM interaction [Eq. (5) in the main text] breaks this symmetry, resulting in the finite thermal Hall effect. A more rigorous justification based on the effective PT symmetry of the BdG Hamiltonian is given in Sec. \ref{sec:absence_thermal_Hall_withoutE}.

\section{The explicit expression of the constant term and the variational equations}\label{sec:variational_eq}
\indent Here, we give the explicit expression of $\Hamzero$ and the variational equations. The constant term $\Hamzero$ is expressed as
\begin{equation}
\begin{split}
    &\Hamzero= -\frac{3J}{2}+\left(J-g\mu_{B}H-\frac{J_{1}+J_{2}}{2}\right)(\sin^{2}\theta_{1}+\sin^{2}\theta_{2})+\frac{3(J_{1}+J_{2})}{4}(\sin^{4}\theta_{1}+\sin^{4}\theta_{2})\\ &+J_{3}[\sin^{2}\theta_{1}\sin^{2}\theta_{2}+2\cos(\phi_{2}-\phi_{1})\sin\theta_{1}\cos\theta_{1}\sin\theta_{2}\cos\theta_{2}]\\ &-\frac{\sin\theta_{1}\cos\theta_{1}}{\sqrt{2}}D^{\mathrm{ext},1}\sin(\phi_{1}+\Phi_{1})-\frac{\sin\theta_{2}\cos\theta_{2}}{\sqrt{2}}D^{\mathrm{ext},2}\sin(\phi_{2}+\Phi_{2}),
\end{split}\label{b0}
\end{equation} 
where $D^{\mathrm{ext},m}$ $(m=1,2)$ is the absolute value of the electric field-induced intra-dimer DM vector projected on the $x$-$y$ plane, and $\Phi_{m}=\arctan(D^{\mathrm{ext}, m}_{y}/D^{\mathrm{ext}, m}_{x})$. Also, the variational equations for $\theta_{1}$ and $\phi_{1}$, which can be obtained by differentiation of the constant term (\ref{b0}), are expressed as
\begin{equation}
\begin{split}
    \frac{\partial \Hamzero}{\partial\theta_{1}} &= 2(J-g\mu_{B}H)\sin\theta_{1}\cos\theta_{1}-(J_{1}+J_{2})\sin\theta_{1}\cos\theta_{1}+3(J_{1}+J_{2})\sin^{3}\theta_{1}\cos\theta_{1}\\ &+2J_{3}(\sin\theta_{1}\cos\theta_{1}\sin^{2}\theta_{2}+\cos(\phi_{2}-\phi_{1})\cos{2\theta_{1}}\sin\theta_{2}\cos\theta_{2})\\ &-\frac{\cos^{2}\theta_{1}-\sin^{2}\theta_{1}}{\sqrt{2}}D^{\mathrm{ext},1}\sin(\phi_{1}+\Phi_{1})\\
    &=0,
\end{split}\label{variationalequation1}
\end{equation}
\begin{equation}\label{variationalequationphi1}
    \begin{split}
\frac{\partial \Hamzero}{\partial\phi_{1}} &= 2J_{3}\sin(\phi_{2}-\phi_{1})\sin{\theta_{1}}\cos{\theta_{1}}\sin{\theta_{2}}\cos{\theta_{2}}-\frac{\sin{\theta_{1}}\cos{\theta_{1}}}{\sqrt{2}}D^{\mathrm{ext,1}}\cos(\phi_{1}+\Phi_{1})\\ &= 0.
    \end{split}
\end{equation}
The variational equations for $\theta_{2}$ and $\phi_{2}$ can also be obtained by exchanging the sublattice indices $1$ and $2$ in Eqs. (\ref{variationalequation1}) and (\ref{variationalequationphi1}). By using the above variational equations (\ref{variationalequation1}) and (\ref{variationalequationphi1}), the variational parameters in the absence of an electric field can be written as follows
\begin{equation}\label{Eq:Variational_withoutE}
    \begin{split}
        \theta_1 = \theta_2 &= \arcsin\sqrt{\frac{J_1 + J_2 + 2J_3 + 2(g\mu_{B}H-J)}{3(J_1 + J_2)+6J_3}},\\
        \phi_2 - \phi_1 &= \pm\pi,
    \end{split}
\end{equation}
which are consistent with the previous studies in the high magnetic field regime \cite{Matsumoto2002, Oosawa2002, Matsumoto2004}.

\section{The explicit expression of the BdG Hamiltonian}
\indent Here, we give the explicit expression for the $4\times 4$ BdG Hamiltonian in Eq. (12) in the main text as follows:
\begin{equation}\label{BdGexplicit}
H_{\mathrm{BdG}}(\bm{k})=
    \begin{pmatrix}
    \Xi(\bm{k}) & \Pi(\bm{k}) \\
    \Pi^{*}(-\bm{k})&\Xi^{*}(-\bm{k})\\
    \end{pmatrix},
\end{equation}
where
\begin{gather}
    \Xi(\bm{k})(i,i)=f_{1}(\theta_{i})+f_{2}(\theta_{i})+g_{1}(\theta_{i})\cos(\bm{k}\cdot\bm{d}_{1})+g_{2}(\theta_{i})\cos(\bm{k}\cdot\bm{d}_{2})+h(\theta_{i})+F_{i}+H_{\mathrm{DM}}(\theta_{i},\phi_{i}),\nonumber\\
    \Xi(\bm{k})(1,2)=G(\theta_{1},\theta_{2},\phi_{1},\phi_{2})+G_{z}(\theta_{1},\theta_{2},\phi_{1},\phi_{2}),\\
    \Xi(\bm{k})(2,1)=G(\theta_{2},\theta_{1},\phi_{2},\phi_{1})-G_{z}(\theta_{2},\theta_{1},\phi_{2},\phi_{1}),\nonumber
\end{gather}

\begin{gather}
    \Pi(\bm{k})(i,i)= h_{1}(\theta_{i})\cos(\bm{k}\cdot\bm{d}_{1})+h_{2}(\theta_{i})\cos(\bm{k}\cdot\bm{d}_{2}),\nonumber\\
    \Pi(\bm{k})(1,2)= H(\theta_{1},\theta_{2},\phi_{1},\phi_{2})+H_{z}(\theta_{1},\theta_{2},\phi_{1},\phi_{2}),\\
    \Pi(\bm{k})(2,1)= H(\theta_{1},\theta_{2},\phi_{1},\phi_{2})+H_{z}(\theta_{1},\theta_{2},\phi_{1},\phi_{2}),\nonumber
\end{gather}
with
\begin{gather}
    h(\theta)=(J-g\mu_{B}H)\cos 2\theta,\nonumber\\
    f_{i}(\theta)=\frac{J_{i}}{2}(5\sin^{2}\theta\cos^{2}\theta-\sin^{4}\theta),\nonumber\\
    g_{i}(\theta)=\frac{J_{i}}{2}[\sin^{2}\theta\cos^{2}\theta-(\sin^{4}\theta+\cos^{4}\theta)],\\  
    h_{i}(\theta)=\frac{3J_{i}}{2}\sin^{2}\theta\cos^{2}\theta, \nonumber\\
    H_{\mathrm{DM}}(\theta_{i},\phi_{i})=\sqrt{2}\sin\theta_{i}\cos\theta_{i} D^{\mathrm{ext},i}\sin(\phi_{i}+\Phi_{i}),\nonumber
\end{gather}

\begin{equation}
\begin{split}
    F_{1}&=J_{3}[-\sin 2\theta_{1}\sin 2\theta_{2}\cos(\phi_{2}-\phi_{1})+\cos^{2}\theta_{1}\sin^{2}\theta_{2} -\sin^{2}\theta_{1}\sin^{2}\theta_{2}],\\ 
    F_{2}&=J_{3}[-\sin 2\theta_{1}\sin 2\theta_{2}\cos(\phi_{2}-\phi_{1})+\cos^{2}\theta_{2}\sin^{2}\theta_{1} -\sin^{2}\theta_{1}\sin^{2}\theta_{2}],\\     G(\theta_{1},\theta_{2},\phi_{1},\phi_{2})&=J_{3}[\cos(\bm{k}\cdot\bm{d_{3+}})+\cos(\bm{k}\cdot\bm{d_{3-}})]\left[e^{i(\phi_{2}-\phi_{1})}\cos^{2}\theta_{1}\cos^{2}\theta_{2}+e^{i(\phi_{1}-\phi_{2})}\sin^{2}\theta_{1}\sin^{2}\theta_{2} +\frac{\sin 2\theta_{1}\sin 2\theta_{2}}{4}\right],\\
    G_{z}(\theta_{1},\theta_{2},\phi_{1},\phi_{2})&=\frac{iD^{\mathrm{int}}}{2}[\cos(\bm{k}\cdot\bm{d_{3+}})-\cos(\bm{k}\cdot\bm{d_{3-}})][e^{i(\phi_{2}-\phi_{1})}\cos^{2}\theta_{1}\cos^{2}\theta_{2}-e^{i(\phi_{1}-\phi_{2})}\sin^{2}\theta_{1}\sin^{2}\theta_{2}],\\
    H(\theta_{1},\theta_{2},\phi_{1},\phi_{2})&=J_{3}[\cos(\bm{k}\cdot\bm{d_{3+}})+\cos(\bm{k}\cdot\bm{d_{3-}})]\left[-e^{i(\phi_{2}-\phi_{1})}\cos^{2}\theta_{1}\sin^{2}\theta_{2}-e^{i(\phi_{1}-\phi_{2})}\cos^{2}\theta_{2}\sin^{2}\theta_{1} +\frac{\sin 2\theta_{1}\sin 2\theta_{2}}{4}\right],\\
    H_{z}(\theta_{1},\theta_{2},\phi_{1},\phi_{2})&=\frac{iD^{\mathrm{int}}}{2}[\cos(\bm{k}\cdot\bm{d_{3+}})-\cos(\bm{k}\cdot\bm{d_{3-}})][e^{i(\phi_{1}-\phi_{2})}\sin^{2}\theta_{1}\cos^{2}\theta_{2}-e^{i(\phi_{2}-\phi_{1})}\cos^{2}\theta_{1}\sin^{2}\theta_{2}],
    \end{split}
\end{equation}
where $\bm{k}\cdot\bm{d}_{1}=\frac{k_{x}+k_{y}}{2}$, $\bm{k}\cdot\bm{d}_{2}=2k_{x}$, $\bm{k}\cdot\bm{d}_{3\pm}=k_{z}\pm k_{x}$. \\

\section{Symmetry condition for the nonzero Berry curvature}\label{sec:absence_thermal_Hall_withoutE}
\indent Here, we explain a symmetry condition for the nonzero Berry curvature. The Berry curvature becomes zero when the BdG Hamiltonian in Eq. (12) in the main text has the following effective PT symmetry \cite{Kawano2019, Fujiwara2022}:
\begin{equation}\label{PTsymmetry}
  P^{\dag}H_{\mathrm{BdG}}^{*}(\bm{k})P = H_{\mathrm{BdG}}(\bm{k}),
\end{equation}
where $P$ is a paraunitary matrix satisfying the paraunitary condition $P^{\dag}\Sigma_{z}P = \Sigma_{z}$. Now we prove that the Berry curvature is zero under the above symmetry (\ref{PTsymmetry}). By acting with the paraunitary matrix $P$ on both sides of Eq. (13) in the main text and using Eq. (\ref{PTsymmetry}) and the paraunitary condition $P^{\dag}\Sigma_{z}P = \Sigma_{z}$, we can rewrite Eq. (13) in the main text as
\begin{equation}
    \Sigma_{z} H_{\mathrm{BdG}}(\bm{k}) P^{*}T^{*}(\bm{k}) = P^{*}T^{*}(\bm{k})\Sigma_{z}E(\bm{k}),
\end{equation}
which indicates that $P^{*}T^{*}(\bm{k})$ is also the paraunitary matrix that diagonalizes $\Sigma_{z}H_{\mathrm{BdG}}(\bm{k})$. Specifically, in the absence of degeneracy, $T(\bm{k})$ should satisfy
\begin{equation}\label{Eq:T_k_condition}
    T(\bm{k}) = P^{*}T^{*}(\bm{k})M_{\bm{k}},
\end{equation}
where $(M_{\bm{k}})_{mn} = \delta_{m,n}\exp[i\theta_{m,\bm{k}}]$ comes from the fact that we can choose the overall phases of eigenvectors arbitrarily. 
Taking into account the above condition (\ref{Eq:T_k_condition}), 
one can rewrite the $y$ component of the  Berry curvature $\Omega^{y}_{n}(\bm{k})$ as \cite{Fujiwara2022}
\begin{equation}\label{Eq:Berry_PT_relation}
    \begin{split}
        \Omega^{y}_{n}(\bm{k}) &=-2\mathrm{Im}\left[\Sigma_{z}\frac{\partial T^{\dag}(\bm{k})}{\partial k_{z}}\Sigma_{z}\frac{\partial T(\bm{k})}{\partial k_{x}}\right]_{nn}\\
        &=-2\mathrm{Im}\left[\Sigma_{z}\frac{\partial M^{\dag}_{\bm{k}}T^{T}(\bm{k})}{\partial k_{z}}P^{T}\Sigma_{z}P^{*}\frac{\partial T^{*}(\bm{k})M_{\bm{k}}}{\partial k_{x}}\right]_{nn}\\
        &=2\mathrm{Im}\left[\Sigma_{z}\frac{\partial T^{\dag}(\bm{k})}{\partial k_{z}}\Sigma_{z}\frac{\partial T(\bm{k})}{\partial k_{x}}\right]_{nn} - 2\mathrm{Im}\left[\Sigma_{z} \frac{\partial M^{\dag}_{\bm{k}}}{\partial k_{z}} T^{T}(\bm{k})\Sigma_{z} T^{*}(\bm{k}) \frac{\partial M_{\bm{k}}}{\partial k_{x}}\right]_{nn}\\
        &-2\mathrm{Im}\left[\Sigma_{z}\frac{\partial M^{\dag}_{\bm{k}}}{\partial k_{z}} T^{T}(\bm{k}) \Sigma_{z} \frac{\partial T^{*}(\bm{k})}{\partial k_{x}} M_{\bm{k}} + \Sigma_{z} M^{\dag} (\bm{k}) \frac{\partial T^{T}({\bm{k}})}{\partial k_{z}} \Sigma_{z} T^{*}(\bm{k}) \frac{\partial M_{\bm{k}}}{\partial k_{x}}\right]_{nn}\\
        &=-\Omega^{y}_{n}(\bm{k})-2\mathrm{Im}\left[\Sigma_{z} \frac{\partial M^{\dag}_{\bm{k}}}{\partial k_{z}} \Sigma_{z} \frac{\partial M_{\bm{k}}}{\partial k_{x}}\right]_{nn}\\
        &-2\mathrm{Im}\left[-i\frac{\partial \theta_{n,\bm{k}}}{\partial k_{z}}\Sigma_{z}T^{T}(\bm{k})\Sigma_{z}\frac{\partial T^{*}(\bm{k})}{\partial k_{x}} + i\frac{\partial \theta_{n,\bm{k}}}{\partial k_{x}}\Sigma_{z}\frac{\partial T^{T}(\bm{k})}{\partial k_{z}}\Sigma_{z}T^{*}(\bm{k})\right]_{nn}\\
        & = -\Omega^{y}_{n}(\bm{k})-2\mathrm{Im}\left[-i\frac{\partial \theta_{n,\bm{k}}}{\partial k_{z}}\frac{\partial}{\partial k_{x}}\left[\Sigma_{z}T^{T}(\bm{k})\Sigma_{z}T^{*}(\bm{k})\right]\right]_{nn}\\
        & = -\Omega^{y}_{n}(\bm{k}),
    \end{split}
\end{equation}
where we used the paraunitary condition $T^{\dag}(\bm{k})\Sigma_z T(\bm{k}) = \Sigma_z$ and the antisymmetry of the Berry curvature, i.e., $-2\mathrm{Im}\left[\Sigma_{z}\frac{\partial T^{\dag}(\bm{k})}{\partial k_{z}}\Sigma_{z}\frac{\partial T(\bm{k})}{\partial k_{x}}\right]_{nn} = 2\mathrm{Im}\left[\Sigma_{z}\frac{\partial T^{\dag}(\bm{k})}{\partial k_{x}}\Sigma_{z}\frac{\partial T(\bm{k})}{\partial k_{z}}\right]_{nn}$. We can also rewrite the other components of the Berry curvature $\Omega^{x}_{n}(\bm{k})$ and $\Omega^{z}_{n}(\bm{k})$ in the same way as in Eq. (\ref{Eq:Berry_PT_relation}). The above relation (\ref{Eq:Berry_PT_relation}) concludes that the Berry curvature is zero under the effective PT symmetry (\ref{PTsymmetry}) if there is no degeneracy.

Next, we suppose $P = I_{2\times 2}\otimes\sigma_{1}$ with $I_{2\times 2} = \bigl(
\begin{smallmatrix}
   1 & 0 \\
   0 & 1
\end{smallmatrix}
\bigl)$ and $\sigma_{1} = \bigl(
\begin{smallmatrix}
   0 & 1 \\
   1 & 0
\end{smallmatrix}
\bigl)$. Then the above condition (\ref{PTsymmetry}) is satisfied if
\begin{equation}\label{PTsymmetrydetails}
    \Xi^{3} (\bm{k}) = \Pi^{3} (\bm{k}) = 0,\ \Pi^{i}\in \mathbb{R}\ (\mu = 0,1,2),
\end{equation}
where $\Xi^{i} (\bm{k})$ and $\Pi^{i}(\bm{k})$ are defined by
\begin{equation}
    \begin{split}
    \Xi(\bm{k})&=\Xi^{0}(\bm{k}) I_{2\times 2}+\sum_{n=1}^{3}\Xi^{n}(\bm{k})\sigma_{n}\\
    \Pi(\bm{k})&=\Pi^{0}(\bm{k}) I_{2\times 2}+\sum_{n=1}^{3}\Pi^{n}(\bm{k})\sigma_{n}.    
    \end{split}
\end{equation}
In the above definitions, $\Xi (\bm{k})$ and $\Pi(\bm{k})$ represent the hopping and the pairing terms of the BdG Hamiltonian (\ref{BdGexplicit}), and $\sigma_{n}\ (n=1,2,3)$ are Pauli matrices.

Now we apply the above symmetry condition to the present case. In the absence of an electric field, the variational parameters in Eq.~(9) in the main text satisfy $\theta_{1} = \theta_{2} \equiv \theta$ and $\phi_{1}-\phi_{2} = \pm\pi$. Hence, the BdG Hamiltonian (\ref{BdGexplicit}) satisfies the above condition (\ref{PTsymmetrydetails}). However, applying an electric field changes the variational parameters from $\theta$ to $\theta_{m} = \theta +\epsilon_m$. Let us estimate the difference between the variational parameters, i.e., $\theta_{2}-\theta_{1}$. By subtracting the variational equations (\ref{variationalequation1}) for $\theta_{1}$ and $\theta_{2}$ and keeping only terms first order in $\epsilon_{1}$ and $\epsilon_{2}$, we obtain  
\begin{equation}
\label{epsilon12}
    \theta_{2}-\theta_{1} = \epsilon_{2}-\epsilon_{1}\simeq \frac{-D^{\mathrm{ext},2}\sin(\phi_{2}+\Phi_{2})+D^{\mathrm{ext},1}\sin(\phi_{1}+\Phi_{1})}{\sqrt{2}F(\theta)},
   \end{equation}
\begin{equation}
\label{ftheta}
    \begin{split}
    F(\theta)&=2(J-g\mu_{B}H)-J_{1}-J_{2}+3(J_{1}+J_{2})\frac{3\sin^{2}\theta\cos^{2}\theta-\sin^{4}\theta}{\cos^{2}\theta-\sin^{2}\theta}+2J_{3}\frac{\cos^{2}\theta-2\sin^{2}\theta-4\sin^{2}\theta\cos^{2}\theta}{\cos^{2}\theta-\sin^{2}\theta},
    \end{split}
\end{equation}
where $\theta$ is given in Eq. (\ref{Eq:Variational_withoutE}). From Eq. (\ref{epsilon12}), we see that the electric field induced DM interaction term in Eq. (5) in the main text gives rise to the difference between $\theta_{1}$ and $\theta_{2}$. In this case, the symmetry condition (\ref{PTsymmetrydetails}) is not satisfied, and thus the Berry curvature can be nonzero.

\section{Symmetry protected nodal lines}
\label{sec:Symmetry_nodal_lines}
\indent Here, we demonstrate that the nodal lines $G_{\pm,j}$ are protected by the effective PT symmetry (\ref{PTsymmetry}). First, we establish that the Berry phase $\gamma_{n,C}$, defined on a closed loop $C$ with no degeneracy, is quantized to either $0$ or $\pi$ under the effective PT symmetry (\ref{PTsymmetry}). By considering the condition (\ref{Eq:T_k_condition}), 
we can rewrite the Berry phase $\gamma_{n,C}$ as 
\begin{equation}\label{Eq:quantized_Berry_phase}
    \begin{split}
        \gamma_{n,C} &= \oint_{C} id\bm{k} \left[\Sigma_{z}T^{\dag}(\bm{k})\Sigma_{z}\nabla T(\bm{k}) \right]_{nn}\\
        &=\oint_{C} id\bm{k} \left[\Sigma_{z}M^{\dag}_{k}T^{T}(\bm{k})P^{T}\Sigma_{z}P^{*}\nabla (T^{*}(\bm{k})M_{\bm{k}})\right]_{nn}\\
        &=\oint_{C} id\bm{k} \left[\Sigma_{z}M^{\dag}_{k}T^{T}(\bm{k})\Sigma_{z}\nabla (T^{*}(\bm{k})M_{\bm{k}}) \right]_{nn}\\
        &=\oint_{C} id\bm{k} \left[\Sigma_{z}M^{\dag}_{k}T^{T}(\bm{k})\Sigma_{z}T^{*}(\bm{k})\nabla M_{\bm{k}} \right]_{nn} + \oint_{C} id\bm{k} \left[\Sigma_{z}M^{\dag}_{k}T^{T}(\bm{k})\Sigma_{z}\nabla T^{*}(\bm{k})M_{\bm{k}} \right]_{nn}\\
        &=\oint_{C} id\bm{k} \left[\Sigma_{z}M^{\dag}_{k}\Sigma_{z}\nabla M_{\bm{k}} \right]_{nn} + \oint_{C} id\bm{k} \left[\Sigma_{z}M^{\dag}_{k}T^{T}(\bm{k})\Sigma_{z}\nabla T^{*}(\bm{k})M_{\bm{k}} \right]_{nn}\\
        &=-\oint_{C} d\bm{k} \nabla\theta_{n,\bm{k}} + \oint_{C} id\bm{k} \left[\Sigma_{z}T^{\dag}(\bm{k})\Sigma_{z}\nabla T(\bm{k}) \right]^{*}_{nn}\\
        &= -\gamma_{n,C}\quad \mathrm{mod}\ 2\pi,
    \end{split}
\end{equation}
where we used the paraunitary condition. From the above relation (\ref{Eq:quantized_Berry_phase}), we can see that the Berry phase can take $0$ or $\pi$ since the Berry phase is defined modulo $2\pi$. The quantization of the Berry phase has also been discussed in various fermionic and non-Hermitian systems \cite{zak1989berry, hatsugai2006quantized, hasan2010colloquium, sun2012topological, hou2013hidden, wagner2017numerical, dangel2018topological, hirayama2018topological, kawabata2019symmetry, tsubota2022symmetry, okuma2023bosonic}.

Next, we 
argue that the nodal lines $G_{\pm,j}$ are protected by the quantized Berry phase under the effective PT symmetry (\ref{PTsymmetry}). To show this, 
consider the Berry phase on a closed loop $C_{\sigma,j}$ ($\sigma = \pm$) around the nodal line $G_{\sigma,j}$. This Berry phase only takes $0$ or $\pi$ as explained above since the BdG Hamiltonian has the effective PT symmetry (\ref{PTsymmetry}). Suppose that the Berry phase on a closed loop $C_{\sigma,j}$ is $\pi$. In this case, the band gap cannot be opened at the nodal lines $G_{\pm,j}$ by any continuous perturbation that preserves the effective PT symmetry (\ref{PTsymmetry}). If such perturbations open the band gap at the nodal lines $G_{\pm,j}$, the Berry curvature is zero in the entire Brillouin zone due to the effective symmetry (\ref{PTsymmetry}). This means that the Berry phase on a closed loop $C_{\sigma,j}$ becomes zero. However, such changes are prohibited since the quantized Berry phase cannot be changed continuously. Thus, we can say that the nodal lines are protected by $\pi$ Berry phase in the presence of the effective PT symmetry (\ref{PTsymmetry}). Indeed, we numerically confirm that the Berry phase on a closed loop $C_{\sigma,j}$ is $\pi$ in the present case \footnote{We used the method in Ref. \cite{fukui2005}.}. Therefore, we can conclude that the nodal lines $G_{\pm,j}$ are protected in the presence of the effective PT symmetry (\ref{PTsymmetry}).

\section{Triplon band structure without and with an electric field}
\begin{figure}[H]
    \centering
    \includegraphics[width = 1.0\linewidth]{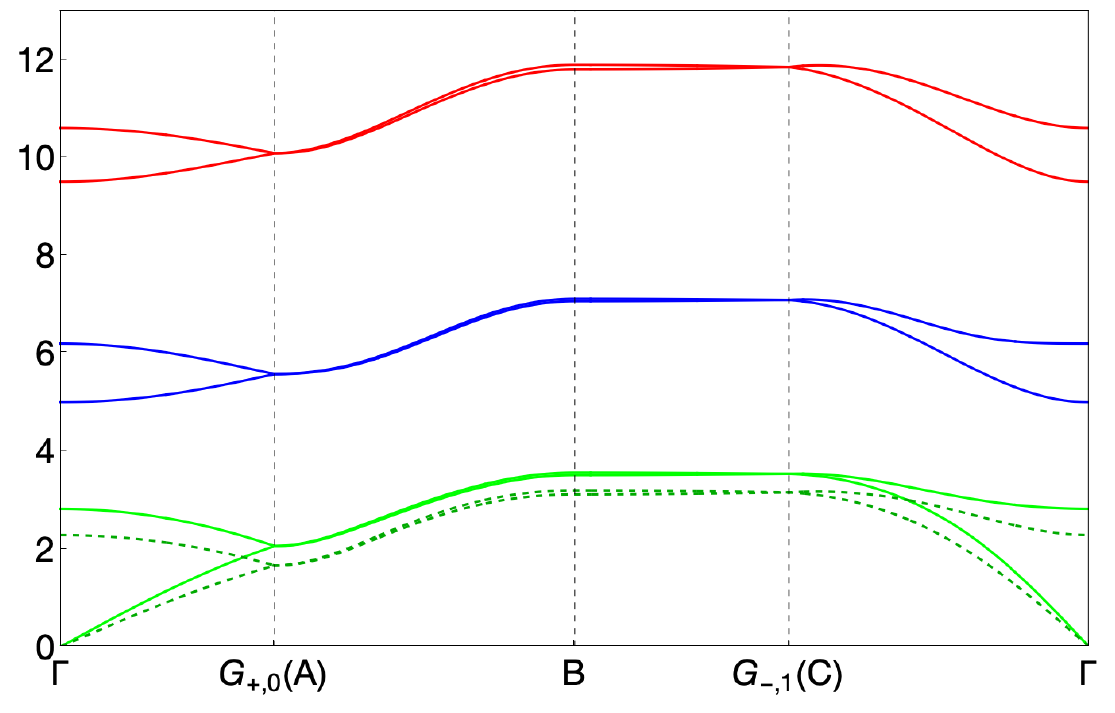}
    \caption{The band structure of $\mathcal{H}^{(2)}$ and BdG Hamiltonian described in the $4\times 4$ matrix without an electric field ($H = \SI{42}{T}$). The energy scale is in $\si{meV}$. The points in reciprocal space are denoted by $\Gamma=(0,0,0)$, $G_{+,0}(\mathrm{A}) = (0,0,\frac{\pi}{2})$, $\mathrm{B}= (\frac{\pi}{2},-\frac{\pi}{2},\frac{\pi}{2})$, and $G_{-,1}(\mathrm{C})=(\frac{\pi}{2},-\frac{\pi}{2},0)$. The solid lines indicate the six energy bands of $\mathcal{H}^{(2)}$, whereas the dashed green lines represent the two energy bands of the BdG Hamiltonian described in the $4\times 4$ matrix, which are also shown in Fig. (\ref{fig:band_effective}).}
    \label{fig:band_12_12}
\end{figure}

\indent Here, we present the triplon band structure from the BdG Hamiltonian. First, we show the band structure of $\mathcal{H}^{(2)}$, incorporating all triplon modes, alongside the BdG Hamiltonian, which takes into account only the lowest two modes (described in the $4\times 4$ matrix [Eq. (12) in the main text]) in the absence of an electric field, as illustrated in Fig. \ref{fig:band_12_12}.

Figure \ref{fig:band_12_12} reveals six bands and a branch of the Nambu-Goldstone mode. It is evident that the energy of the lowest two excitation modes and those of the upper four modes are sufficiently separated, and thus our approximation, which takes into account only the lowest two energy modes works well in the high magnetic field regimes.

\begin{figure}[H]
    \centering
    \includegraphics[width=1.0\linewidth]{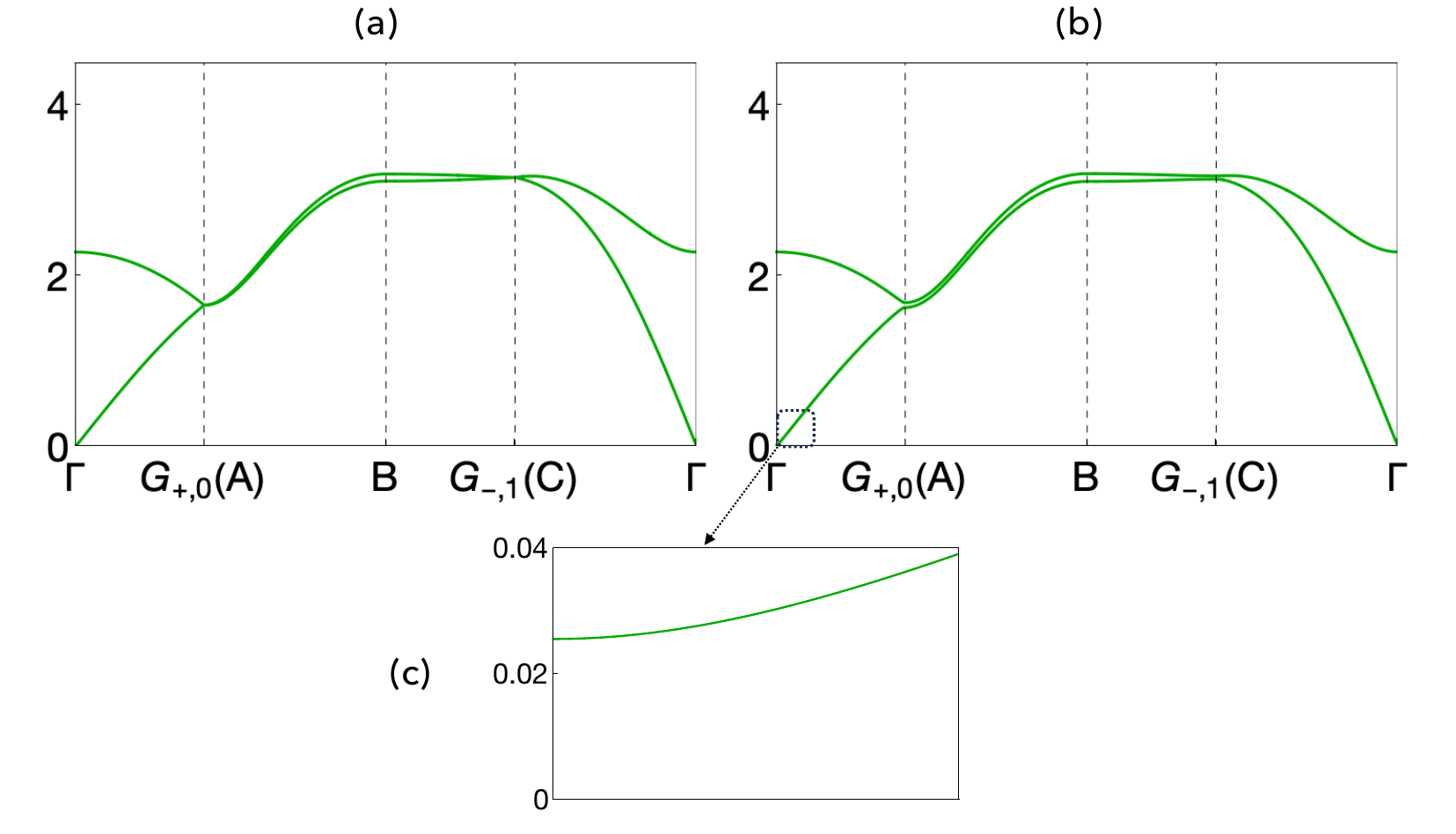}
    \caption{The band structure of the low-energy effective Hamiltonian described in the $4\times 4$ matrix without (a) and with (b) an electric field ($\bm{E} = (0,0.016,1.6)~\si{MV/cm}$). The magnetic field $H$ is $\SI{42}{T}$ in (a) and (b). The energy scale is in $\si{meV}$. The points in reciprocal space are denoted by $\Gamma=(0,0,0)$, $G_{+,0}(\mathrm{A}) = (0,0,\frac{\pi}{2})$, $\mathrm{B}= (\frac{\pi}{2},-\frac{\pi}{2},\frac{\pi}{2})$, and $G_{-,1}(\mathrm{C})=(\frac{\pi}{2},-\frac{\pi}{2},0)$. 
    In (b), the band gaps open at $G_{\pm,1}$. We provide a magnified view of the spectrum near the $\Gamma$ point in (c), which clearly shows that there is a spin gap at $\Gamma$ due to the rotational symmetry breaking in the $x$-$y$ plane by an electric field. 
    }
    \label{fig:band_effective}
\end{figure}

Next, we show the band structure of the BdG Hamiltonian described in the $4\times 4$ matrix [Eq. (12) in the main text] without and with an electric field [see Fig. \ref{fig:band_effective}]. From Fig. \ref{fig:band_effective}, we can find that an electric field opens the band gaps at the nodal lines $G_{\pm,j}$. This is consistent with the discussions based on the symmetry and analytical calculation. See Sec. \ref{sec:absence_thermal_Hall_withoutE}, \ref{sec:Symmetry_nodal_lines}, and \ref{sec:E_dependence} for more details.

\section{Electric field dependence of the band gap}\label{sec:E_dependence}
\indent Here, we give the derivation of the electric field dependence of $E_{\mathrm{gap},\pm,j}(k_{y})$. The energy of the upper ($n=1$) and the lower ($n=2$) bands $E_{1}$ and $E_{2}$ at $G_{\pm,j}$ can be written as 
\begin{equation}
    \begin{split}\label{Eq:E1E2}
    E_{1}&=\sqrt{\xi^{2}_{0}+\xi^{2}_{3}-\pi^{2}_{0}-\pi^{2}_{3}+2|\xi_{0}\xi_{3}-\pi_{0}\pi_{3}|},\\
    E_{2}&=\sqrt{\xi^{2}_{0}+\xi^{2}_{3}-\pi^{2}_{0}-\pi^{2}_{3}-2|\xi_{0}\xi_{3}-\pi_{0}\pi_{3}|},
    \end{split}
\end{equation}
where $\xi_{0}$, $\xi_{3}$, $\pi_{0}$, and $\pi_{3}$ are defined by
\begin{equation}
    \begin{split}\label{Eq:paulicoefficients}
    \Xi&=\xi_{0} I_{2\times 2}+\xi_{3}\sigma_{3},\\
    \Pi&=\pi_{0} I_{2\times 2}+\pi_{3}\sigma_{3},    
    \end{split}
\end{equation}
where $\Xi$ and $\Pi$ represent the hopping and the pairing terms of the BdG Hamiltonian (\ref{BdGexplicit}) at $G_{\pm,j}$, $I_{2\times 2}$ is the $2\times 2$ identity matrix, and $\sigma_{n}\ (n=1,2,3)$ are Pauli matrices. From the above equations (\ref{Eq:E1E2}) and (\ref{Eq:paulicoefficients}), the band gap $E_{\mathrm{gap},\pm,j}$ is given as
\begin{equation}\label{Eq:Egap}
    E_{\mathrm{gap},\pm,j}(k_{y})=\frac{4|\xi_{0}\xi_{3}-\pi_{0}\pi_{3}|}{E_{1}+E_{2}}.
\end{equation}
Also, approximate expressions for $\xi_{3}$ and $\pi_{3}$ at $G_{\sigma,j}$ are given as 
\begin{equation}
    \begin{split}\label{Eq:xipi}
    \xi_{3}&\simeq v+t_{1}\cos(\frac{j\pi}{2}-\sigma \frac{\pi}{4}+\frac{k_{y}}{2})\sigma t_{2},\\
    \pi_{3}&\simeq u_{1}\cos(\frac{j\pi}{2}-\sigma\frac{\pi}{4}+\frac{k_{y}}{2})\sigma u_{2},
    \end{split}
\end{equation}
where $\sigma = \pm $, $j=0,1$, and 
\begin{equation}
    \begin{split}\label{Eq:vt12}
        v &=-2[[(J_{1}+J_{2})(5\sin\theta\cos\theta(\cos^{2}\theta-\sin^{2}\theta)-2\sin^{3}\theta\cos\theta)-2J_{3}\sin^{3}\theta\cos\theta-4(J-g\mu_{B}H)\sin\theta\cos\theta](\epsilon_{2}-\epsilon_{1})\\ &-\sqrt{2}(D^{\mathrm{ext},2}\sin(\phi_{2}+\Phi_{2})-D^{\mathrm{ext},1}\sin(\phi_{1}+\Phi_{1}))]\sin\theta\cos\theta,\\ 
        t_{1,2} &= 6J_{1,2}(\cos^{2}\theta-\sin^{2}\theta)\sin^{2}\theta\cos^{2}\theta(\epsilon_{2}-\epsilon_{1}),\\
        u_{1,2}&=\frac{3J_{1,2}}{2}\sin{2\theta}\cos{2\theta}.
        \end{split}
\end{equation}
In the above expressions, $\theta$ is the variational parameter in the system without the electric field, which is given in Eq. (\ref{Eq:Variational_withoutE}) and $\epsilon_{m}=\theta_{m}-\theta$ $(m=1,2)$. From Eqs. (\ref{epsilon12})-(\ref{Eq:vt12}), we find that $\xi_{3}$ and $\pi_{3}$ are proportional to $D^{\mathrm{ext},2}\sin(\phi_{2}+\Phi_{2})-D^{\mathrm{ext},1}\sin(\phi_{1}+\Phi_{1})$. In addition, $\xi_{0}$ and $\pi_{0}$ depend only on $\theta$ when we keep only terms that are zeroth order in $\epsilon_{1}$ and $\epsilon_{2}$. Therefore, from Eqs. (6) in the main text, (\ref{Eq:E1E2}), and (\ref{Eq:Egap}), we can see that $E_{\mathrm{gap},\pm,j}(k_{y})$ is proportional to $|\bm{E}|$ for a fixed direction of the electric field \footnote{We can usually ignore the effect of $|\bm{E}|$ on the phase factor $\phi_{m}+\Phi_{m}$ for a fixed direction of the electric field.}.


\end{document}